\documentclass[aps,pra,twocolumn,showpacs,superscriptaddress,groupedaddress]{revtex4}  
\pdfoutput=1
\usepackage{graphicx}  
\usepackage{dcolumn}   
\usepackage{bm}        
\usepackage{amssymb}   

\usepackage{dsfont}
\usepackage{amsmath}
\usepackage{bbm}
\usepackage{here}
\usepackage{hyperref}
\usepackage{qcircuit}
\usepackage{setspace}

\begin{document}

\newcommand{\ket}[1]{\left| #1 \right\rangle}
\newcommand{\bra}[1]{\left\langle #1 \right|}
\newcommand{\cA}{\mathcal{P}}
\newcommand{\cH}{\mathcal{H}}
\newcommand{\cL}{\mathcal{L}}
\newcommand{\cK}{\mathcal{K}}
\newcommand{\cM}{\mathcal{M}}
\newcommand{\cP}{\mathcal{P}}
\newcommand{\cR}{\mathcal{R}}
\newcommand{\cS}{\mathcal{S}}
\newcommand{\cU}{\mathcal{U}}

\newcommand{\bracket}[2]{\langle {#1}|{#2} \rangle}
\newcommand{\ketbra}[2]{\ket{#1}\negmedspace\bra{#2}}
\newcommand{\id}{\mathbf{1}}


\title{Mesoscopic Spin Systems as Quantum Entanglers}
                            
\author{Maryam Sadat Mirkamali}
 \email{msmirkam@uwaterloo.ca}
\affiliation{Institute for Quantum Computing, University of Waterloo, Waterloo, Ontario  N2L 3G1, Canada}
\affiliation{Department of Physics and Astronomy, University of Waterloo, Waterloo, Ontario N2L 3G1, Canada}
\author{David G. Cory}%
\affiliation{Institute for Quantum Computing, University of Waterloo, Waterloo, Ontario N2L 3G1, Canada}
\affiliation{Department of Chemistry, University of Waterloo, Waterloo, Ontario  N2L 3G1, Canada}
\affiliation{Canadian Institute for Advanced Research, Toronto, Ontario  M5G 1Z8, Canada}
\affiliation{Perimeter Institute for Theoretical Physics, Waterloo, Ontario N2L 2Y5, Canada}

\date{\today}

\begin{abstract}
We quantify the resources required for entangling two uncoupled spin qubits through an intermediate mesoscopic spin system (MSS) by indirect joint measurement. 
Indirect joint measurement benefits from coherent magnification of the target qubits' state in the collective magnetization of the MSS; such that a low-resolution collective measurement on the MSS suffices to prepare post-selected entanglement on the target qubits. 
 A MSS consisting of two non-interacting halves, each coupled to one of the target qubits 
 is identified as a 
   geometry that allows implementing the magnification process with experimentally available control tools. 
It is proved that the requirements on the amplified state of the target qubits and the MSS perfectly map to the specifications of micro-macro entanglement between each target qubit and its nearby half of the MSS. 
In the light of this equivalence, the effects of experimental imperfections are explored; in particular, bipartite entanglement between the target qubits is shown to be robust to imperfect preparation of the MSS. 
Our study provides a new approach for using an intermediate spin system for connecting separated qubits. It also opens a new path in exploring entanglement between microscopic and mesoscopic spin systems.
\end{abstract}

\maketitle

\section{Introduction}
Coherent control via a mesoscopic system is an emerging tool in quantum information processing \cite{Sorensen04,Trifunovic13,Elman17,Benito16,Yang16,Szumniak15,Trifunovic12,Friesen07}.
Using a mesoscopic system to indirectly measure a joint property of two noninteracting qubits through a coarse-grained collective measurement has recently been introduced as a new approach for entangling uncoupled qubits \cite{Mirkamali18}. 
Here, we analyze creating micro-macro entanglement between target spin qubits and a mesoscopic spin system as 
a robust strategy for implementing indirect joint measurement on spin qubits. 
Micro-macro entangled states have two main characteristics, bipartite entanglement between a microscopic system e.g., a qubit and a many-body system e.g., a mesoscopic system and macroscopic distinctness
 between the states of the many-body system that are correlated with different states of the microscopic system \cite{Andersen13,Lvovsky13,Leggett02}. 

Interest in micro-macro entangled states dates back to Schrodinger's well-known thought cat experiment \cite{Schrodinger35} which was designed to formulate fundamental questions such as to what extent the quantum mechanics laws apply? Or what causes quantum to classic transition? \cite{Leggett02,Zurek03}. It took several decades for quantum technology to reach the capability to allow realizing purely quantum correlations at macroscopic scales (of course not as macroscopic as a cat).
 Micro-macro entangled states have been produced with Rydberg atoms as the microscopic system coupled to photons confined in a cavity \cite{Brune96,Haroche98}, transmon qubit coupled to photons in a waveguide cavity resonator \cite{Vlastakis13}, 
path degree of freedom of a single photon and optical coherent states with different phases \cite{Lvovsky13,Bruno13}  
and internal state of trapped ions entangled to their  motional degrees of freedom  \cite{Monroe96,Johnson17}.  
These experiments pave the way for the application of micro-macro entangled states in quantum processing.

Here, we study the requirements for generating micro-macro entangled states between individual spin qubits and mesoscopic spin systems and for using such states to entangle two uncoupled spin qubits by indirect joint measurement.
In particular, we show that with the experimentally available control on the mesoscopic spin system including collective rotations and internal magnetic dipole-dipole interactions, local coupling between a target spin qubit and 
the mesoscopic spin system 
suffices for generating  an extended micro-macro entangled state. 
Moreover, we show 
 micro-macro entangled states facilitate creating post-selected entanglement between uncoupled spin qubits through indirect joint measurement that needs only a coarse-grained collective measurement on the MSS.

Bipartite entanglement between separated qubits is equivalent to quantum state transfer (QST) up to local operations and classical communications \cite{Bennett00,Bose3}. An entangled pair of qubits can be used to transfer a quantum state using quantum teleportation protocols \cite{Bennett93}.
 On the other hand, two separated qubits can be entangled by first entangling one of them with a nearby ancilla qubit with local operations then transferring the state of the ancilla to the second qubit through QST. 
There are 
extensive studies on  QST through a (hypothetical) 1D spin chain
\cite{Bose3,Christandl4,Kay6,Avellino6,Burgarth7, Cappellaro07,Franco8,Ramanathan11,Cappellaro11,Yao11,Ajoy13}. 
These studies usually consider spin preserving interaction Hamiltonians and fully polarized initial state, which allows restricting the dynamics to the first excitation manifold \cite{Bose3,Christandl4,Kay6,Avellino6,Burgarth7}. Nearest-neighbor coupling is also widely assumed, which enables finding analytical solutions through Jordan-Wigner transformation \cite{Jordan28}.
Although these simplified models are very insightful, when it comes to physical systems, such as dipolarly coupled spin systems,  
they do not provide a complete enough description of the dynamics.
Here, we focus on analyzing a fair model of the intermediate MSS. We consider the experimentally available grade-raising Hamiltonian not the spin preserving flip-flop (XY) or Heisenberg Hamiltonian and all-to-all dipolar coupling and not only nearest-neighbor interaction.
Thus the many-body dynamics of the spin system neither is limited to the first excitation manifold nor can be solved analytically. We simulate the dynamics for up to 20 spins and extrapolate the results for larger sizes of the MSS. 
We also do not limit the geometry to a 1D spin chain. In fact, we observe significantly faster responses in higher dimensions. 
Comparing to QST proposals, high fidelity bipartite entanglement between separated qubits is anticipated without
assuming single spin addressability, 
 engineering the interaction between the spins in the chain \cite{Christandl4,Franco8} or adaptive two-qubit gates at the end of the spin chain \cite{Burgarth7}  given that a coarse-grained collective 
 non-destructive magnetization measurement on the MSS is available.
The difference in the requirements is because our approach 
is based on magnification of the state of the target qubits and global measurement of the MSS, compared to directional information transfer from one qubit to the other, needed in the QST procedures.

\section{Statement of the problem}
\label{sec:problem}
Consider two uncoupled spin qubits and an intermediate MSS.
The target spin qubits are spin-half particles that can be initialized, controlled, and measured individually.
The MSS is an ensemble of identical electron spins or spin half nuclei that can be controlled and measured collectively. The spins in the MSS interact with each other according to the two-body magnetic dipole coupling,
\begin{equation}
\label{eq:Hdip}
H_{\text{dip}}=\sum_{i,j;i<j}d_{ij}(2\sigma_z^i\sigma_z^j-\sigma_x^i\sigma_x^j-\sigma_y^i\sigma_y^j)
\end{equation}
 where $\sigma_x,\sigma_y$ and $\sigma_z$ are the Pauli operators and the interaction strength is proportional to the inverse cube of the distance between the spins, $d_{ij}\propto1/|\vec{r}_{ij}|^3$. 
 Each target qubit is locally coupled to the MSS. To be specific, we consider that each qubit, $q_i$,  is interacting with one nearby spin within the MSS, $s_i$,  and universal control over the pair is available.
  An example of such a set-up consists of two nitrogen-vacancy (NV) centers in a diamond 
  as the target qubits and electronic $P_1$ defects in the diamond or 
  nuclear (or electron) spins of phosphorous defects in a silicon lattice attached to the surface of the diamond as the MSS.

\begin{figure}[hbt!]
        \centering
        \includegraphics[scale=0.4]{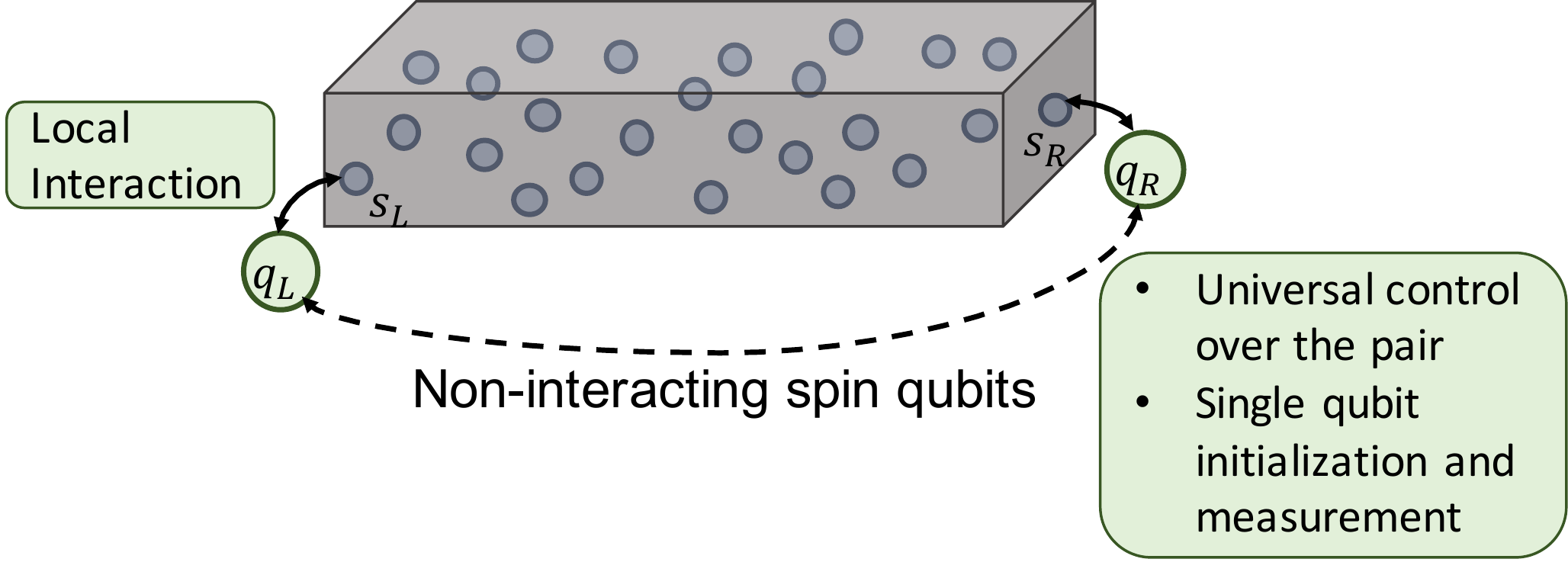}
        \caption{\label{fig:drawing} A schematic of a MSS in local contact with two non-interacting individual qubits}
\end{figure}

The goal is to evaluate the resources required for entangling the target qubits by indirect joint measurement through the MSS. 
The analyzed resources of the MSS are the purity of the initial state, the size, control and internal dynamics, measurement, and robustness to noise.
The approach is to limit the coherent control tools to experimentally available ones, (including collective rotations, internal dipolar interaction among the spins in the MSS and local coupling between each target qubit and the MSS), and find the requirement on the other 
resources.

This paper is organized as follows. In section \ref{sec:Indirect}, the general scheme for entangling two non-interacting spin qubits through indirect joint measurement is reviewed, the role of micro-macro entangled states is highlighted, and the measurement requirements are identified. In section \ref{sec:magnification}, we present a scheme that can generate a mesoscopic superposition state with micro-macro entanglement between a spin qubit and a mesoscopic spin system using experimentally available control including collective rotations and internal dipole-dipole interaction in the MSS and local coupling between the qubit and the MSS.  The scaling of the magnification time with the size of the MSS and its dependency on the geometry and dimension are discussed. In section \ref{sec:measurement}, the entanglement of the target qubits is quantified based on the magnification procedure of section \ref{sec:magnification}
and a general collective measurement through a two-level apparatus. In sections \ref{sec:Mixed} and \ref{sec:noise}, the sensitivity of the scheme 
 to limited initial polarization of the MSS and particle loss is analyzed. In particular, it is shown that limited initial polarization can be compensated for by enlarging the MSS.
We summarize the required resources for entangling two uncoupled spin qubits through a MSS 
and conclude the paper in section \ref{sec:conclusion}.

\section{Indirect joint measurement}
\label{sec:Indirect}
\begin{figure}[ht]
\centerline{
\Qcircuit @ C=1.2em @ R=0.2em {
&\lstick {q_L, \ket{0}}  & \gate{H}&\multigate{8}{\rule{0.4em}{0em}U_{q,\text{MSS}}\rule{0.4em}{0em}} \ar@{.}[]+<2.8em,0.5em>;[d]+<2.8em,-5.5em> & \qw & \multigate{8}{\rule{0.4em}{0em}(U^{\dagger}_{q,\text{MSS}})\rule{0.4em}{0em}} & \qw \\
&& & & & & \\
&& \qw & \ghost{\rule{0.4em}{0em}U_{q,\text{MSS}}\rule{0.4em}{0em}} & \multimeasureD {4}{M} & \ghost{\rule{0.4em}{0em}(U^{\dagger}_{q,\text{MSS}})\rule{0.4em}{0em}}& \qw \\
\push{\rule{2em}{0em}} 
&\lstick {\text{MSS}, \hspace{0.1cm}} & \qw & \ghost{\rule{0.4em}{0em}U_{q,\text{MSS}} \rule{0.4em}{0em}} & \ghost{M} & \ghost{\rule{0.4em}{0em}(U^{\dagger}_{q,\text{MSS}})\rule{0.4em}{0em}} & \qw \\
\push{\rule{2em}{0em}} 
& \lstick{\rho_{in}\hspace{0.1cm}}& \vdots & & & &\\
 && & & & & \\
 & &\qw & \ghost{\rule{0.4em}{0em}U_{q,\text{MSS}}\rule{0.4em}{0em}} & \ghost{M} & \ghost{\rule{0.4em}{0em}(U^{\dagger}_{q,\text{MSS}})\rule{0.4em}{0em}} & \qw \\
& && & & & & \\
&\lstick {q_R, \ket{0}} &\gate{H}& \ghost{\rule{0.4em}{0em}U_{q,\text{MSS}}\rule{0.4em}{0em}} & \qw  & \ghost{\rule{0.4em}{0em}(U^{\dagger}_{q,\text{MSS}})\rule{0.4em}{0em}} & \qw
\gategroup{3}{2}{7}{2}{.8em}{\{}
\gategroup{3}{7}{7}{7}{.8em}{\}}\\
&&&&\hspace{.5em}\ket{\psi}_{q,\text{MSS}} &&
}
}
\caption{\label{fig:general} The general circuit of indirect joint measurement on two separated qubits through an intermediate MSS. \cite{Mirkamali18} 
}
\end{figure}
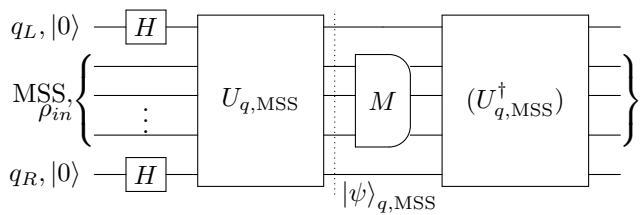

Two non-interacting qubits can be entangled either by creating an indirect interaction between them or by projectively measuring a joint property of them. 
Measuring the parity of two qubits each prepared in a superposition state, $\ket{\pm}=\frac{1}{\sqrt{2}}\left(\ket{0}\pm\ket{1}\right)$ projects their state into a maximally entangled state with odd, $\ket{o_{\pm}}=\frac{1}{\sqrt{2}}\left(\ket{01}\pm\ket{10}\right)$, or even, $\ket{e_{\pm}}=\frac{1}{\sqrt{2}}\left(\ket{00}\pm\ket{11}\right)$, parity.  Similarly, total magnetization measurement of the qubits projects their state into the maximally entangled state $\ket{m_0}=\frac{1}{\sqrt{2}}\left(\ket{01}\pm\ket{10}\right)$ or separable states $\ket{m_{-1}}=\ket{11}$ and $\ket{m_{+1}}=\ket{00}$ with the probabilities of $\frac{1}{2}, \frac{1}{4}$ and $\frac{1}{4}$, respectively. Here, the qubit states $\ket{0}$ and $\ket{1}$ represent the spin states  $\ket{\uparrow}$ and  $\ket{\downarrow}$. Entangling two spin qubits by projective measurement needs a very high-resolution joint measurement able to detect a single spin flip. 
Indirect joint measurement through a MSS relaxes this criterion by first coherently amplifying the state of the target qubits in the collective magnetization of the MSS along a known direction (called $z$) (gate $U_{q,\text{MSS}}$ in FIG. \ref{fig:general}), then measuring the MSS by a coarse-grained collective magnetization measurement that is capable to detect only many spin flips (operation $M$ in FIG. \ref{fig:general}). 

The coherent magnification process, represented by gate $U_{q,\text{MSS}}$ in FIG. \ref{fig:general}, changes the state of the MSS conditioned on the state of the target qubits, $U_{q,\text{MSS}}=\ketbra{00}{00}_q \otimes U_{00}+\ketbra{01}{01}_q \otimes U_{01}+\ketbra{10}{10}_q \otimes U_{10}+\ketbra{11}{11}_q \otimes U_{11}$. With a pure initial state over the MSS, $\ket{\psi_{in}}$, the state of the qubits and the MSS after applying this gate is,
\begin{eqnarray}
\label{eq:QMSS1}
\ket{\psi}_{q,\text{MSS}}&=&\frac{1}{2}\left(\ket{00}_{q}\ket{\psi_{00}}+\ket{01}_{q} \ket{\psi_{01}}\right. \nonumber\\
&+& \left. \ket{10}_{q}\ket{\psi_{10}}+\ket{11}_{q}\ket{\psi_{11}}\right) 
\end{eqnarray}
where $\ket{\psi_{ij}}=U_{ij}\ket{\psi_{in}}$, for $i,j=0,1$. 
To indirectly measure the joint magnetization of the target qubits, the collective coarse-grained magnetization measurement over the MSS needs to distinguish the pair of states $\{\ket{\psi_{01}},\ket{\psi_{10}}\}$ from the pair $\{\ket{\psi_{00}},\ket{\psi_{11}}\}$ but should not discern between the states $\ket{\psi_{01}}$ and $\ket{\psi_{10}}$. With these criteria, the state of the qubits and the MSS after the measurement and post-selection ideally is,
\begin{equation}
\label{eq:QMSS-measured0}
\ket{\psi^{m_0}}_{q,\text{MSS}}=\frac{1}{\sqrt{2}}\left(\ket{01}_q \ket{\psi_{01}}+\ket{10}_q\ket{\psi_{10}}\right)
\end{equation}  
In general the states $\ket{\psi_{01}}$ and $\ket{\psi_{10}}$ are not equal, thus $\ket{\psi^{m_0}}_{q,\text{MSS}}$ is an entangled state between the target qubits and the MSS.
To prepare the target qubits in the maximally entangled triplet zero state, $\ket{m_0}=\frac{1}{\sqrt{2}}(\ket{01} +\ket{10})$, 
 they need to be disentangled from the MSS by undoing the magnification step (gate $U^{\dagger}_{q,\text{MSS}}$ in FIG. \ref{fig:general}) \cite{Mirkamali18}. 
 In the quantum eraser language, the MSS is like a tagging particle and the target qubits' entanglement needs to be restored similar to the reversible eraser scheme \cite{Garisto99}.

\subsection{Micro-macro Entanglement}
\label{sec:mic-mac}
 With the experimentally available control tools, an interesting and potentially implementable geometry consists of a MSS with a barrier in the middle; such that there is no internal interaction, and thus no flow of information, between the two sides of the barrier. The state of each target qubit is magnified in the collective state of its nearby side. However, the collective measurement is implemented on the whole MSS.
 
 Here, we show that, within this geometry, the conditions on the magnified state of the qubits and the MSS entirely maps to the specifications of micro-macro entangled states between each target qubit and its nearby half of the MSS. In section \ref{sec:magnification}, we will show that creating  micro-macro entanglement between each target qubit and half the MSS needs only experimentally available control tools including 
local interaction between the qubit and the MSS, collective rotations on the MSS and internal magnetic dipole interaction between the spins in the MSS.
  
\begin{figure}[t!h]
    \centerline{
        \Qcircuit @ C=1.5em @ R=0.4em {
   &&\hspace{3.0em}\frac{\ket{0}+\ket{1}}{\sqrt{2}}&&\hspace{.5em}\ket{\phi}_{q_L,\text{MSS}_L} &&\\
      &  \lstick{q_L,\ket{0}}  
         & \gate{H}&  \multigate{5}{U_{L}}  \ar@{.}[]+<1.5em,1.0em>;[d]+<1.5em,-4.0em>& \qw & \multigate{5}{U^{\dagger}_L}&\qw \\
      &  & \qw & \ghost{U_{L}}& \multimeasureD{10}{M} &\ghost{U^{\dagger}_L}  &\qw \\
\push{\rule{4em}{0em}} &     \lstick {\text{MSS}_L,\rho_{in}^{L} \hspace{0.2cm}}   
     &\qw & \ghost{U_{L}}& \ghost{M} &\ghost{U^{\dagger}_{L}}  &\qw \\
       && \vdots & &  & & \\
         && & & & & \\
         & &\qw &\ghost{U_{R}}&  \ghost{M} &\ghost{U^{\dagger}_{R}} &\qw \\       
       &&&&&&\\
         & &\qw & \multigate{5}{U_{R}}\ar@{.}[]+<1.5em,0.4em>;[d]+<1.5em,-5.0em>& \ghost{M}& \multigate{5}{U^{\dagger}_{R}}   &\qw \\
           && \vdots & & & & \\
  \push{\rule{4em}{0em}} &         \lstick {\text{MSS}_R,\rho_{in}^{R} \hspace{0.2cm}}
          & &&&  & & \\
        & &\qw & \ghost{U_{R}}&  \ghost{M} &\ghost{U^{\dagger}_{R}}&\qw \\
         && \qw & \ghost{U_{R}}& \ghost{M} &\ghost{U^{\dagger}_{R}}  &\qw \\
   &\lstick{q_R,\ket{0}} & \gate{H} & \ghost{U_{R}}& \qw & \ghost{U^{\dagger}_{R}} &\qw  
   \gategroup{3}{2}{7}{2}{1 em}{\{} 
   \gategroup{9}{2}{13}{2}{1 em}{\{}
   \gategroup{3}{7}{13}{7}{.8em}{\}} \\
   &&&&&& \\
   &&\hspace{3.5em}\frac{\ket{0}+\ket{1}}{\sqrt{2}}&&\hspace{.5em}\ket{\phi}_{q_R,\text{MSS}_R} &&\\
   }
   }     
    \caption{\label{fig:TWOMSS} Indirect joint measurement with a MSS consisting of two non-interacting halves}
\end{figure}
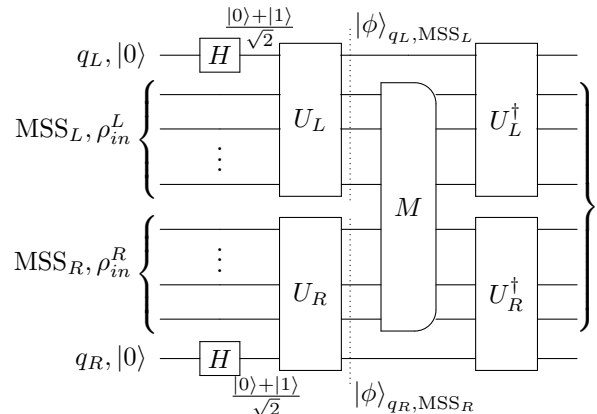

\begin{figure}[h]
        \centering
        \includegraphics[scale=0.45]{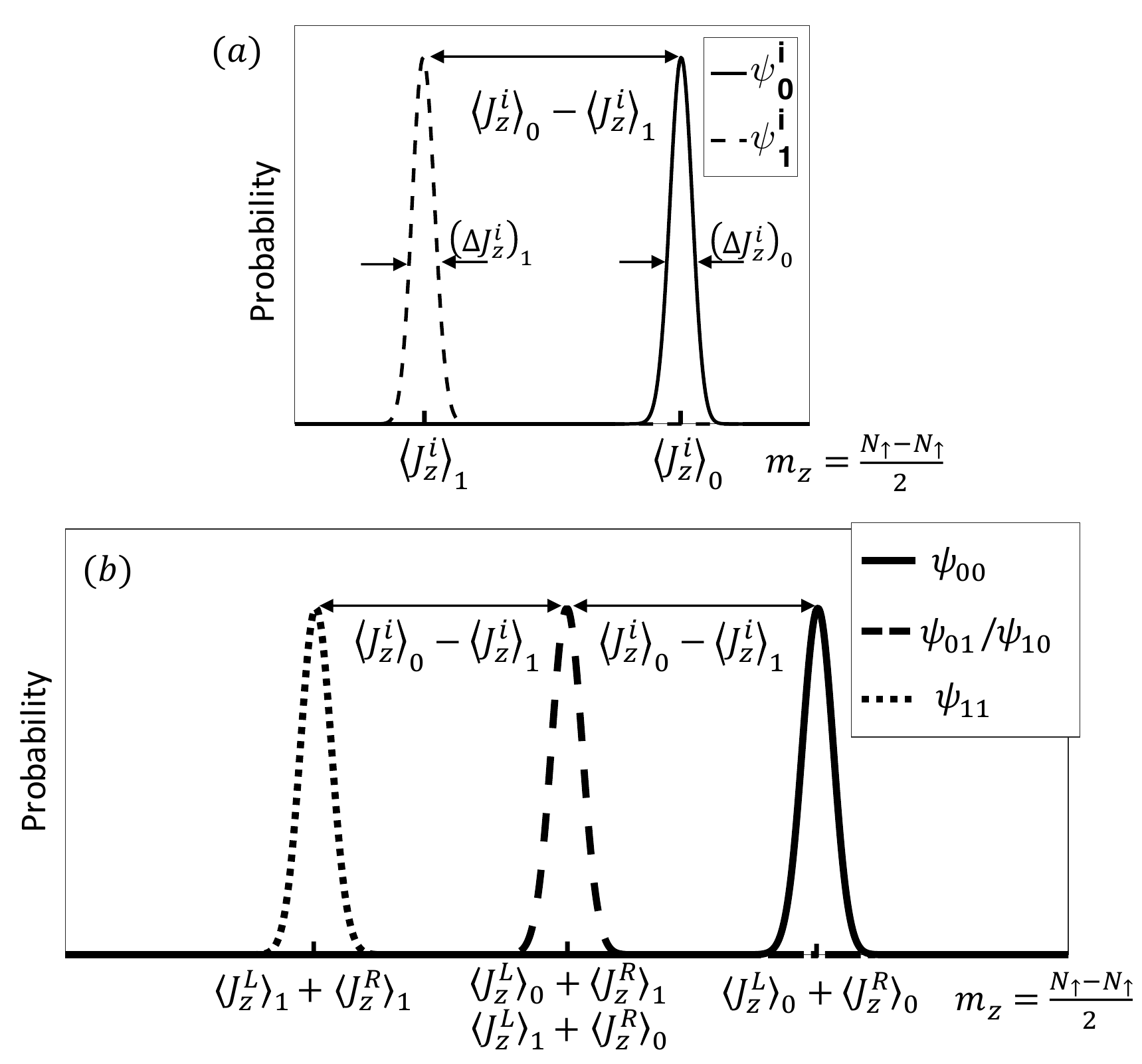}
        \caption{\label{fig:SeparatedSpectraBoth} Magnetization spectrum of (a) An example of two macroscopically distinct states of half of the MSS and (b) the corresponding states of the whole MSS.}
\end{figure}

The general circuit for this geometry, depicted in FIG. \ref{fig:TWOMSS}, is a subset of the generic indirect joint measurement circuit in FIG. \ref{fig:general}, in which the magnification gate is decomposed into two parts $U_{q,\text{MSS}}=U_L\otimes U_R$, each being a conditional gate on half of the MSS controlled by its nearby target qubit, 
\begin{equation}
U_{i}=\ketbra{0}{0}_{q_i}\otimes U_{0}^{\text{MSS}_i}+\ketbra{1}{1}_{q_i}\otimes U_{1}^{\text{MSS}_i}, i=L,R
\end{equation} 
With an ideal pure separable initial state of the MSS, $\ket{\psi_{in}}=\ket{\psi_{in}^L}\otimes \ket{\psi_{in}^R}$, the general state of each target qubit and its nearby half of the MSS after applying this gate and before the measurement is,
\small
\begin{equation}
\label{eq:micmac}
\ket{\phi}_{q_i,\text{MSS}_i}=\frac{1}{\sqrt{2}}\left(\ket{0}_{q_i}\ket{\psi_0^{i}}+\ket{1}_{q_i}\ket{\psi_1^{i}}\right),  i=L,R
\end{equation}
\normalsize
This state is a micro-macro entangled state if  $\ket{\psi_0^{i}}$ and $\ket{\psi_1^{i}}$ are orthogonal and macroscopically distinct i.e., distinguishable by a coarse-grained collective measurement \cite{Andersen13,Lvovsky13,Leggett02}. 
Macroscopic distinctness between the states $\ket{\psi_0^{i}}$ and $\ket{\psi_1^{i}}$ mathematically means that the difference in the expectation value of a particular 
collective observable e.g. the collective magnetization along z, $J_z=\sum_{j}\sigma_z^j$, for these two states is large compared both to the quanta of the collective observable (e.g. $\hbar$ for collective magnetization) and to the sum of their standard deviation \cite{Andersen13,Lvovsky13,Leggett02}.,
 \begin{eqnarray}
 \label{eq:macroscopicDis}
 \dfrac{|\left\langle J_z^{i}  \right\rangle_0- \left\langle J_z^{i}  \right\rangle_1|}{\max((\delta J_z^{i})_0+(\delta J_z^{i})_1,\hbar)}\gg 1
 \end{eqnarray}
 Taking the maximum between $(\delta J_z^{i})_0+(\delta J_z^{i})_1$ and $\hbar$ ensures meaningful answer when both $(\delta J_z^{i})_0$ and $(\delta J_z^{i})_1$ are zero.   
In addition, to effectively use all the spins in the MSS,
 the difference in the expectation value of the collective magnetization observable preferably should be proportional to the size of the MSS,
\begin{equation}
 \left\langle J_z^{i} \right\rangle_0- \left\langle J_z^{i}  \right\rangle_1\propto N
\end{equation}
The collective magnetization observable, $J_z$, for $N$ spins follows the spectral decomposition,
\begin{equation}
J_z=\sum_{m_z=-\frac{N}{2}}^{\frac{N}{2}} m_z \Pi^{N}(m_z)
\end{equation}
where the operator $\Pi^{N}(m_z)$ projects onto the subspace with total magnetization of $m_z$ and $\hbar$ is set to one. The magnetization spectrum of an arbitrary state, $\ket{\phi}$, is,
\begin{equation}
P_{\phi}(m_z)=\text{Tr}(\Pi^{N}(m_z)\ketbra{\phi}{\phi})
\end{equation}
Macroscopic distinctness between the states $\ket{\psi_0^{i}}$ and $\ket{\psi_1^{i}}$ requires  them to have well separated magnetization spectra, as depicted in 
FIG. \ref{fig:SeparatedSpectraBoth}a.

The states of the whole MSS associated with different states of the target qubits are, $\ket{\psi_{00}}=\ket{\psi_0^{L}}\ket{\psi_0^{R}}$, $\ket{\psi_{01}}=\ket{\psi_0^{L}}\ket{\psi_1^{R}}$, $\ket{\psi_{10}}=\ket{\psi_1^{L}}\ket{\psi_0^{R}}$ and $\ket{\psi_{11}}=\ket{\psi_1^{L}}\ket{\psi_1^{R}}$.
To implement indirect magnetization measurement on the qubits,
the states $\{\ket{\psi_{10}},\ket{\psi_{10}}\}$ not only need to be orthogonal to the states $\{\ket{\psi_{00}},\ket{\psi_{11}}\}$ but also must be distinguishable from them by a coarse-grained collective magnetization measurement. In addition, the states $\ket{\psi_{10}}$ and $\ket{\psi_{10}}$ must not be distinguishable from each other \cite{Mirkamali18}. 
These three conditions are satisfied if and only if each qubit and its nearby half of the MSS are prepared in (similar) micro-macro entangled states.

The pair of the states  $\{\ket{\psi_{10}},\ket{\psi_{10}}\}$ and  $\{\ket{\psi_{00}},\ket{\psi_{11}}\}$ are orthogonal if and only if the states $\ket{\psi_0^{i}}$ and $\ket{\psi_1^{i}}$ are orthogonal to each other for $i=L,R$.
The second criterion requires macroscopic distinctness between the states $\ket{\psi_0^{i}}$ and $\ket{\psi_1^{i}}$.  
The magnetization spectra of the states $\ket{\psi_{00}}$, $\ket{\psi_{01}}$, $\ket{\psi_{10}}$ and $\ket{\psi_{11}}$ of the whole MSS
 are the convolution of the spectra of the corresponding states of the halves. 
 Consequently, their means and variances are sum of the means and variances of the spectra of the corresponding states of the halves,
\begin{eqnarray}
\label{eq:conv}
\left\langle J_z \right\rangle_{kj}&=&\left\langle J_z^{L}\right\rangle_{k}+\left\langle J_z^R \right\rangle_{j} \\
(\delta J_z)_{kj}&=&\sqrt{((\delta J_z^L)_{k})^2+((\delta J_z^R)_{j})^2} \nonumber
\end{eqnarray} 
   for $k,j=0,1$. 
 The equivalence between macroscopic distinctness between the states $\ket{\psi_0^{i}}$ and $\ket{\psi_1^{i}}$ and the distinguishablity between the pairs $\{\ket{\psi_{01}},\ket{\psi_{10}}\}$ and $\{\ket{\psi_{00}},\ket{\psi_{11}}\}$ follows from these relations.
 The distinguishablity between $\{\ket{\psi_{01}},\ket{\psi_{10}}\}$ and $\{\ket{\psi_{00}},\ket{\psi_{11}}\}$, by a course-grained collective magnetization measurement along a particular axis e.g., $z$-axis, requires that,
 \begin{eqnarray}
|\left\langle J_z \right\rangle_{01}-\left\langle J_z \right\rangle_{00}|\gg (\delta J_z)_{01}+(\delta J_z)_{00} \nonumber \\
|\left\langle J_z \right\rangle_{01}-\left\langle J_z \right\rangle_{11}|\gg (\delta J_z)_{01}+(\delta J_z)_{11} \nonumber \\
|\left\langle J_z \right\rangle_{10}-\left\langle J_z \right\rangle_{00}|\gg (\delta J_z)_{10}+(\delta J_z)_{00} \nonumber \\
|\left\langle J_z \right\rangle_{10}-\left\langle J_z \right\rangle_{11}|\gg (\delta J_z)_{10}+(\delta J_z)_{11}
\end{eqnarray}
Replacing the means and the standard deviations according to Eq. (\ref{eq:conv}) and
 assuming that the two target qubits and their nearby sides of the MSS are prepared in similar states i.e., $\left\langle J_z^R \right\rangle_{k}\approx\left\langle J_z^L \right\rangle_{k}$ and $(\delta J_z^L)_{k}\approx(\delta J_z^R)_{k}$, 
the above conditions are met if \footnote{For any $a\geq 0$ and $b\geq 0$, $\sqrt{a^2+b^2}\leq a+b$},
\begin{eqnarray}
|\left\langle J_z^i \right\rangle_{1}-\left\langle J_z^i \right\rangle_{0}|\gg(\delta J_z^i)_{1} +(1+\sqrt{2})(\delta J_z^i)_{0}&&\nonumber \\
|\left\langle J_z^i \right\rangle_{1}-\left\langle J_z^i \right\rangle_{0}|\gg(\delta J_z^i)_{0} +(1+\sqrt{2})(\delta J_z^i)_{1}&&
\end{eqnarray}
for $i=L,R$. 
or more simply if, 
\begin{equation}
\label{eq:macDis2}
|\left\langle J_z^i \right\rangle_{1}-\left\langle J_z^i \right\rangle_{0}|\gg (1+\sqrt{2})((\delta J_z^i)_{1} +(\delta J_z^i)_{0})
\end{equation}
which is the same as the macroscopic distinctness condition in Eq. (\ref{eq:macroscopicDis}) up to a small coefficient $(1+\sqrt{2})\approx 2.41$.
Satisfaction of relation (\ref{eq:macDis2}) clearly requires the macroscopic distinctness condition given in Eq. (\ref{eq:macroscopicDis}) to be fulfilled. 
Thus, each target qubit and its nearby half of the MSS need to be in a micro-macro entangled state prior to the  measurement step. 
On the other hand, preparing each qubit and its nearby half of the MSS in similar micro-macro entangled states 
guarantees that the states $\{\ket{\psi_{01}},\ket{\psi_{10}}\}$ have similar magnetization spectra separated from the spectra of the states $\{\ket{\psi_{00}},\ket{\psi_{11}}\}$. Thus, the two pairs can be distinguished by a coarse-grained collective measurement while the states $\ket{\psi_{01}}$ and $\ket{\psi_{10}}$ will not be discerned due to their similar spectra.

\subsection{Coarse-grained Collective Measurement}

The measurement and post-selection must project the state of the qubits into zero magnetization subspace along with the states of the MSS. 
In other words, the collective magnetization measurement on the MSS and post-selection not only need to discern the MSS's states correlated with zero magnetization of the qubits from states associated with $\pm 1$ magnetizations; but also must update the MSS's state according to the measurement outcome, 
with minimum disturbance on the selected states.
The state of the qubits and the MSS after the measurement ideally is,
\begin{equation}
\label{eq:QMSS-measured}
\ket{\psi}_{q,\text{MSS}}=\frac{1}{\sqrt{2}}\left(\ket{01}_q \ket{\psi_0^{L}} \ket{\psi_1^{R}}+\ket{10}_q\ket{\psi_1^{L}}\ket{\psi_0^{R}}\right)
\end{equation}  
 Bipartite entangled state between the qubits separable from the MSS can be created from this state simply by reversing the magnification gate, similar to reversible quantum eraser protocol \cite{Garisto99,Mirkamali18},
\begin{equation}
\ket{\psi}_{q,\text{MSS}}=\frac{1}{\sqrt{2}}\left(\ket{01}_q +\ket{10}_q\right)\otimes \ket{\psi_{in}}
\end{equation}  
The desired coarse-grained collective magnetization measurement is mathematically represented by a Positive-Operator Valued Measure (POVM) with measurement operators, $\{E_{\alpha}\}$, satisfying two conditions: positivity, $E_{\alpha} \geq 0$, and trace-preserving, $\sum_{\alpha} E_{\alpha} = \id$.
Since the measurement is collective, the POVM operators can be expanded in terms of the collective magnetization projection operators, $\Pi^N(m_z)$,
\begin{equation}
\label{eq:POVMexp}
E_{\alpha}=\sum_{m_z} a_{\alpha,m_z} \Pi^N(m_z)
\end{equation}
The expansion coefficients, $a_{\alpha,m_z}$, satisfy two conditions $0\leq a_{\alpha, m_z} \leq 1$ and $\sum_{\alpha} a_{\alpha,m_z} =1$ following the positivity and trace-preserving of the $E_{\alpha}$ operators.
The probability of each measurement outcome, $\alpha$, upon measuring the MSS in a general state $\rho_{\text{MSS}}$ is,
\begin{equation}
P_{\alpha}=\text{Tr}(E_{\alpha}\rho_{\text{MSS}})
\end{equation}
and the state of the MSS after the measurement is,
\begin{equation}
\label{eq:update}
\rho_{\text{MSS},\alpha}=\dfrac{M_{\alpha}\rho_{\text{MSS}}M_{\alpha}^{\dagger}}{P_{\alpha}}
\end{equation}
where the operator $M_{\alpha}$ satisfies the relation $M_{\alpha}M_{\alpha}^{\dagger}=E_{\alpha}$. Following the expansion of $E_{\alpha}$ in Eq. (\ref{eq:POVMexp}), the operators $M_{\alpha}$ are expanded in terms of collective projectors as,
\begin{equation}
\label{eq:MalphaExp}
M_{\alpha}=\sum_{m_z} e^{i\phi_{\alpha,m_z}}\sqrt{a_{\alpha,m_z} }\Pi^N(m_z)
\end{equation}
The phase factor, $e^{i\phi_{\alpha,m_z}}$, depends on the details of the measurement implementation. The operator $M_{\alpha}$ simplifies to $\sqrt{E_{\alpha}}$ if $\phi_{\alpha,m_z}$ 
does not depend on 
$m_z$, $\phi_{\alpha,m_z}:=\phi_{\alpha}$.

The measurement requirements can be specified 
by the necessity that the measurement and post-selection updates the qubits and the MSS's state from the state in Eq. (\ref{eq:QMSS1}) into the state in Eq. (\ref{eq:QMSS-measured0}).
There should exist as least one measurement operator, $M_{\beta}$, that overlaps with the states  $\ket{\psi_0^{L}} \ket{\psi_1^{R}}$ and $\ket{\psi_1^{L}} \ket{\psi_0^{R}}$ but does not overlap with the states $\ket{\psi_0^{L}} \ket{\psi_0^{R}}$ and $\ket{\psi_1^{L}} \ket{\psi_1^{R}}$. Moreover, this measurement operator ideally must preserve the amplitude and the phase of the 
spectral expansion of the states $\ket{\psi_0^{L}} \ket{\psi_1^{R}}$ and $\ket{\psi_1^{L}} \ket{\psi_0^{R}}$ i.e. in the expansion of the measurement operators in Eq. (\ref{eq:MalphaExp}) the amplitudes, $a_{\beta,m_z}$, and the phases, $e^{i\phi_{\beta,m_z}}$, should be equal for all the collective magnetizations that the spectra of the states $\ket{\psi_0^{L}} \ket{\psi_1^{R}}$ and $\ket{\psi_1^{L}} \ket{\psi_0^{R}}$ contain.
The former condition guarantees that $\pm 1$ magnetizations of the target qubits i.e. the states $\ket{00}_q$ and $\ket{11}_q$ are not selected by the measurement and the latter ensures that the coherence between $\ket{01}_q$ and $\ket{10}_q$ states of the qubits can be restored 
 by disentangling the MSS through reversing the magnification gate.

If these two measurement requirements are not perfectly satisfied,  the final entangled state of the target qubits, $\rho_{q}$, deviates from the maximally entangled state $\ket{m_{0}}=\frac{1}{\sqrt{2}}(\ket{01}_q+\ket{10}_q)$. However, $\rho_{q}$ is an entangled state and can be distilled towards the state $\ket{m_{0}}$, if the fidelity defined as the overlap of these two states is greater than $0.5$  \cite{Bennett96PRL,Bennett96PRA},
\begin{equation}
\label{eq:fidelity}
F_{m_{0}}(\rho_{q}):=\text{Tr}(\rho_q \ketbra{m_{0}}{m_{0}})
\end{equation}
Fidelity ranges between $0$ and $1$ and  if $F_{m_{0}}(\rho_{q})>(2+3\sqrt2)/8\approx 0.78$, $\rho_{q}$ is entangled enough to violate Clauser-Horne-Shimony-Holt (CHSH) inequality \cite{Bennett96PRL,CHSH69} \footnote{Note that we use the fidelity, defined in Eq.  (\ref{eq:fidelity}), as a measure for entanglement since we know a priori what the 
 expected maximally entangled state is. However fidelity is not a measure for entanglement in general e.g. the maximally entangled states $\frac{1}{\sqrt{2}}(\ket{00}\pm\ket{11})$ have zero overlap with $\ket{m_0}$ state.}.

Entangling the target spin qubits by first entangling each with the nearby half of MSS and then measuring the whole MSS might remind one of entanglement swapping \cite{swapping93}. One main difference is the measurement process. In the entanglement swapping procedure, measurement of two qubits, each from an entangled pair, in the Bell basis entangles the two other qubits. The analogy in our case is measuring an observable that $\frac{1}{\sqrt{2}}(\ket{\psi_0^{L}} \ket{\psi_1^{R}}\pm\ket{\psi_1^{L}}\ket{\psi_0^{R}})$ and $\frac{1}{\sqrt{2}}(\ket{\psi_0^{L}} \ket{\psi_0^{R}}\pm\ket{\psi_1^{L}}\ket{\psi_1^{R}})$ are four of its eigenstates with different eigenvalues. Such an observable, in general, is not a collective observable in contrast to the observable in the indirect joint measurement procedure.

\section{Creation of micro-macro entanglement}
\label{sec:magnification}
In this section, we discuss producing a mesoscopic superposition state with micro-macro entanglement between one target spin qubit and a MSS, half the size of the whole MSS, as the first step towards implementing indirect joint measurement on two non-interacting target qubits. 
The focus is on using experimentally available control elements namely interaction between the target qubit, $q$, and one nearby spin within the MSS, $s$, collective rotations on the MSS and magnetic dipole-dipole interaction among the spin in the MSS. 
The qubit is prepared in the superposition state, $\ket{+}=\frac{1}{\sqrt{2}}(\ket{0}+\ket{1})$, and the MSS is 
 ideally prepared in the polarized state, $\ket{\uparrow}^{\otimes N_h}$, where $N_h\approx N/2$ is the number of spins in the MSS.

 We start with an intuitive approach 
 based on repetitive application of a conditional local gate 
on the MSS controlled by the qubit and a duration of internal interaction  between the spins of the MSS. 
The internal evolution of the MSS redistributes the magnetization between the spins in the MSS but preserves the total magnetization. The collective magnetization is only changed conditioned on the state of the target qubit.
After enough repetitions, on the order of $N_h$, the states of the MSS correlated with different states of the target qubit become macroscopically distinct and a micro-macro entangled state 
 is produced.

Next we present a different scheme in which the target qubit interacts with the MSS only once. 
The key feature of this approach is that the MSS is prepared in a globally correlated state prior to its interaction with the target qubit such that a local change in the MSS conditioned on the state of the target qubit has a global conditional effect. 
The maximally entangled GHZ state, $\frac{1}{\sqrt{2}}\left(\ket{\uparrow}^{\otimes N_h}+\ket{\downarrow}^{\otimes N_h}\right)$, is an ideal state for this purpose  \cite{Mirkamali18}. However preparing the GHZ state 
is challenging for a mesoscopic size system \footnote{It needs either accessing individual spins in the MSS or synthesizing N-body interaction among all the spins.}; 
 we show that 
micro-macro entanglement between the target qubit and the MSS can be produced by preparing less demanding correlated states, created through the experimentally available two-body dipolar coupling and collective rotations.

After presenting these two approaches, the magnification time and its relation to the size of the MSS and its dimensionality are discussed.

\subsection{Repeated interactions}
\begin{figure}[t!h]
    \centerline{
        \Qcircuit @ C=1.5em @ R=0.4em {
         & &  &  \mbox{$U_{XY}$} & \mbox{$\times r\hspace{.1cm}$}  \\
      \lstick{q,\ket{+}}  & \qw  & \ctrl{1} & \qw & \qw \\
        \lstick{s}  & \qw &  \targ & \multigate{4}{H_{XY},dt}& \qw  \\
    \lstick {\text{MSS},\ket{\uparrow}^{\otimes N_h}\hspace{0.1cm}}  & \qw & \qw & \ghost{H_{XY},dt} & \qw \\
        & \vdots &   & &  \\
        & & & &  \\
         & \qw & \qw & \ghost{H_{XY},dt}&\qw      
       \gategroup{3}{1}{7}{1}{0.8em}{\{}
       \gategroup{2}{3}{7}{3}{1.0em}{(}
              \gategroup{2}{4}{7}{4}{1em}{)}
        }  
        }
    \caption{\label{fig:HXY} Magnification process based on repetitive interaction between the external qubit and its nearby spin from the MSS intervened by internal evolution of the MSS under the magnetization preserving 
    $H_{XY}$ Hamiltonian}
\end{figure}
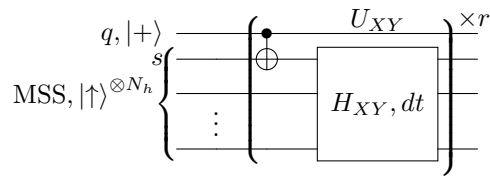

The circuit in FIG. \ref{fig:HXY} shows 
an intuitive approach for making a macroscopic global change in the collective magnetization of the MSS conditioned on the state of the qubit using only local interactions between the two. 
The CNOT gate, 
controlled by the qubit, $q$, on its nearby spin within the MSS, $s$, $CNOT=\ketbra{0}{0}_q \otimes \id_s + \ketbra{1}{1}_q \otimes \sigma_x^s$, changes the magnetization of the MSS locally conditioned on the state of the qubit and evolving under zero-quantum flip-flop Hamiltonian, 
\begin{equation}
\label{eq:HXY}
H_{XY}=\sum_{i,j; i< j} a_{ij} \left(\sigma_{+}^i \sigma_{-}^j+ \sigma_{-}^i \sigma_{+}^j\right) \hspace{0.3cm} a_{ij}\propto \frac{1}{|\vec{r}_{ij}|^3}
\end{equation}
passes this change to the rest of the spins in the MSS while preserving the total magnetization.
These two processes are repeated $r$ times to create a macroscopic effect. The Hamiltonian $H_{XY}$ is widely used in QST proposals usually with only nearest-neighbor interactions, $a_{ij}=0$ for $|i-j|\neq 1$. Here we consider all-to-all interactions with the coefficients $a_{ij}$ proportional to inverse cube of the distance between the two spins, consistent with the magnetic dipolar interaction among the spins in the MSS. The important feature of $H_{XY}$ is that it only redistributes the magnetization among the spins while preserving the collective magnetization of the MSS. The collective magnetization of the MSS varies only conditioned on the state of the qubit by the CNOT gate. 
Hence, in each repetition, the total magnetization of the MSS either is preserved 
 or varies by $\delta m_z\in [-1,1]$, depending on the qubit's state. 

\begin{figure}[t!h]
\centering
         \includegraphics[scale=0.6]{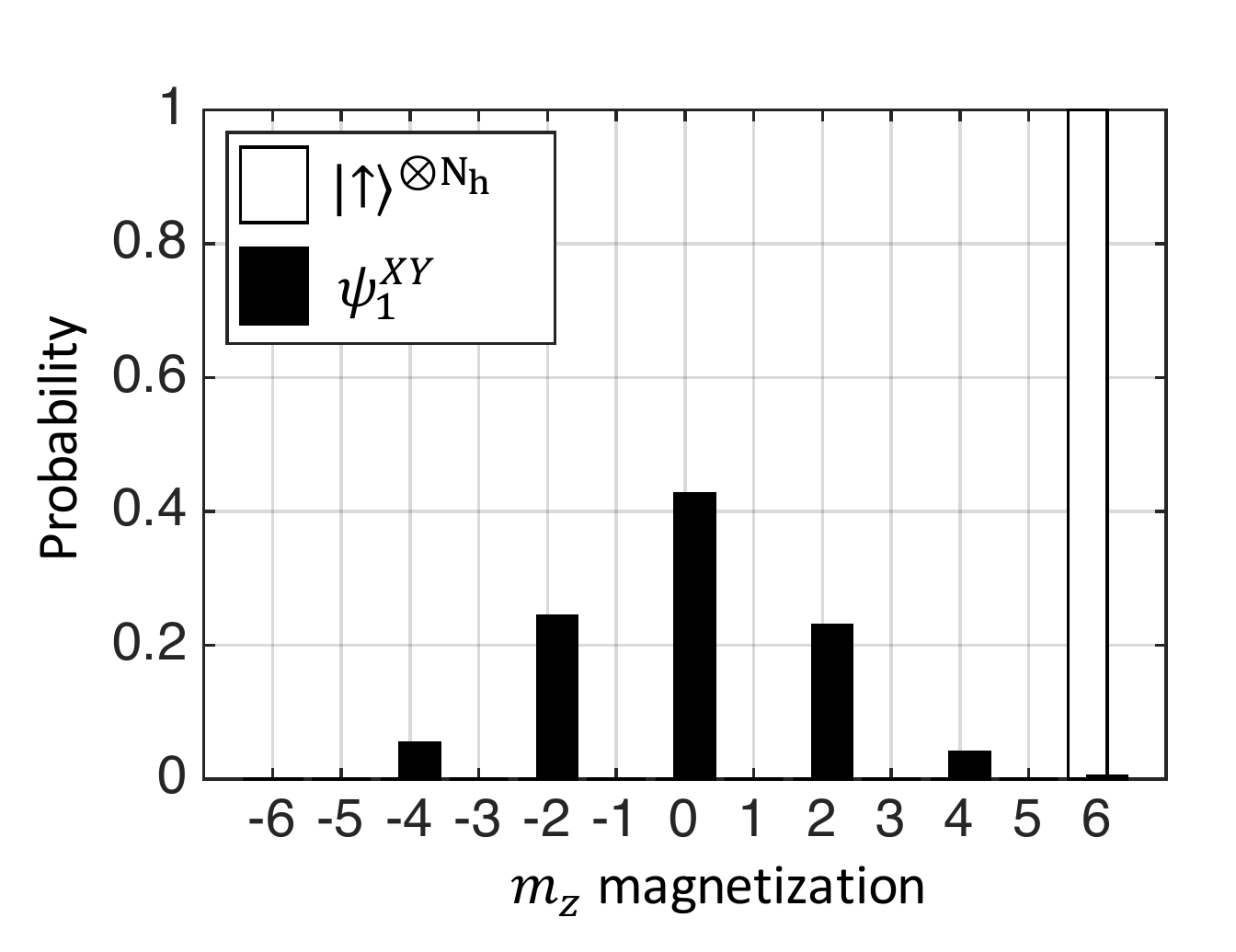}
                 \caption{\label{fig:SpecXY} The distinct magnetization spectra of the MSS's states, $\ket{\uparrow}^{\otimes N_h}$ and $\ket{\psi_1^{XY}}$ correlated to $\ket{0}$ and $\ket{1}$ states of the target qubit simulated based on the circuit in FIG. \ref{fig:HXY} with $dt=\pi/a_{12}$ and $r=2N_h$ for $N_h=12$ spins in a 1D chain geometry.}    
\end{figure}

With an initial superposition state of the qubit, $\ket{+}=\frac{1}{\sqrt{2}}(\ket{0}+\ket{1})$, and polarized state of the MSS, $\ket{\uparrow}^{\otimes N_h}$, the general output state for the circuit \ref{fig:HXY} is,
\small
\begin{equation}
\label{eq:stateHXY}
\ket{\phi^{XY}(dt,r)}_{q,\text{MSS}}=\frac{1}{\sqrt{2}}\left(\ket{0}_q \ket{\uparrow}^{\otimes N_h}+\ket{1}_q\ket{\psi_1^{XY}(dt,r)}\right).
\end{equation}
\normalsize
For appropriate choice of the evolution time, $dt$, and large enough repetitions, $r\propto N_h$, the state $\ket{\psi_1^{XY}(dt,r)}$ is macroscopically distinct from the state $\ket{\uparrow}^{\otimes N_h}$                  
upon collective magnetization measurement along $z$.
Figure \ref{fig:SpecXY} shows the simulation results of the magnetization spectra of these 
two states for a MSS in a 1D spin chain geometry 
with number of repetitions $r=2N_h$ and $dt=\pi/a_{12}$, where $a_{12}=a_{i i+1}$ is the nearest neighbor interaction strength of the Hamiltonian in Eq. (\ref{eq:HXY}) \footnote{All the numerical simulations are conducted using the open source "Expokit" software package \cite{Sidje98}}.
The spectrum of the polarized state $\ket{\uparrow}^{\otimes N_h}$ is a peak at $m_z=N_h/2$; whereas, the spectrum of the state $\ket{\psi_1^{XY}}$ is distributed around $m_z=0$ and has nonzero values for $m_z=N_h/2, N_h/2-2, ..., -N_h/2$. 

\begin{figure}[t!h]
\centering
         \includegraphics[scale=0.7]{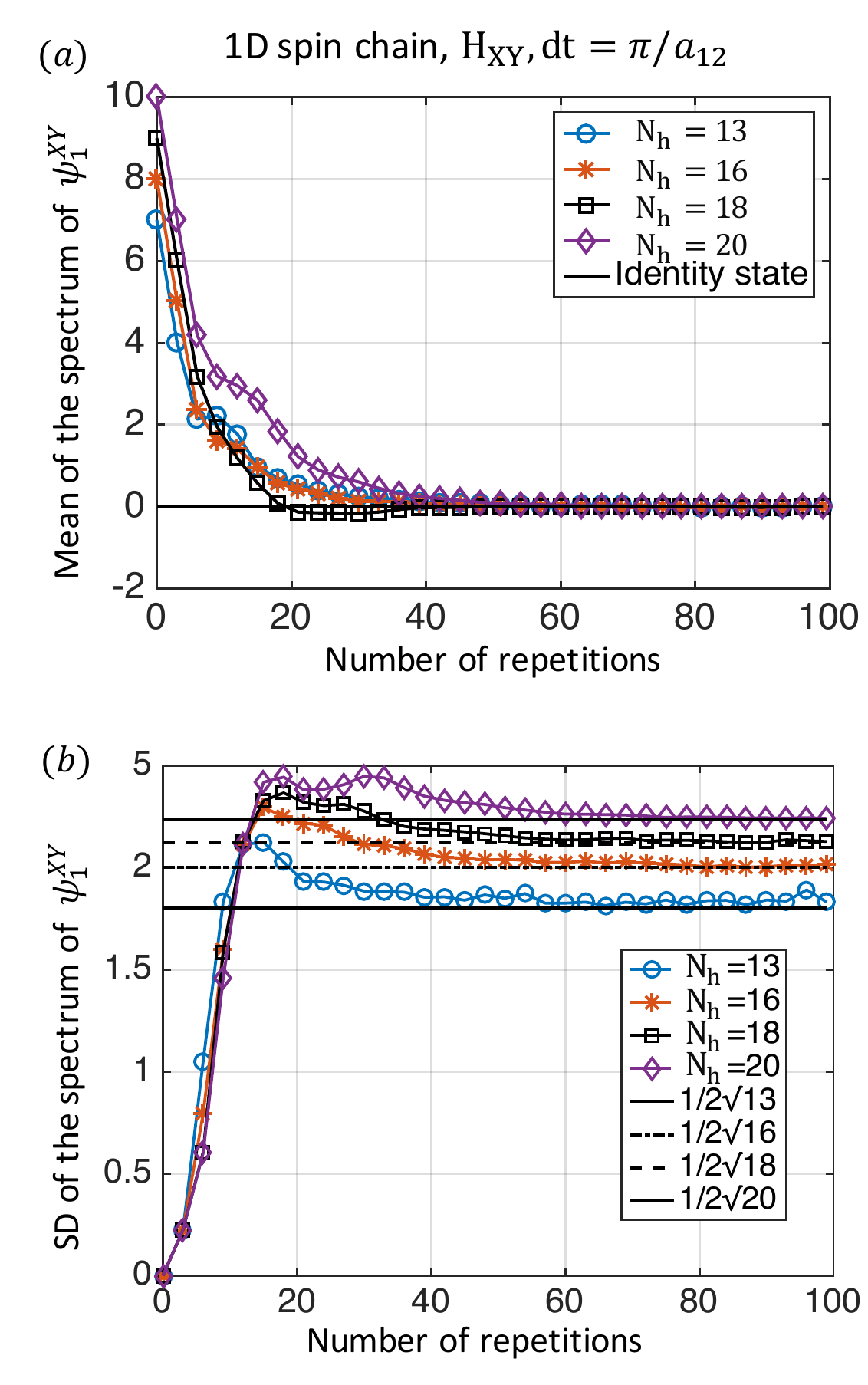}
                 \caption{\label{fig:MSDXY} (a) The mean and (b) the standard deviation (SD) of the magnetization spectrum of  $\ket{\psi_1^{XY}}$ as a function of the number of repetitions for $N_h=13, 15, 16, 18$ qubits. After a transient time both the mean and the SD of the spectrum approach those of the identity state with the same size.}    
\end{figure}

 To characterize the spectrum of the state $\ket{\psi_1^{XY}}$, we simulate the mean and the standard deviation (SD) of its distribution as a function of the number of repetitions, $r$, with $dt=\pi/a_{12}$ for up to $N_h=20$ spins. As FIG. \ref{fig:MSDXY} shows, after a transient time 
 the mean of the spectrum of $\ket{\psi_1^{XY}}$  approaches zero and its SD approaches $\sqrt{N_h}/2$, which are the same as the mean and the SD of a fully mixed state with $N_h$ spins, $(\id_2/2)^{\otimes N_h}$, or an equal superposition state, $\left(\left(\ket{0}+\ket{1}\right)/\sqrt{2}\right)^{\otimes N_h}$ \footnote{One difference is that the spectrum of the state $\ket{\psi_1^{XY}(t)}$ 
 has nonzero values for every other magnetization 
 whereas the spectrum of a fully mixed state or $\left(\left(\ket{0}+\ket{1}\right)/\sqrt{2}\right)^{\otimes N_h}$ state includes all magnetization. But what is important is the extend of the two spectra which is quantified by their mean and SD and is similar for the two cases}. 
 This result can be extrapolated to larger systems; 
 the mean and the SD of the spectrum of $\ket{\psi_1^{XY}}$ are expected to be $\approx 0$ and $\approx\sqrt{N_h}/2$, respectively.
On the other hand, the spectrum of $\ket{\uparrow}^{\otimes N_h}$ is focussed at $N_h/2$. 
As a result, the 
 macroscopic distinctness between the states  $\ket{\psi_1^{XY}}$ and $\ket{\uparrow}^{\otimes N_h}$ 
 scales as $\sqrt{N_h}$, 
\begin{equation}
\dfrac{\left\langle J_z\right\rangle_0^{XY}- \left\langle J_z \right\rangle_1^{XY} }{(\delta J_z)_0^{XY}+(\delta J_z)_1^{XY}}\approx \dfrac{N_h/2-0}{0+\sqrt{N_h}/2}\propto \sqrt{N_h}
\end{equation}
It should be mentioned that the two states are not necessarily orthogonal; nevertheless, for the proper choices of $dt$ and $r$ their overlap is small.
Thus, for a large enough MSS, $\sqrt{N_h}\gg 1$,  and with appropriate $dt$ and $r$ the state in Eq. (\ref{eq:stateHXY}) is a micro-macro entangled state.

The introduced procedure provides a reasonable process for creating a  
micro-macro entangled state using only local interactions. However, it is hard to implement experimentally in a spin system with dipolar coupling. It needs the XY Hamiltonian which can not be synthesized out of the natural dipole-dipole interaction using collective rotations \footnote{Synthesis of XY Hamiltonian out of dipolar coupling needs $\pi-$pulses on every other qubit \cite{Cappellaro07}. Depending on the geometry it might be achieved using field gradients.}\footnote{Secular dipolar-dipole interaction, $H_{dip}$ in Eq. (\ref{eq:Hdip}), preserves the collective magnetization similar to $H_{XY}$; but according to our simulations replacing $H_{XY}$ by $H_{dip}$ in the circuit in FIG. \ref{fig:HXY} does not yield the desired response}.  Moreover, the number of the CNOT gates between the target qubit and the MSS 
is proportional to the number of spins in the MSS which is challenging for large systems. 

Next, we will introduce a different procedure that requires only a one-time interaction between the qubit and the MSS. It also uses a Hamiltonian that can be engineered from the dipolar coupling using only collective control.

\subsection{One-time interaction}
\label{sec:GR}
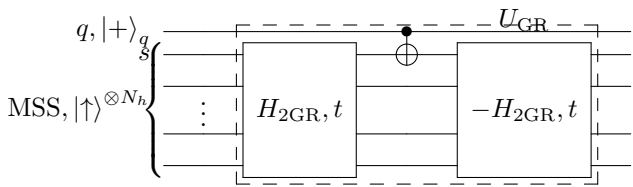
\begin{figure}[t!h]
        \Qcircuit @ C=1.5em @ R=0.3em {
      &  & & & & \mbox{$U_{\text{GR}}$} & & &\\
       &\lstick{q,\ket{+}_q} & \qw & \qw & \ctrl{1} & \qw & \qw  \\
     &\lstick{s}  & \qw & \multigate{5}{H_{2\text{GR}},t}& \targ & \multigate{5}{-H_{2\text{GR}},t}& \qw \\
        && \qw & \ghost{H_{2\text{GR}},t}& \qw & \ghost{-H_{2\text{GR}},t}& \qw \\
    \push{\rule{4em}{0em}} &  \lstick{\text{MSS},\ket{\uparrow}^{\otimes N_h}}   & \vdots & &  & & \\
      &  & & & & & & & \\
       & & \qw & \ghost{H_{2\text{GR}},t}& \qw & \ghost{-H_{2\text{GR}},t}  &\qw \\
       & & \qw & \ghost{H_{2\text{GR}},t}& \qw & \ghost{-H_{2\text{GR}},t} &\qw 
        \gategroup{2}{4}{8}{6}{0.5em}{--} 
        \gategroup{3}{2}{8}{2}{0.8em}{\{}      
        }  
        \caption{\label{fig:H2GR} 
       This circuit creates micro-macro entanglement between the target qubit and the MSS with a one-time interaction between the two and using experimentally available control.}
    \end{figure}

Here we show that the circuit in FIG. \ref{fig:H2GR} coherently magnifies the state of the target qubit in the collective magnetization of the MSS and creates a micro-macro entangled state using only  one CNOT gate.
The internal dynamics of the MSS is 
governed by the reversible grade-raising Hamiltonian,
\begin{equation}
\label{eq:GR}
H_{2\text{GR}}=\sum_{i,j; i< j} a_{ij} \left(\sigma_{+}^i \sigma_{+}^j+ \sigma_{-}^i \sigma_{-}^j\right)  \hspace{0.3cm} a_{ij}\propto \frac{1}{|\vec{r}_{ij}|^3}
\end{equation} 
which is a well-known Hamiltonian within the nuclear magnetic resonance (NMR) community. Both $\pm H_{2\text{GR}}$ can be synthesized out of the naturally occurring 
magnetic dipole-dipole interaction at high field 
by applying appropriate sequences of collective rotations \cite{Warren79}.

The circuit in FIG. \ref{fig:H2GR} works as follows. 
First, evolution under the grade-raising Hamiltonian correlates the spins in the MSS. For long enough evolution times a globally correlated state is created; specifically the spin of the MSS that is in contact with the external target qubit becomes correlated with the rest of the spins in the MSS. Next, the CNOT gate controlled by target qubit, $q$, on its nearby spin in the MSS, $s$, perturbs the state of the MSS \footnote{Controlled-Z gate has similar effect.}. This local conditional gate has a global conditional effect due to correlations established in the MSS prior to its local interaction with the target qubit. Finally, applying the reverse of the first gate 
makes this global conditional effect observable in the collective magnetization spectrum of the MSS along the quantization axis.  
The unperturbed state of the MSS returns back to the initial polarized state $\ket{\uparrow}^{\otimes N}$ while the perturbed one evolves to a state with a \textit{very} different collective magnetization.

The state of the target qubit and the MSS after the evolution follows the general form of a micro-macro entangled state in Eq. (\ref{eq:micmac}) with $\ket{\psi_0^i}=\ket{\uparrow}^{\otimes N_h}$ and $\ket{\psi_1^i}=\ket{\psi_1^{\text{GR}}}$,
\small
\begin{equation}
\label{eq:micmacGR}
\ket{\phi^{\text{GR}}(t)}_{q,\text{MSS}}=\frac{1}{\sqrt{2}}\left(\ket{0}_q \ket{\uparrow}^{\otimes N_h}+\ket{1}_q\ket{\psi_1^{\text{GR}}(t)}\right).
\end{equation}
\normalsize
The states $\ket{\psi_1^{\text{GR}}(t)}$ and $\ket{\uparrow}^{\otimes N_h}$ are not only orthogonal but also 
macroscopically distinct
given that the evolution time, $t$, is long enough.

\begin{figure}[t!h]
        \centering
        \includegraphics[scale=0.6]{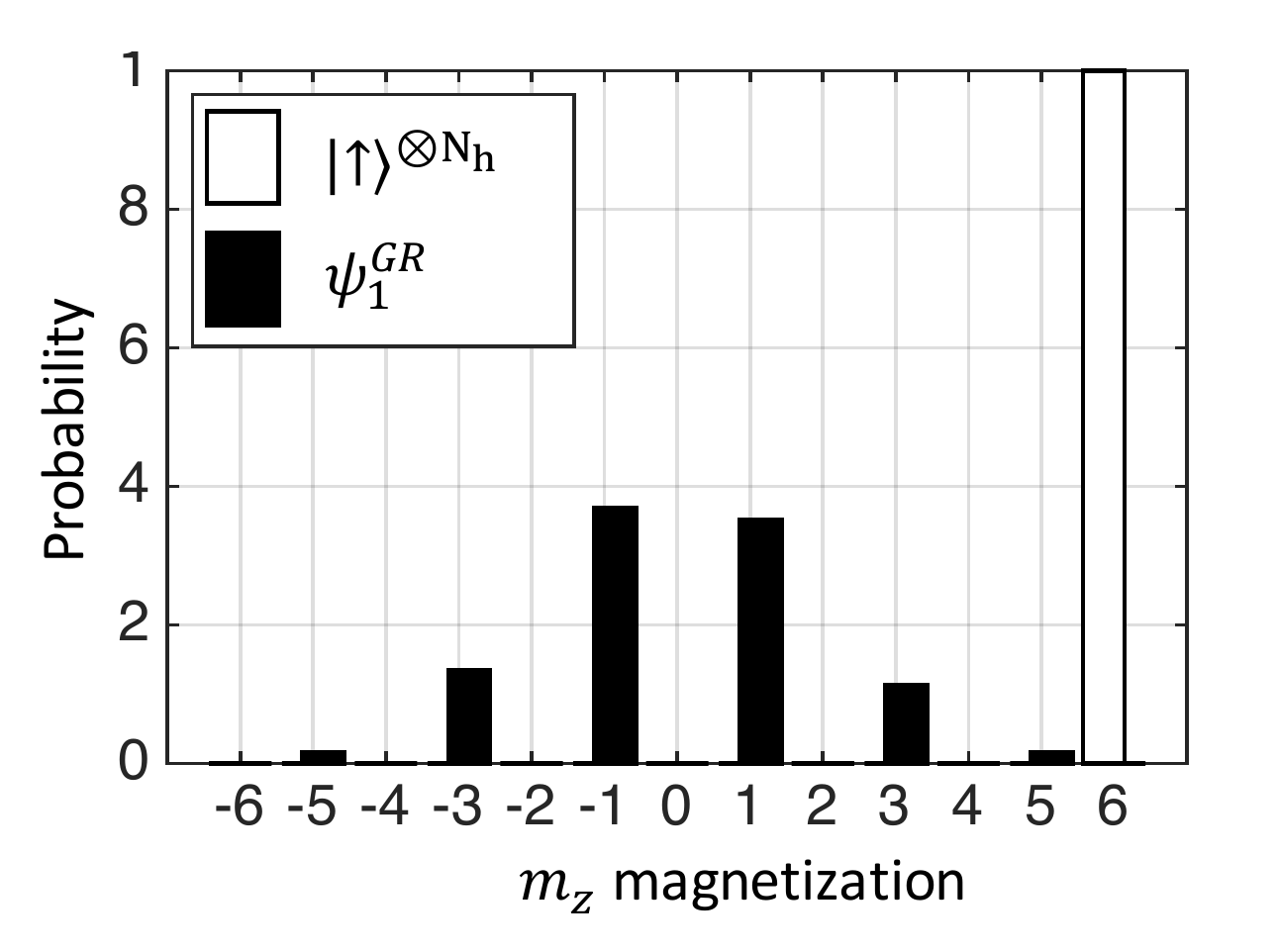}
        \caption{\label{fig:oneMSSH2GR} The distinct magnetization spectra of the MSS's states, $\ket{\uparrow}^{\otimes N_h}$ and $\ket{\psi_1^{\text{GR}}}$ correlated to $\ket{0}$ and $\ket{1}$ states of the target qubit simulated based on the circuit in FIG. \ref{fig:H2GR} with $t=2\pi N_h/a_{12}$ for $N_h=12$ spins in a 1D chain geometry. 
        }
\end{figure}

\begin{figure}[t!h]
\centering
         \includegraphics[scale=0.7]{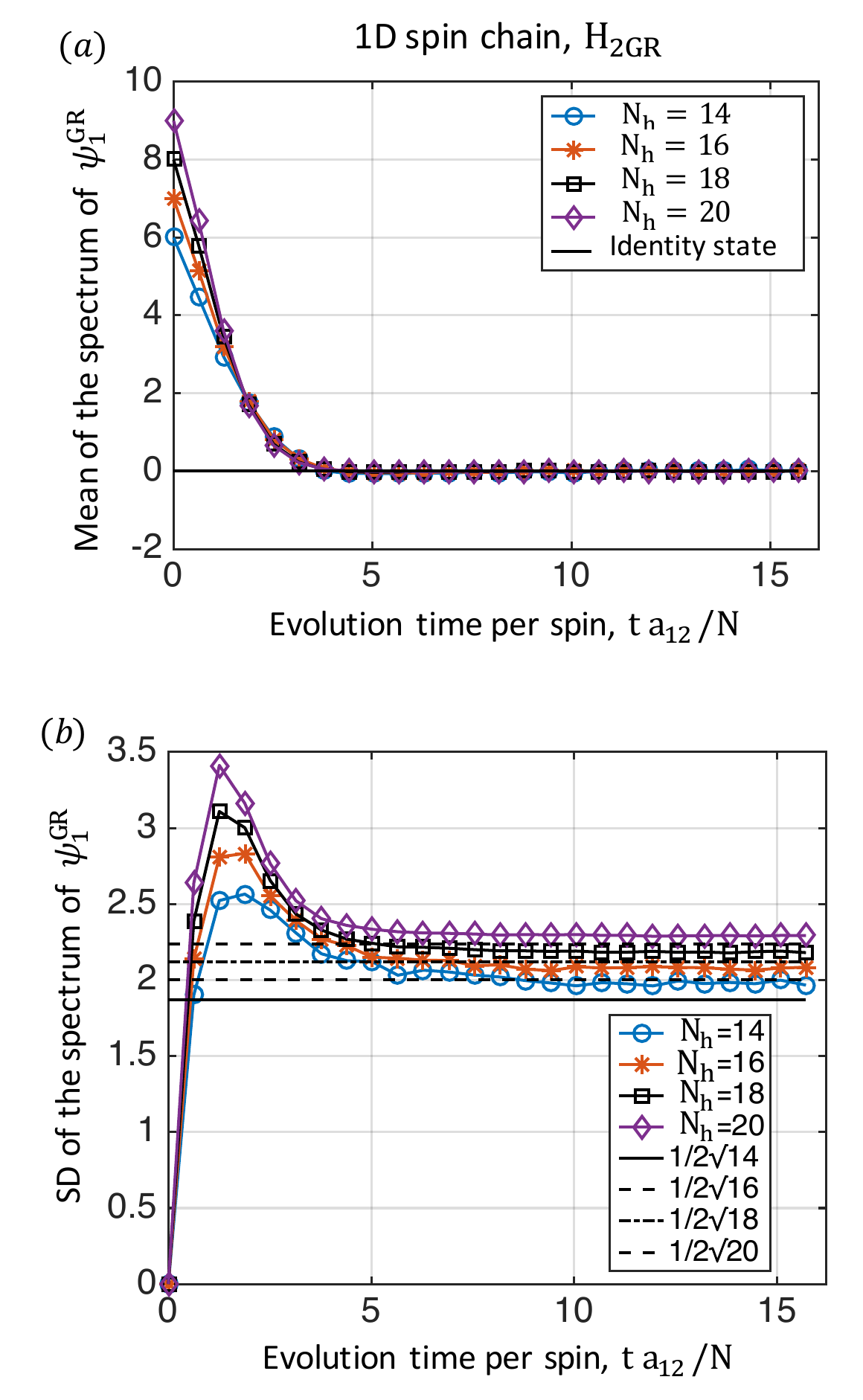}
         \caption{\label{fig:MSD1} (a) The mean and (b) the SD of the magnetization spectrum of  $\ket{\psi_1^{\text{GR}}}$ as a function of the normalized evolution time for MSSs with $N_h=12, 16, 18, 20$ spins. After a transient time both the mean and the SD of the spectrum approach those of the identity state with the same size.}   
\end{figure}

 Figure \ref{fig:oneMSSH2GR} shows the separation in the collective magnetization spectra of the states $\ket{\uparrow}^{\otimes N_h}$ and $\ket{\psi_1^{\text{GR}}(t)}$ 
 simulated for a MSS in a 1D chain geometry with $N_h=12$ spins and evolution time $t=2\pi N_h/a_{12}$, where $a_{12}$ is the nearest neighbor coupling strength of the grade-raising Hamiltonian represented in Eq. (\ref{eq:GR}). 
Figure \ref{fig:MSD1} displays the mean and the SD of the magnetization spectrum of the state $\ket{\psi_1^{\text{GR}}}$ as a function of the normalized evolution time, $t/N_h$, for up to $N_h=20$ spins. 
After a transient time, 
 the mean of the spectrum approaches zero and the SD approaches $\sqrt{N_h}/2$ similar to the steady-state behaviour of the state $\ket{\psi_1^{XY}}$.
Thus, the macroscopic distinctness of the states  $\ket{\psi_1^{\text{GR}}}$ and $\ket{\uparrow}^{\otimes N_h}$, upon collective $J_z$ measurement, scales as $\sqrt{N_h}$,
\begin{equation}
\dfrac{\left\langle J_z\right\rangle_0^{\text{GR}}- \left\langle J_z \right\rangle_1^{\text{GR}} }{(\delta J_z)_0^{\text{GR}}+(\delta J_z)_1^{\text{GR}}}\approx \dfrac{N_h/2-0}{0+\sqrt{N_h}/2}\propto \sqrt{N_h}
\end{equation}
 
After applying the introduced magnification process on both target qubits and their nearby halves of the MSS, the states of the whole MSS correlated with different states of the target qubits are: $\ket{\psi_{00}^{\text{GR}}}=\ket{\uparrow}^{\otimes N}$, $\ket{\psi_{01}^{\text{GR}}}=\ket{\uparrow}^{\otimes N_L}\ket{\psi_1^{R,\text{GR}}}$, $\ket{\psi_{10}^{\text{GR}}}=\ket{\psi_1^{L,\text{GR}}}\ket{\uparrow}^{\otimes N_R}$ and $\ket{\psi_{11}^{\text{GR}}}=\ket{\psi_1^{L,\text{GR}}}\ket{\psi_1^{R,\text{GR}}}$.  According to the relations in Eq. (\ref{eq:conv}), the mean and the SD of the collective magnetization spectra of these states scale as,
\small
\begin{eqnarray}
\left\langle J_z\right\rangle_{00}^{\text{GR}}&\approx & \frac{N}{2}, \hspace{0.2cm} (\delta J_z)_{00}^{\text{GR}}\approx 0 \\
\left\langle J_z\right\rangle_{01}^{\text{GR}}&\approx & \left\langle J_z\right\rangle_{01}^{\text{GR}} \approx \frac{N}{4},\hspace{0.2cm} (\delta J_z)_{01}^{\text{GR}}\approx (\delta J_z)_{10}^{\text{GR}} \approx \frac{\sqrt{N/2}}{2} \nonumber \\
\left\langle J_z\right\rangle_{11}^{\text{GR}}&\approx & 0, \hspace{0.2cm} (\delta J_z)_{11}^{\text{GR}}\approx  \frac{\sqrt{N}}{2} \nonumber
\end{eqnarray}
\normalsize
where $N$ is the size of the whole MSS and $N_L\approx N_R \approx N_h \approx N/2$. %
Macroscopic distinctness of the states $\ket{\psi_{01}^{\text{GR}}}$ and $\ket{\psi_{10}^{\text{GR}}}$ from both of the states $\ket{\psi_{00}^{\text{GR}}}$ and $\ket{\psi_{11}^{\text{GR}}}$ imposes a lower bound on the size of the MSS, 
\begin{equation}
\frac{N}{4}\gg \left(\frac{\sqrt{N/2}}{2} +\frac{\sqrt{N}}{2} \right) \Rightarrow N \gg 12
\end{equation}

Comparing to the circuit based on XY Hamiltonian, the coherent control elements of this circuit 
meshes better with the experimentally available tools. It needs only one CNOT gate. Additionally, the grade-raising Hamiltonian can be synthesized from dipolar interaction with only collective pulses in contrast to the XY Hamiltonian that requires both collective pulses and rotations on every other spin \cite{Cappellaro07}. 
Thus, in the rest of this paper we will consider the circuit in FIG. \ref{fig:H2GR},  based on the grade-raising Hamiltonian as the magnification process. 

\subsection{Dimensionality}
The simulations in section \ref{sec:GR} were all set in a 1D geometry. Here, generating micro-macro entanglement between a target qubit and a MSS that has a 2D structure is studied. 

\begin{figure}[t!h]
\centering
         \includegraphics[scale=0.7]{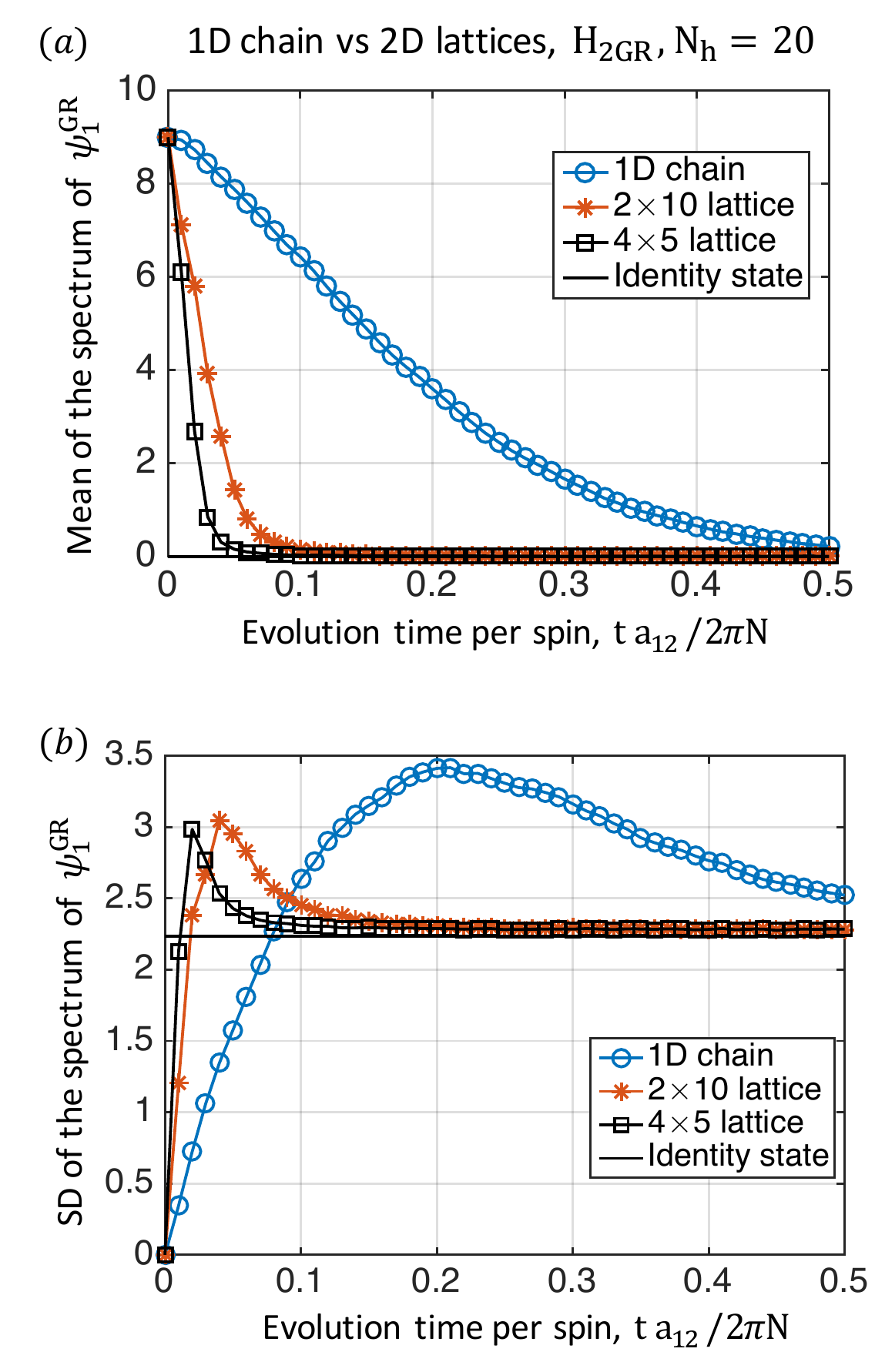}
         \caption{\label{fig:MSD1D2D} Comparing (a) the mean and (b) the SD of $\ket{\psi_1^{\text{GR}}}$ with the circuit in FIG. \ref{fig:H2GR} for a 1D chain and 2D lattices. 2D lattice structures have much shorter transient times than a 1D chain with the same number of spins but the steady state responses are similar.}   
\end{figure}

Figure \ref{fig:MSD1D2D} compares 
(a) the mean and (b) the SD of the spectrum of $\ket{\psi_1^{\text{GR}}}$ 
simulated for $N_h=20$ spins when in a 1D chain versus 2 by 10 and 4 by 5 2D lattices. 
The asymptotic behaviours of 2D lattices are similar to that of a 1D chain; however, the transition times of the 2D structures are much shorter meaning that the information flows much faster. One simple explanation for this difference is that 
information flows over just one path  in a 1D structure compared to multiple paths in 2D (or 3D) structures. 
One-directional information flow is crucial in quantum state transfer proposals; in contrast, our method relies on amplification of the qubit's state in the whole system rather than propagation of information in a specific direction. Therefore, it benefits from faster response in 2D (and 3D) structures.

To conclude, all the previous steady-state results apply to higher dimensions with an essential advantage of shorter transient times and faster responses.

\subsection{Magnification time}
\begin{figure}[t!h]
\centering
         \includegraphics[scale=0.6]{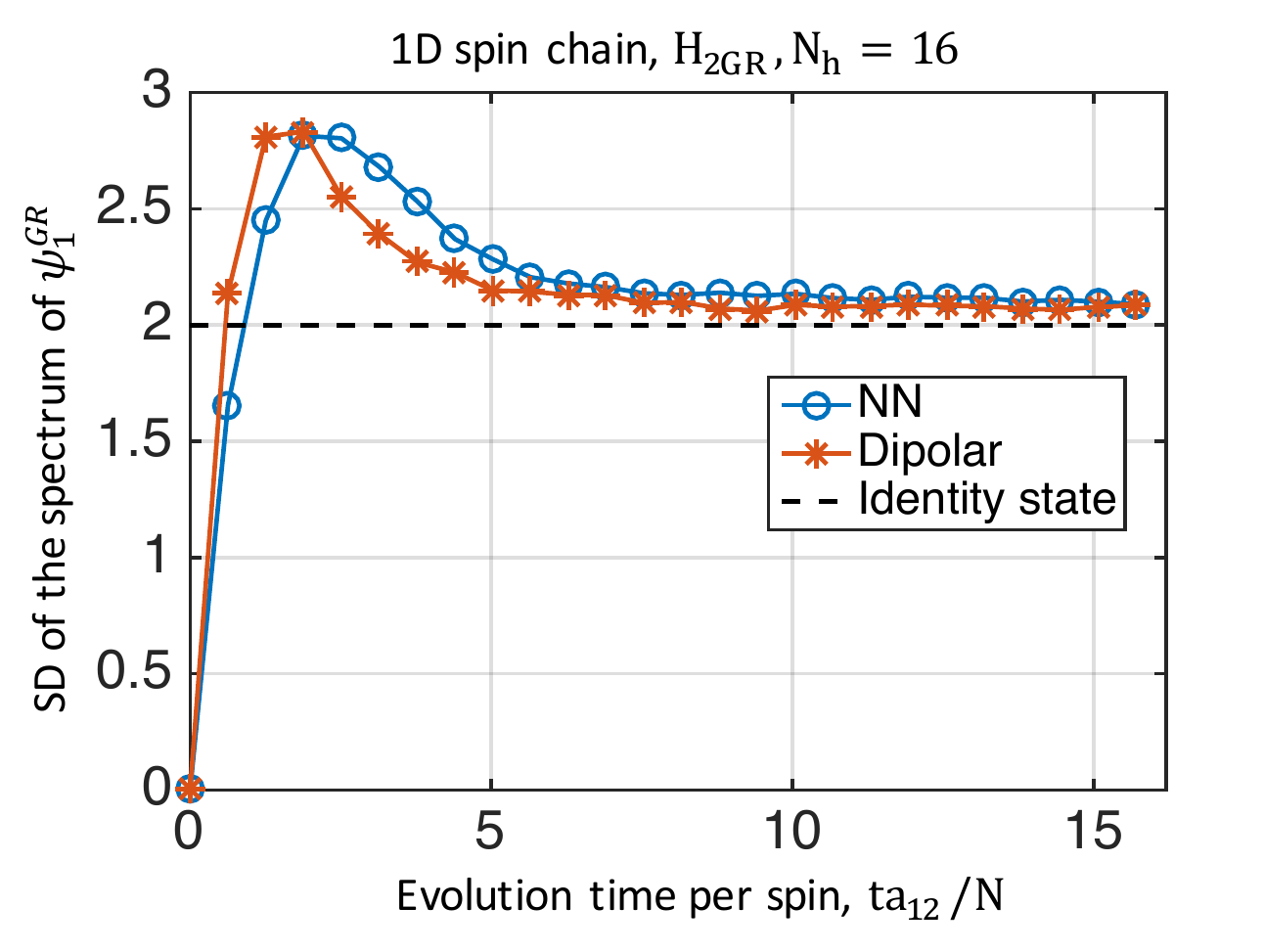}
         \caption{\label{fig:NNvsDip} Comparing the transient times for MSSs with NN coupling and long range dipolar coupling. Information flows faster in a system with full dipolar compared to truncating to only NN interactions.}   
\end{figure}

An important consideration moving forward is determining the magnification process's time. Of particular interest is how the magnification time scales with the size of the MSS and what its relation is to the dimension of the MSS.
 This question is in general hard to answer
  because it depends on the many-body dynamics of the MSS.
 Nevertheless, we have some clues to the answer. We have shown that the dimension of the MSS 
 significantly affects the response time. The magnification process has a much shorter transient time if the MSS has a 2D structure, compared to a 1D chain of the same size. 
 Moreover, as depicted in FIG. \ref{fig:NNvsDip},
 the long range magnetic dipole interaction 
 entails shorter transient times compared with truncating to only nearest-neighbor (NN) interactions. 
Furthermore, comparing the SD vs normalized time for different numbers of spins in a 1D chain in FIG. \ref{fig:MSD1} shows that as the size of the MSS increases, the peak is shifted towards shorter normalized times; 
 indicating that the transient time has a sub-linear relation with the size of the MSS even in a 1D geometry.

The magnification time in our protocol is closely related to the rate of information flow in a system with dipolar coupling. In 1972, Lieb and Robinson showed that there is a constant group velocity for the flow of information in a system with local interactions, e.g., nearest-neighbor interactions (or exponentially decaying interaction strength), known as the Lieb-Robinson bound \cite{Lieb72}. Our results show that the dynamics of MSS violates the Lieb-Robinson bound,  a finding consistent with long-range dipolar interaction in the system.
 Recently, numerous attempts have been made to find the rate of information flow in systems with long-range interactions decaying with power law, $a_{ij}\propto \frac{1}{r_{ij}^{\alpha}}$ \cite{Hastings06,Hauke13,Gong14,Hazzard14,Jurcevic14,Richerme14,Foss-Feig15}.
Based on these studies, different relations between the magnification time and the size of the MSS are expected depending on the MSS's dimension.
It has been shown that the correlation times for a system with power-law interaction, $a_{ij}\propto \frac{1}{r_{ij}^{\alpha}}$, grow as $T\propto r^{\zeta}$, with $\frac{1}{\zeta}=1+\frac{1+D}{\alpha-2D}$ when $\alpha>2D$ \cite{Foss-Feig15}.
Thus, for a 1D chain, the magnification time is expected to scale as $t_{mag}^{1D}\propto l^{\frac{1}{3}}\propto N_h^{\frac{1}{3}}$ where $l$ is the 
length of the spin chain.
For 2D and 3D lattices with dipolar coupling, no bound tighter than an exponential information flow is found \cite{Hastings06}. We also know that the information flow is faster in 2D and 3D structures than in 1D chains. Thus, in 2D and 3D structures the respective range of the magnification times are expected to be $(\sim log(l)\propto log(\sqrt{N_h}) )\leq t_{mag}^{2D} < (\sim l^{\frac{1}{3}}\propto  N_h^{\frac{1}{6}})$ and $(\sim log(l)\propto log(\sqrt[3]{N_h}) )\leq t_{mag}^{3D} < (\sim l^{\frac{1}{3}}\propto  N_h^{\frac{1}{9}})$.
 It is worth mentioning that recently an algorithm has been proposed that saturates the logarithmic bound for a 3D structure with dipolar coupling. It needs  $t\propto \log(r)$ to transfer a state through a system with $\frac{1}{r^{\alpha}}$ interaction if $\alpha=D$ \cite{Eldredge17}.

\section{Measurement and Fidelity}
\label{sec:measurement}

The requirements of an ideal measurement procedure were discussed in section \ref{sec:Indirect}.
Here we estimate the fidelity of the target qubits' post-selected entangled state using the magnification process introduced in section \ref{sec:GR} and 
a collective measurement on the MSS through a two-level apparatus (or a fair model for the measurement on the MSS).

\begin{figure}[t!h]
\centering
         \includegraphics[scale=0.5]{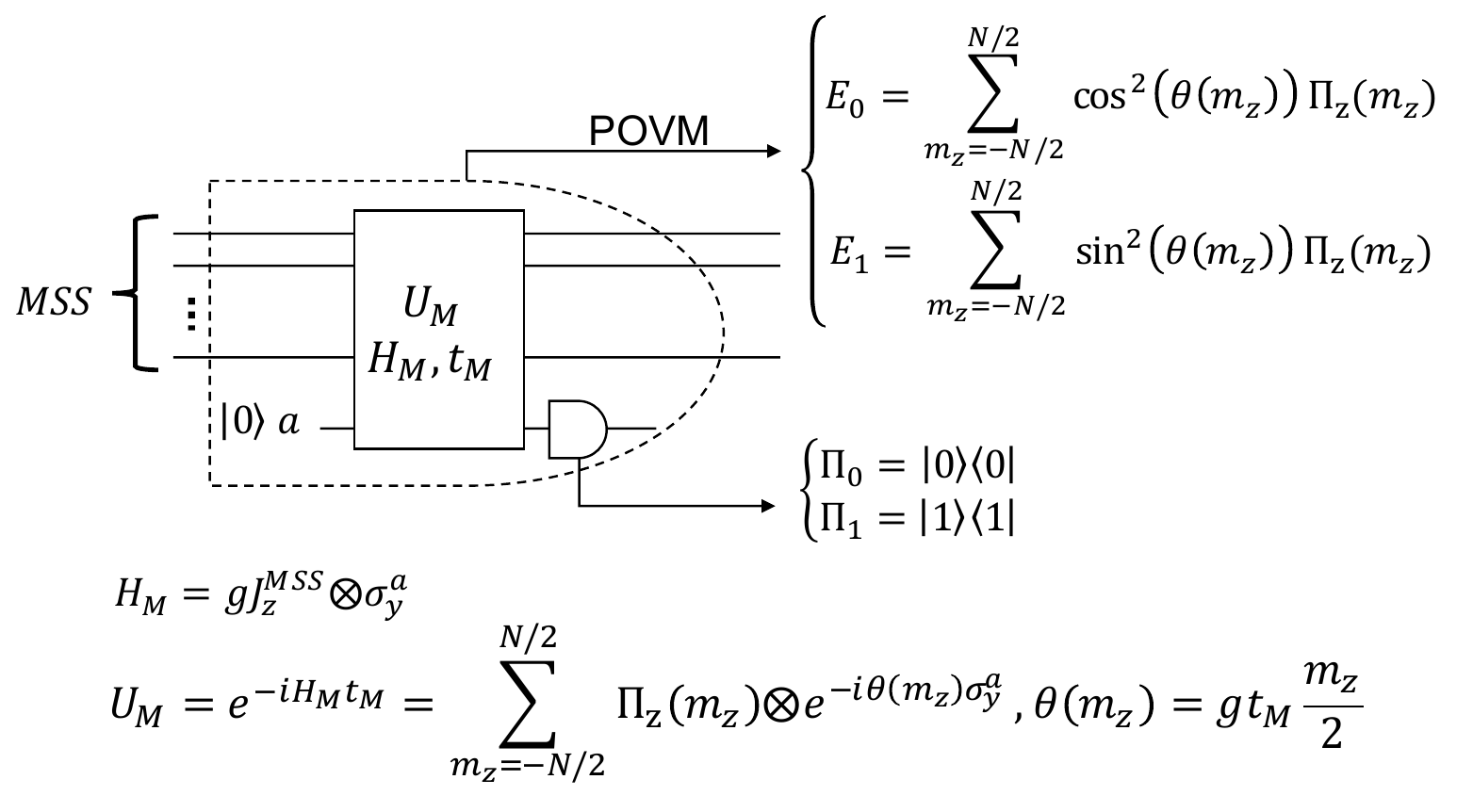}
         \caption{\label{fig:MeasureProc}Two outcome POVM on a MSS implemented through an apparatus qubit.}   
\end{figure}

The measurement model is based on the general collective two-outcome POVM on a mesoscopic system suggested in our previous work \cite{Mirkamali18}. Any two outcome collective POVM can be parametrized with a phase function $\theta(m_z)$,
\begin{eqnarray}
E_0&=&\sum_{m_z} \cos^2(\theta(m_z)) \Pi^N(m_z) \\
E_1&=& \id-E_0 = \sum_{m_z} \sin^2(\theta(m_z)) \Pi^N(m_z) \nonumber
\end{eqnarray}
Such a measurement is equivalent to a projective measurement on a two-level apparatus system after it interacts with the MSS according to the interaction gate \cite{Neumann32,Neumark43},
\begin{equation}
U_M=\sum_{m_z=-\frac{N}{2}}^{\frac{N}{2}}\Pi^N(m_z)\otimes e^{-i\theta(m_z)\sigma_y^a}
\end{equation}
Linear collective interaction between the MSS and the apparatus qubit, $H_M=g J_z\otimes \sigma_y$, conveniently creates $U_M$ with a phase function proportional to the collective magnetization, $\theta(m_z)\propto m_z$.
 See FIG. \ref{fig:MeasureProc}. In this measurement process, the state of the MSS is updated \cite{Giedke06} according to Eq. (\ref{eq:update}), with the measurement operators,
 \begin{eqnarray}
 \label{eq:Mmeasure}
 M_0&=&\sum_{m_z} \cos(\theta(m_z)) \Pi^N(m_z) \\
 M_1&=& i \sum_{m_z} \sin(\theta(m_z)) \Pi^N(m_z)\nonumber
 \end{eqnarray}
 
\begin{figure}[t!h]
\centering
         \includegraphics[scale=0.36]{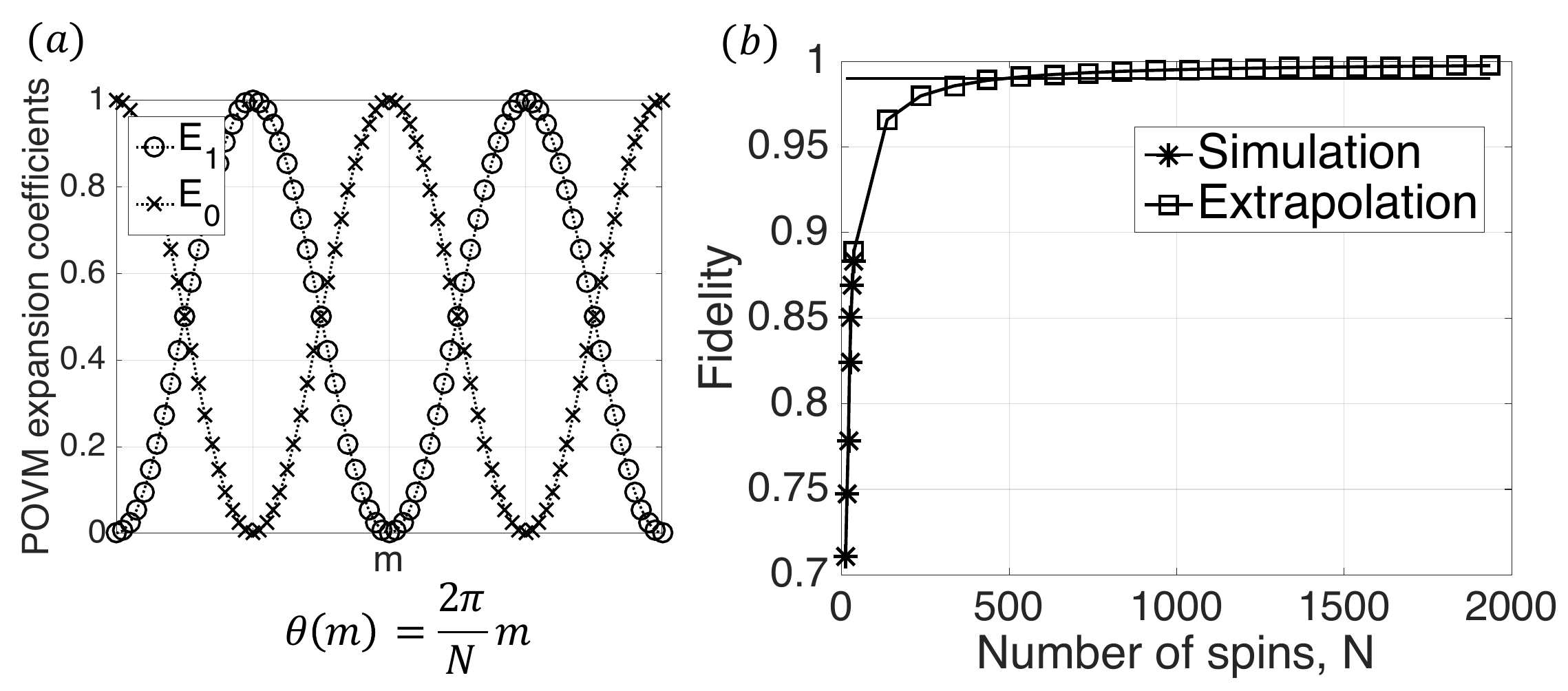}
         \caption{\label{fig:fidelity_SimExt} (a) The expansion coefficients of the two POVM operators based on the measurement procedure shown in FIG. \ref{fig:MeasureProc} with $\theta(m_z)=\frac{2\pi}{N}m_z$, chosen to distinguish between $\{\ket{\psi_{01}^{\text{GR}}},\ket{\psi_{10}^{\text{GR}}}\}$ and $\{\ket{\psi_{00}^{\text{GR}}},\ket{\psi_{11}^{\text{GR}}}\}$ with highest probability. (b) The corresponding fidelity of the entangled state of the target qubits with the maximally entangled state $\ket{m_0}$, after applying the measurement in (a) on the MSS, post-selecting on outcome $1$ and disentangling from the MSS. The fidelity is computed based on simulation of the spectra of the states $\{\ket{\psi_{01}^{\text{GR}}},\ket{\psi_{10}^{\text{GR}}},\ket{\psi_{00}^{\text{GR}}},\ket{\psi_{11}^{\text{GR}}}\}$ for $N=12,16, 20, 24, 28, 32, 36$ spins and extrapolation of their spectra according to binomial distribution for larger systems.}   
\end{figure}

In order to select $\ket{\psi_{01}^{\text{GR}}}$ and $\ket{\psi_{10}^{\text{GR}}}$ over $\ket{\psi_{00}^{\text{GR}}}$ and $\ket{\psi_{11}^{\text{GR}}}$, the linear phase function is chosen to be $\theta(m_z)=\frac{2\pi}{N}m_z$. 
Figure \ref{fig:fidelity_SimExt} depicts the corresponding expansion coefficients of the POVM operators and the fidelity of the target qubits' state with the maximally entangled state $\ket{m_{0}}$, upon measurement, post-selection on outcome $1$ and disentangling from the MSS. 
The fidelity increases with the size of the MSS and asymptotically approaches its maximum value, one. 
This increase has two origins. First, the macroscopic distinctness between the states $\{\ket{\psi_{01}^{\text{GR}}},\ket{\psi_{10}^{\text{GR}}}\}$ and $\{\ket{\psi_{00}^{\text{GR}}},\ket{\psi_{11}^{\text{GR}}}\}$ grows with the size of the MSS. Second, for larger MSSs, the measurement coefficients become closer to uniform distribution over the expansion of the spectrum of $\ket{\psi_{01}^{\text{GR}}}$ and $\ket{\psi_{10}^{\text{GR}}}$; thus these states get less distorted by the measurement and the following disentangling gate will restore more coherence between the qubits' states $\ket{01}_q$ and $\ket{10}_q$.
It should be mentioned that 
we have simulated an ideal noise-free process. In practice, the fidelity is not expected to increase with the size of the MSS, indefinitely.
Including noise effect imposes an upper bound on size of the MSS, as will be discussed in section \ref{sec:noise}.
 
Figure \ref{fig:fidelity_SimExt} shows that the fidelity, $F_{m_{0}}(\rho_{q})$,exceeds $0.5$ 
 and the target qubits are entangled for all simulated sizes of the MSS, although the assumed measurement model is not an ideal measurement procedure.
Moreover, for $N\geq 24$, the fidelity is greater than $0.78$, 
enough to violate the CHSH inequality \cite{CHSH69,Bennett96PRL}.

\section{Mixed initial state}
\label{sec:Mixed}

So far, a pure polarized state, $\ket{\uparrow}^{\otimes N}$, has been considered as the initial state of the MSS. 
In this section, we prove robustness of the introduced indirect joint measurement procedure  to limited initial polarization of the MSS. 
In particular, we will show that 
micro-macro entanglement between each target qubit and its nearby half of the MSS and 
the subsequent 
bipartite entanglement of the non-interacting target qubits are robust to 
deviations of the MSS's initial state from fully polarized state;
when the MSS is initially in the experimentally relevant mixed state,

\begin{eqnarray}
\rho_{in}(N,\epsilon)&=&\left(\frac{\id+(1-\epsilon)\sigma_z}{2}\right)^{\otimes N} \\
&=&\left((1-\frac{\epsilon}{2})\ketbra{\uparrow}{\uparrow}+\frac{\epsilon}{2}\ketbra{\downarrow}{\downarrow}\right)^{\otimes N} \nonumber
\end{eqnarray}
The polarization parameter, $\epsilon$, ranges from $0$, for a fully polarized pure state, to $1$, for the maximally mixed state. We are particularly interested in highly polarized states, i.e., $\epsilon$ close to 0.

The magnification gate in FIG. \ref{fig:H2GR} 
can be written as,
\begin{equation}
U_{\text{GR}}=\ketbra{0}{0}_q\otimes \id + \ketbra{1}{1}_q\otimes V_1
\end{equation}
with $V_{1} \ket{\uparrow}^{\otimes N_h}=\ket{\psi_1^{\text{GR}}}$. The state of one target qubit and its nearby half of the MSS after applying 
gate $U_{\text{GR}}$ to the initial state $\ketbra{+}{+}\otimes \rho_{in}(N_h,\epsilon)$ is,
\begin{eqnarray}
\label{eq:rhoGR}
\rho_{q,\text{MSS}}^{\text{GR}}&=&\frac{1}{2}\left(\ketbra{0}{0}_q\otimes \rho_{in} + \ketbra{1}{1}_q\otimes (V_1\rho_{in}V_1^{\dagger}) \right. \nonumber \\
 &+& \left. \ketbra{0}{1}_q\otimes (\rho_{in} V_1^{\dagger}) + \ketbra{1}{0}_q\otimes (V_1\rho_{in}) \right)
\end{eqnarray}
Micro-macro entanglement 
of state $\rho_{q,\text{MSS}}^{\text{GR}}$ 
requires bipartite entanglement between the qubit and the MSS 
 and macroscopic distinctness between the state $\rho_0^{\text{GR}}=\rho_{in}$  and $\rho_1^{\text{GR}}=V_1\rho_{in} V_1^{\dagger}$. We investigate how these two characteristics  change when the initial state of the MSS deviates from the ideal polarized state. 
Direct verification of bipartite entanglement between a microscopic and a mesoscopic system experimentally is a challenging task \cite{Sekatski14PRL,Wang13}. 
Nevertheless, it can be simulated for small sizes of the mesoscopic system.
A computable measure of bipartite entanglement 
for a general state, $\rho_{AB}$, regardless of the size of each party and the purity of the overall state,
is negativity, which is defined as the sum of the absolute values of the negative eigenvalues of the partially transposed density matrix, $\rho_{AB}^{T_A}$, \cite{Vidal02}
\begin{equation}
\text{Neg}(\rho_{AB}):=\sum_{i}|\lambda_i|
\end{equation}
Negativity ranges from zero for separable states to $0.5$ for maximally entangled states \footnote{Based on the PPT criteria \textbf{cite} all separable states have zero negativity but not all entangled states have nonzero negativity except for $2\times 2$ and $2\times 3$ systems. In other words nonzero negativity guarantees entanglement but there are entangled states with zero negativity.
}. This measure is specifically helpful 
in quantifying bipartite entanglement of a mixed state, when one or both of the parties have more than two levels; where, other computable measures for mixed state entanglement such as concurrence can not be applied.
A related measure is logarithmic negativity, defined as,
\begin{eqnarray}
\text{Lneg}(\rho_{AB})&:=&\log_2 ||\rho_{AB}^{T_A}||_1 \nonumber \\
&=&\log_2(2\text{Neg}(\rho_{AB})+1)
\end{eqnarray} 
 where $||\rho_{AB}^{T_A}||_1$ is the trace norm of the partially transposed density matrix, $\rho_{AB}^{T_A}$. Logarithmic negativity ranges from $0$, for separable states, to $1$, for maximally entangled states \cite{Vidal02}. 
\begin{figure}[hbt!]
        \centering
        \includegraphics[scale=0.6]{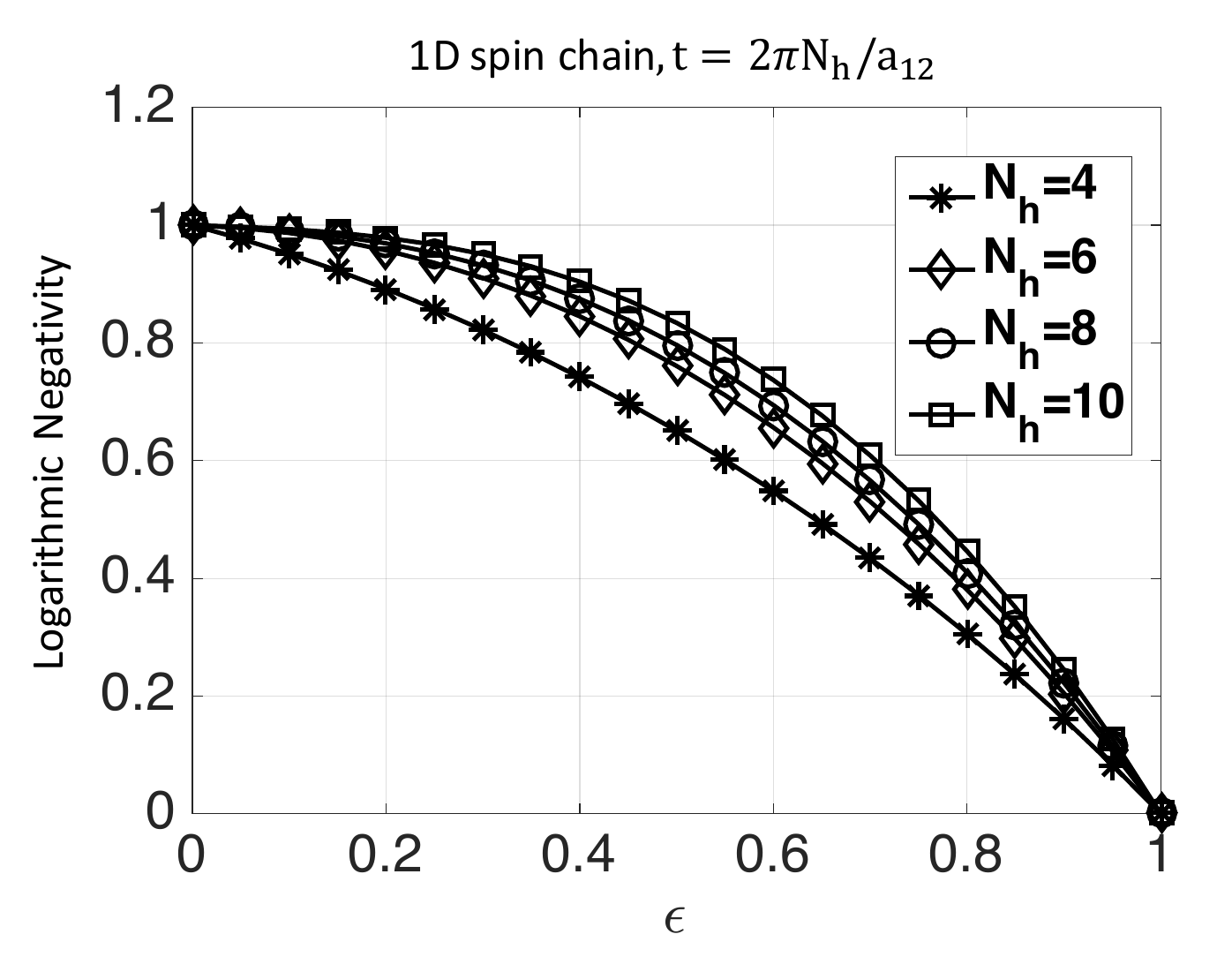}
        \caption{\label{fig:LogNegMixed} Entanglement between one target qubit and its nearby half of the MSS as a function of deviation of the initial state of the MSS from fully polarized state simulated for different sizes of the MSS. The simulation is based on the circuit in FIG. \ref{fig:H2GR} with the evolution time $t=2\pi N_h/a_{12}$. For larger MSSs, bipartite entanglement between the target qubit and the MSS is more robust to polarization reduction.}
\end{figure}

Figure \ref{fig:LogNegMixed} shows logarithmic negativity of the bipartite entangled state of the qubit and the MSS, $\rho_{q,\text{MSS}}^{\text{GR}}$, as a function of the polarization parameter, $\epsilon$, simulated 
for different sizes of the MSS up to 10 spins.
We are particularly interested in highly polarized states, where $\epsilon$ is close to $0$. For the fully polarized initial state ($\epsilon=0$), the state $\rho_{q,\text{MSS}}^{\text{GR}}$ is maximally entangled as already discussed. 
The entanglement reduces with decrease in the polarization (increase in $\epsilon$) with a slow initial pace. 
The larger the MSS, the slower the initial drop in entanglement, 
i.e. for larger MSS, bipartite entanglement of state $\rho_{q,\text{MSS}}^{\text{GR}}$ is more robust to the polarization reduction of the MSS's initial state. 

The macroscopic distinctness between the states $\rho_{0}^{\text{GR}}$ and $\rho_{1}^{\text{GR}}$ can be quantified according to  Eq. (\ref{eq:macroscopicDis}). 
The collective magnetization spectrum of the state $\rho_{0}^{\text{GR}}=\rho_{in}(N_h,\epsilon)$ is a shifted \footnote{The distribution ranges from $-N_h/2$ to $N_h/2$ rather than $0$ to $N_h$.} binomial distribution with the probability of success $p=1-\epsilon/2$, and number of trials $N_h$. Its mean and SD are $\left\langle J_z\right\rangle^{\text{GR}}_0(\epsilon)=(1-\epsilon)N_h/2$ and $(\delta J_z)^{\text{GR}}_0 (\epsilon)=\sqrt{N_h\frac{\epsilon}{2}(1-\frac{\epsilon}{2})}$.

\begin{figure}
\centering
         \includegraphics[scale=0.7]{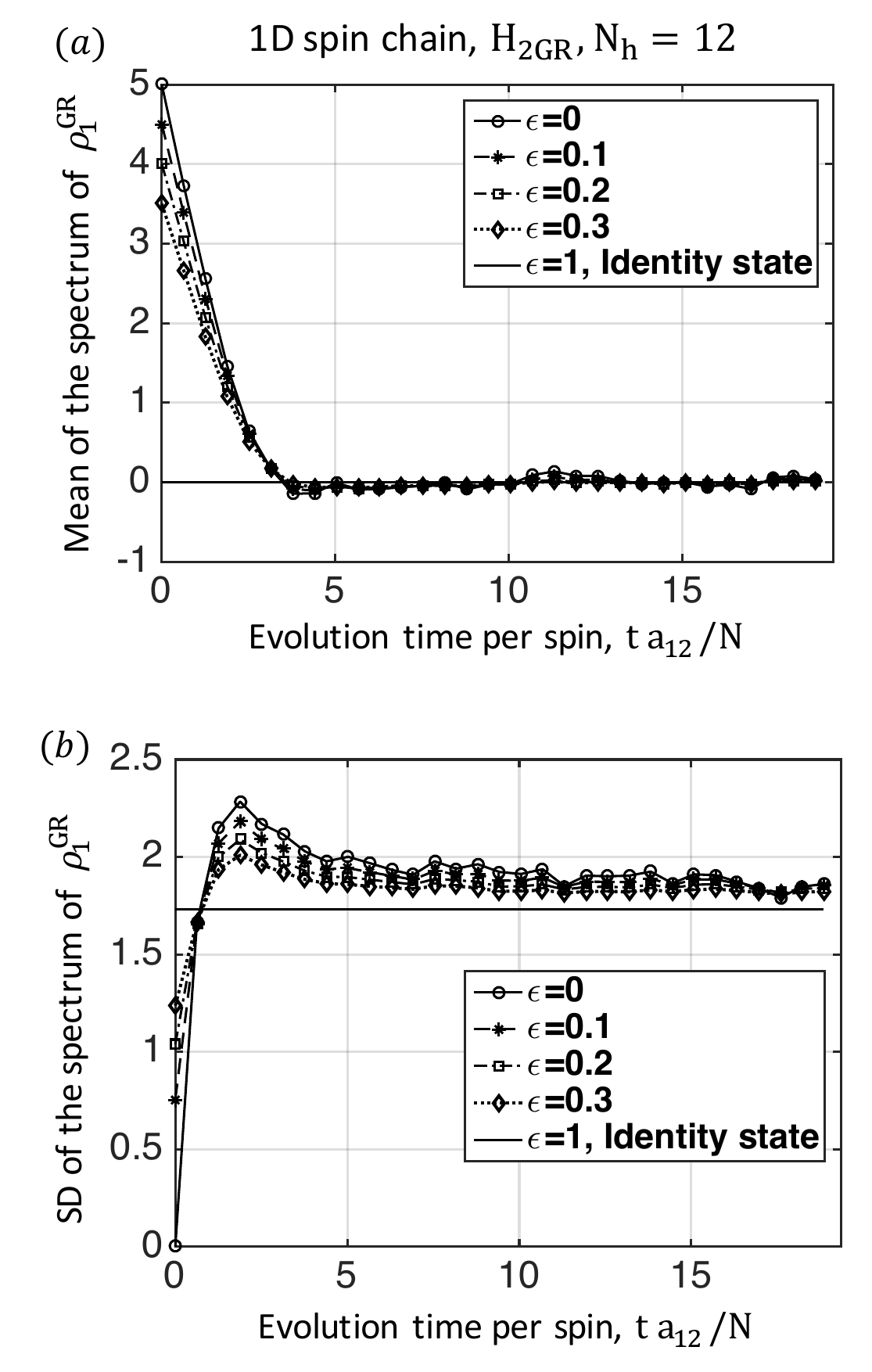}
         \caption{\label{fig:MSDMixed24} (a) The mean and (b) the SD of the spectrum of $\rho_1^{\text{GR}}$ as a function of time simulated  with $N_h=12$ spins in a 1D chain for different initial polarization of the MSS.}   
\end{figure}

The mean and SD of the spectrum of  state $\rho_{1}^{\text{GR}}=V_1\rho_{in} V_1^{\dagger}$ are known for the two extreme cases; $\epsilon=0$ and $\epsilon=1$. It was shown that with a polarized initial state,  $\epsilon=0$, the mean of the spectrum is $\left\langle J_z \right\rangle^{\text{GR}}_1(\epsilon=0)\approx 0$ and its SD scales as $(\delta J_z)^{\text{GR}}_1 (\epsilon=0)\approx \sqrt{N_h}/2$. On the other side of the range, when $\epsilon=1$, the initial state, $\rho_{in}$, is a fully mixed state; thus $\rho_{1}^{\text{GR}}$ is also a fully mixed state and the mean and SD of its spectrum are, $\left\langle J_z \right\rangle^{\text{GR}}_1(\epsilon=1)=0$ and $(\delta J_z)^{\text{GR}}_1 (\epsilon=1)=\sqrt{N_h}/2$. Similar mean and SD for the two extreme cases suggests the same scaling for all other polarizations, $0< \epsilon <1$. Simulation results for different polarization with $N_h=12$ spins, shown in FIG. \ref{fig:MSDMixed24}, confirms this prediction. Thus, the mean and SD of the spectrum of $\rho_{1}^{\text{GR}}$ are $\left\langle J_z \right\rangle^{\text{GR}}_1(\epsilon)\approx 0$ and $(\delta J_z)^{\text{GR}}_1 (\epsilon)\approx \sqrt{N_h}/2$, for all initial polarizations.
Consequently, the macroscopic distinctness between the states $\rho_0^{\text{GR}}$  and $\rho_1^{\text{GR}}$ requires that,
\small
\begin{equation}
\dfrac{\left\langle J_z\right\rangle_0^{\text{GR}}(\epsilon)- \left\langle J_z \right\rangle_1^{\text{GR}}(\epsilon) }{(\delta J_z)_0^{\text{GR}}(\epsilon)+(\delta J_z)_1^{\text{GR}}(\epsilon)}\approx \dfrac{(1-\epsilon)N_h/2-0}{\sqrt{\epsilon(2-\epsilon)}\sqrt{N_h}/2+\sqrt{N_h}/2}\gg 1
\end{equation}
\normalsize
For $0\leq \epsilon \leq 1$, the maximum of $\sqrt{\epsilon(2-\epsilon)}$ is $1$ at $\epsilon=1$ and 
the above relation is lower bounded by,
\begin{equation}
\dfrac{\left\langle J_z\right\rangle_0^{\text{GR}}(\epsilon)- \left\langle J_z \right\rangle_1^{\text{GR}}(\epsilon) }{(\delta J_z)_0^{\text{GR}}(\epsilon)+(\delta J_z)_1^{\text{GR}}(\epsilon)} > \dfrac{(1-\epsilon)N_h/2}{\sqrt{N_h}}\propto \frac{1-\epsilon}{2}\sqrt{N_h}
\end{equation}
Thus, $\frac{1-\epsilon}{2}\sqrt{N_h}\gg 1$ assures macroscopic distinctness between the states $\rho_0^{\text{GR}}$  and $\rho_1^{\text{GR}}$. 
This condition along with robustness of entanglement to decreases in polarization, shown in FIG. \ref{fig:LogNegMixed}, prove that the micro-macro entanglement between each target qubit and its nearby half of the MSS is robust to polarization loss when $\epsilon$ is close to $0$ and 
$N_h (1-\epsilon)^2\gg 4$.


Two copies of the state in Eq. (\ref{eq:rhoGR}) 
represent  the state of the two non-interacting target qubits and the uncoupled halves of the intermediate MSS,
\begin{eqnarray}
\rho_{q,\text{MSS}}&=&\frac{1}{4}\left(\ketbra{00}{00}_q\otimes \rho^{\text{GR}}_{00}+\ketbra{01}{01}_q\otimes \rho^{\text{GR}}_{01}\right. \nonumber \\
&+&  \ketbra{10}{10}_q\otimes \rho^{\text{GR}}_{10}+\ketbra{11}{11}_q\otimes \rho^{\text{GR}}_{11} \nonumber \\
&+&  \ketbra{01}{10}_q\otimes (\rho_{in}^L V_1^{L\dagger})\otimes (V_1^R \rho_{in}^R) \nonumber \\
&+&  \ketbra{10}{01}_q\otimes (V_1^{L}\rho_{in}^L )\otimes (\rho_{in}^R V_1^{R\dagger} ) \nonumber \\
&+&\left. \text{other off-diagonal terms} \right)
\end{eqnarray}
where $\rho_{ij}^{\text{GR}}=\rho_i^{\text{GR}}\otimes \rho_j^{\text{GR}}$ for $i,j=0,1$. 
The normalized state of the target qubits after the measurement, post-selection on zero magnetization and disentangling from the MSS is,
\begin{widetext}
\begin{equation}
\rho_{q}=\text{Tr}_{\text{MSS}}\left( U_{\text{GR}}^{L\dagger}\otimes U_{\text{GR}}^{R\dagger}\frac{(\id_q \otimes M_1) \rho_{q,\text{MSS}} (\id_q \otimes M_1^{\dagger})}{\text{Tr}((\id_q \otimes E_1)\rho_{q,\text{MSS}})} U_{\text{GR}}^L\otimes U_{\text{GR}}^R\right)
\end{equation}
\end{widetext}
The measurement on the MSS and post-selection, $\id \otimes M_1$, defined in Eq. (\ref{eq:Mmeasure}), 
selects 
 $\rho^{\text{GR}}_{01}$ and $\rho^{\text{GR}}_{10}$, correlated with $\ket{01}_q$ and $\ket{10}_q$ states of the qubits, and the disentangling gate, $U_{\text{GR}}^{L\dagger}\otimes U_{\text{GR}}^{R\dagger}$, 
 restores the coherence between the target qubits. 
 
 Success of the measurement process relies on distinguishability of the states $\rho^{\text{GR}}_{01}$ and $\rho^{\text{GR}}_{10}$ from the states $\rho^{\text{GR}}_{00}$ and $\rho^{\text{GR}}_{11}$, which requires,
\begin{eqnarray}
\frac{N}{4}(1-\epsilon) &\gg &  \left(\frac{\sqrt{N}}{2}\sqrt{\frac{1+\epsilon(2-\epsilon)}{2}}+\frac{\sqrt{N}}{2}\right) \sim \sqrt{N} \nonumber \\
&\Rightarrow &  N(1-\epsilon)^2 \gg 16
\end{eqnarray}
Restoring the coherence between the states $\ket{01}_q$ and $\ket{10}_q$ requires each qubit to be entangled with its nearby half of the MSS prior to the measurement on the MSS.

The target qubits' state can be expanded in the computational basis as: $\rho_q=\sum_{i,j,k,l=0}^1 c_{ij,kl} \ketbra{ij}{kl}$, with the normalization condition $c_{00,00}+c_{01,01}+c_{10,10}+c_{11,11}=1$.
The amplitude of the states $\ket{01}_q$ and $\ket{10}_q$ ($c_{01,01}$ and $c_{10,10}$) and the coherence between them ($c_{01,10}$ and $c_{10,01}$) equally contribute to the
fidelity of the target qubits' state with the maximally entangled state $\ket{m_0}=\frac{1}{\sqrt{2}}\left(\ket{01}+\ket{10}\right)$, 
\begin{eqnarray}
\label{eq:FidExp}
F_{m_0}(\rho_q)&=&\frac{c_{01,01}+c_{10,10}+c_{01,10}+c_{10,01}}{2}\\
&=&c_{01,01}+c_{01,10}=c_{01,01}\left(1+\frac{c_{01,10}}{c_{01,01}}\right)  \nonumber 
\end{eqnarray}
where $0\leq c_{01,10} \leq c_{01,01} \leq 0.5$ and the second line follows the equalities: $c_{01,01}=c_{10,10}$ and $c_{01,10}=c_{10,01}$ that hold assuming  identical states on the two qubits and their nearby halves of the MSS. 

Equation \ref{eq:FidExp} shows that 
reduction of the fidelity, as the polarization decreases, originates from two sources: leakage from the subspace spanned by $\{\ket{01}_q,\ket{10}_q\}$ to the subspace spanned by $\{\ket{00}_q,\ket{11}_q\}$ and loss of coherence between the states  $\ket{01}_q$ and $\ket{10}_q$, 
which are associated to losing macroscopic distinctness and bipartite entanglement in the micro-macro entangled state in Eq. (\ref{eq:rhoGR}), respectively.

\begin{figure}[t!h]
\centering
         \includegraphics[scale=0.75]{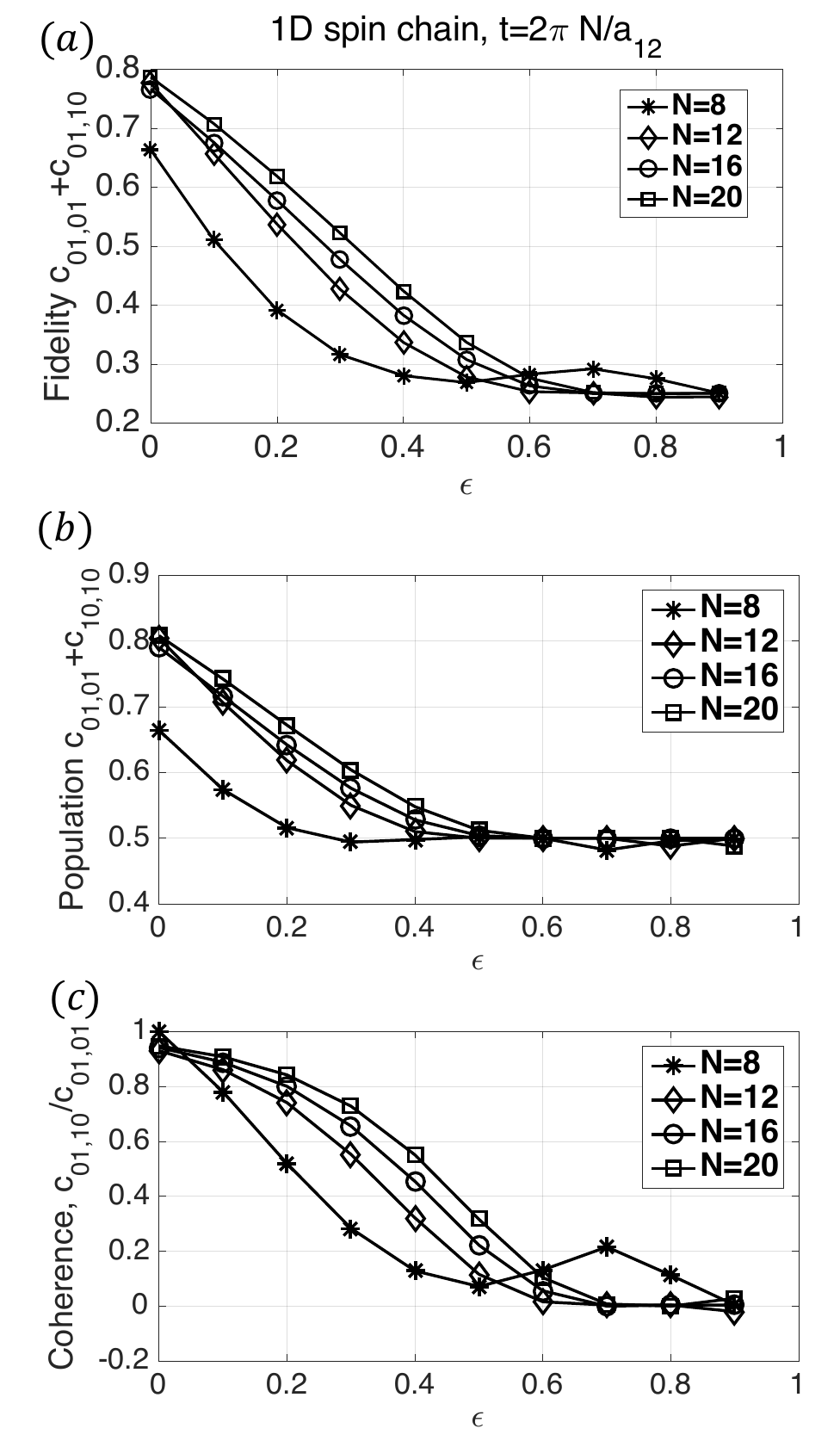}
         \caption{\label{fig:Fidelity_mixed} Simulation results of (a) the fidelity (b) the population (the diagonal terms of the density matrix) and (c) the coherence (the off-diagonal terms of the density matrix relative to diagonal terms) of the target qubits' states as a function of initial polarization of the MSS for small number of spins. Slow initial drop in the coherence follows robustness in the bipartite entanglement between each qubit and its nearby half of the MSS to polarization reduction. Fast decrease in the population and the fidelity results from small sizes of the simulated system that do not satisfy macroscopic distinctness condition.}   
\end{figure}

Figure \ref{fig:Fidelity_mixed} shows (a) the fidelity in Eq. \ref{eq:FidExp}, (b) the population in $\{\ket{01}_q,\ket{10}_q\}$ subspace ($c_{01,01}+c_{10,10}$) and (c) the coherence between the states  $\ket{01}_q$ and $\ket{10}_q$ relative to the population ($c_{01,10}/c_{01,01}$), simulated as a function of $\epsilon$ for MSSs with $N=8, 12, 16, 20$ spins.
 In these simulations, the measurement model of section \ref{sec:measurement} 
is used with the measurement parameter $\theta(m_z)=\frac{2\pi}{N(1-\epsilon)}m_z$, modified as a function of the polarization such that the measurement operator $M_1$ selects $\rho^{\text{GR}}_{01}$ and $\rho^{\text{GR}}_{10}$ over $\rho^{\text{GR}}_{00}$ and $\rho^{\text{GR}}_{11}$ with the highest probability.

These plots show that for the simulated sizes of the MSS, the fidelities drop fast with decrease in the polarization, as a result of the fast decreases in the populations in $\{\ket{01}_q,\ket{10}_q\}$ subspace.
The coherence losses happen at a slow rate 
consistent with the slow entanglement losses in the corresponding micro-macro entangled states, depicted in FIG. \ref{fig:LogNegMixed}. 

\begin{figure}[t!h]
\centering
         \includegraphics[scale=0.65]{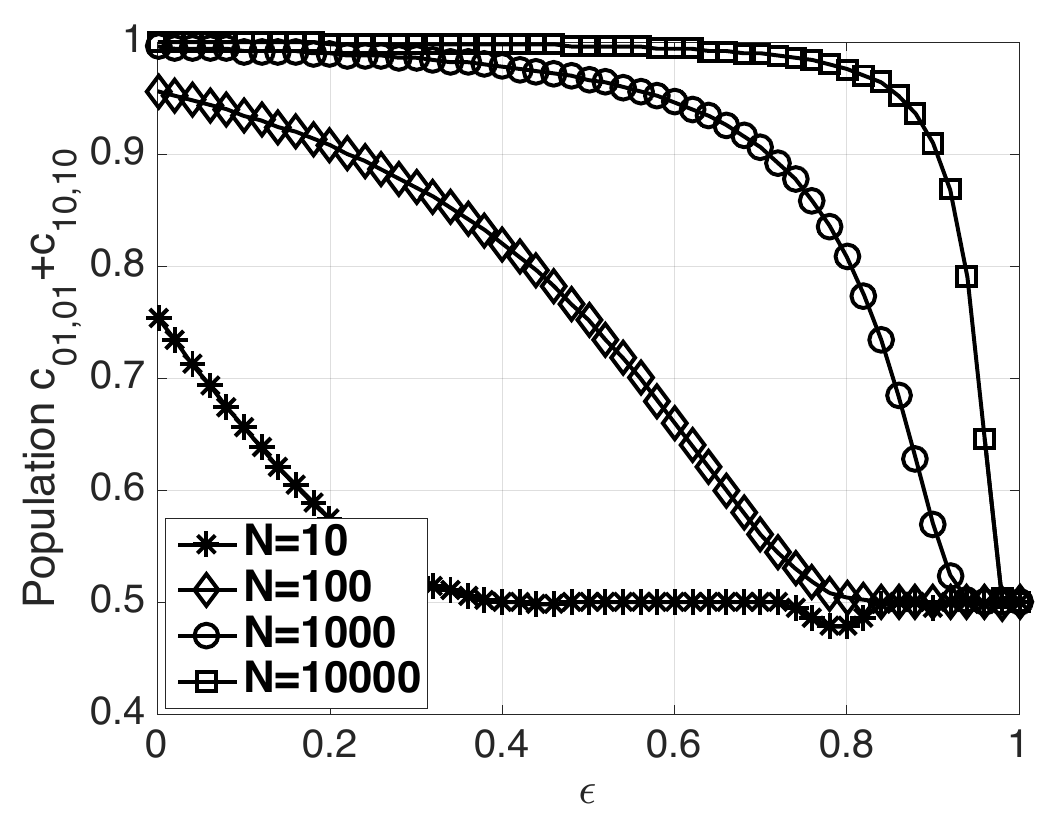}
         \caption{\label{fig:PopExt} Simulation of diagonal terms of the qubits' state, $c_{01,01}+c_{10,10}$, based on the extrapolation of the spectra of $\rho_{00}, \rho_{01}, \rho_{10}$ and $\rho_{11}$ according to the binomial distribution and using the measurement model of section \ref{sec:measurement} with $\theta(m)=\frac{2\pi}{N(1-\epsilon)}$. }   
\end{figure}

 Fast decreases in the populations, observed in FIG. \ref{fig:Fidelity_mixed}(b), are not generic effects 
 and result from the small sizes of the simulated systems, that do not satisfy the distinguishability condition: $N(1-\epsilon)^2\gg 16$. 
 For large enough systems, $N\gg 16$, 
 the population drops with a slow rate
 as $\epsilon$ grows, up to a point where the macroscopic distinctness condition 
 is not satisfied, $\epsilon\approx 1-\frac{8}{\sqrt{N}}$, as shown in FIG. \ref{fig:PopExt}. 
 Thus, the population in  $\{\ket{01}_q,\ket{10}_q\}$ subspace is close to 1 when $N(1-\epsilon)^2\gg 16$.
In addition, as FIG. \ref{fig:Fidelity_mixed}(c) shows, 
the larger the MSS, the slower the rate of the coherence loss.
   Thus, for large MSSs, $N(1-\epsilon)^2\gg 16$, both population in  $\{\ket{01}_q,\ket{10}_q\}$ subspace and coherence between the states $\ket{01}_q$ and $\ket{10}_q$, and  consequently the fidelity, $F_{m_0}(\rho_q)$, are robust to decrease in polarization of the MSS's initial state.

 In conclusion,
  bipartite entanglement between the target qubits is robust to deviation of the initial state of the MSS from the fully polarized state, as long as $N(1-\epsilon)^{2}\gg 16$. With a fixed  measurement resolution limited initial polarization of the MSS needs to be compensated for by enlarging the MSS, $N_{\epsilon}=N/(1-\epsilon)$.

\section{Sensitivity to noise} 
\label{sec:noise}
A common feature of micro-macro entangled states and more generally macroscopic superposition states is their sensitivity to 
noise \cite{Frowis18}; to the extent that the rate of coherence loss has been suggested as a measure of the macroscopicity of quantum superposition states \cite{Dur02,Kwon17}. 
This section discusses
sensitivity of the micro-macro entangled state $\ket{\phi^{\text{GR}}}_{q,\text{MSS}}$, in Eq. (\ref{eq:micmacGR}) and bipartite entanglement of the target qubits
to single particle loss\footnote{Particle loss is a common noise in photonic systems. For a spin system, in which particles are preserved, loss of a single spin models a generalized amplitude damping channel on the spin with an arbitrary fixed point and the damping probability of one.}.  

Among the two characteristics of micro-macro entanglement, namely microscopic distinctness and bipartite entanglement, 
macroscopic distinctness is, by definition, robust to single particle loss. The states of the MSS associated with $\ket{0}_q$ and $\ket{1}_q$ states of the qubit differ by many spin flips; thus, single particle loss does not significantly affect their distinctness.

Before studying the sensitivity of bipartite entanglement of the state  $\ket{\phi^{\text{GR}}}_{q,\text{MSS}}$ to particle loss, we first analyse a class of symmetric entangled states between the target qubit and the MSS, 
\begin{equation}
\label{eq:SymetricMicMac}
\ket{S_k}_{q,\text{MSS}}=\dfrac{1}{\sqrt{2}}\left(\ket{0}_q\otimes \ket{\uparrow}^{\otimes N_h}+\ket{1}_q \otimes \ket{D_k}\right)
\end{equation}
The parameter $k$ ranges from $1$ to $N_h$, and $\ket{D_k}$ is the symmetric pure state 
with $k$ spins $\ket{\downarrow}$ and $N_h-k$ spins $\ket{\uparrow}$,
\begin{equation}
\ket{D_k}=\dfrac{1}{\sqrt{\binom{N_h}{k}}}\sum_{i=1}^{\binom{N_h}{k}}P_i(\ket{\downarrow}^{\otimes k}\ket{\uparrow}^{\otimes N_h-k})
\end{equation}
where $P_i$ is the permutation operator and the summation is over all permutations.
 State $\ket{D_k}$ is an eigenstate of the collective magnetization operator, $J_z$, with the eigenvalue $m_z=\frac{N_h-2k}{2}$; hence, the (macroscopic) distinctness between the states $ \ket{\uparrow}^{\otimes N_h}$ and $\ket{D_k}$  is
proportional to $k$
  and for 
 $k\gg1$, $\ket{S_k}_{q,\text{MSS}}$ is a micro-macro entangled state.
 We show that sensitivity of bipartite entanglement between the qubit and the MSS in 
 state  $\ket{S_k}_{q,\text{MSS}}$ to single particle loss 
 increases with $k$.
  Thus, there is a trade off 
 between macroscopic distinctness of a micro-macro entangled state and robustness of its bipartite entanglement to particle loss.

The state of the target qubit and the MSS after loss of any single particle is, 
 \small
 \begin{eqnarray}
  \rho^k_{q,\text{MSS}-1}&=&p^k _{\uparrow} \ketbra{\psi^k _{\uparrow}}{\psi^k _{\uparrow} }+p^k_{\downarrow}  \ketbra{\psi^k_{\downarrow} }{\psi^k_{\downarrow}} \\
  \sqrt{p^k_{\uparrow}}\ket{\psi^k_{\uparrow}}&=&\frac{1}{\sqrt{2}}\ket{0}_q\otimes \ket{\uparrow}^{\otimes N_h-1}\nonumber \\
  &+&\frac{1}{\sqrt{2}\sqrt{\binom{N_h}{k}}}\ket{1}_q\otimes\sum_{i=1}^{\binom{N_h-1}{k}}P_i(\ket{\downarrow}^{\otimes k}\ket{\uparrow}^{\otimes N_h-k-1}) \nonumber \\
\sqrt{p^k_{\downarrow}}\ket{\psi^k_{\downarrow}}&=&\frac{1}{\sqrt{2}\sqrt{\binom{N_h}{k}}}\ket{1}_q\otimes\sum_{i=1}^{\binom{N_h-1}{k-1}}P_i(\ket{\downarrow}^{\otimes k-1}\ket{\uparrow}^{\otimes N_h-k})\nonumber
\end{eqnarray}
\normalsize 
where the states $\ket{\psi^k _{\uparrow}}$ and $\ket{\psi^k _{\downarrow}}$ are normalized and orthogonal to each other and $p^k _{\uparrow}+p^k _{\downarrow}=1$. The entanglement of projection between the target qubit and the MSS in state $\rho^k_{q,\text{MSS}-1}$ is defined as \cite{Garisto99},
\begin{equation}
\label{eq:Ep_def}
E_{p}( \rho^k_{q,\text{MSS}-1})=p_{\uparrow}^k E(\ket{\psi_{\uparrow}^k })+p^k _{\downarrow} E(\ket{\psi^k _{\downarrow}})
\end{equation} 
where $E(\ket{\psi_{AB}})$ is the von Neumann entropy of the pure bipartite state $\ket{\psi_{AB}}$, defined as $E(\ket{\psi_{AB}})=-\text{Tr}[\rho_A \log_2(\rho_A)]$ with $\rho_A=\text{Tr}_B(\rho_{AB})$. 
Entanglement of projection ranges between $0$ to $1$ and is an upper bound for entanglement of formation \cite{Hill97,Wootters98}.

State $\ket{\psi^k _{\downarrow}}$ is a separable state ,thus $E(\psi^k _{\downarrow})=0$. 
The von Neumann entropy of state  $\ket{\psi^k _{\uparrow}}$ is,
\begin{eqnarray}
E(\ket{\psi^k _{\uparrow}})=\frac{1}{1+r_{\uparrow}}\log_2(\frac{1}{1+r_{\uparrow}})+\frac{r_{\uparrow}}{1+r_{\uparrow}}\log_2(\frac{r_{\uparrow}}{1+r_{\uparrow}})
\end{eqnarray}
where $r_{\uparrow}(k)=\binom{N_h-1}{k}/\binom{N_h}{k}= 1-\frac{k}{N_h}$ is the probability of finding one spin $\ket{\uparrow}$ in the state $\ket{D_k}$ of the MSS, 
 and ranges from $r_{\uparrow}=0$, for $k=N_h$, to $r_{\uparrow}=1-\frac{1}{N_h}$, for $k=1$. 
The entanglement of projection of state $\rho^k_{q,\text{MSS}-1}$, according to Eq. (\ref{eq:Ep_def}), is,
\begin{eqnarray}
\label{eq:Ep}
E_{p}(\rho_{q,\text{MSS}-1}^k)&=& \nonumber \\
E_{p}^r(r_{\uparrow})&:=& -\frac{1+r_{\uparrow}}{2}\left(\frac{1}{1+r_{\uparrow}}\log_2(\frac{1}{1+r_{\uparrow}}) \right. \nonumber\\
&+&\left. \frac{r_{\uparrow}}{1+r_{\uparrow}}\log_2(\frac{r_{\uparrow}}{1+r_{\uparrow}})\right) 
\end{eqnarray}
Note that $E_{p}( \rho^k_{q,\text{MSS}-1})$ depends on the ratio $k/N_h$ and not on $k$ and $N_h$, independently.

\begin{figure}[t!h]
\centering
         \includegraphics[scale=0.5]{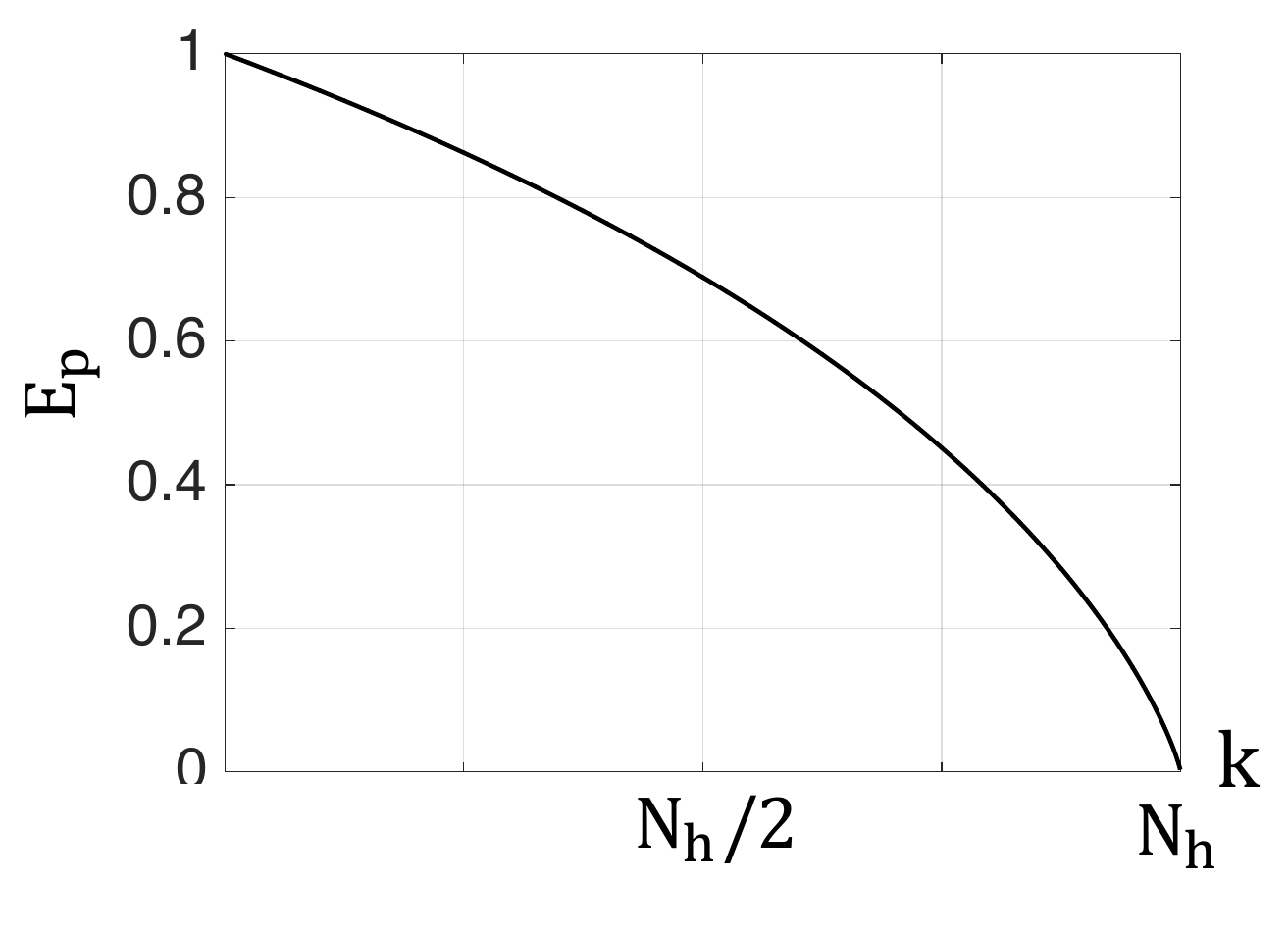}
         \caption{\label{fig:Ep_sym} Entanglement of projection of the symmetric bipartite entangled state $\ket{S_k}_{q,\text{MSS}}$ upon single particle loss as a function of macroscopic distinctness between $\ket{\uparrow}^{N_h}$ and $\ket{D_k}$. The more macroscopically distinct the states $\ket{\uparrow}^{N_h}$ and $\ket{D_k}$ are, the more fragile the bipartite entanglement of $\ket{S_k}_{q,\text{MSS}}$ is.}   
\end{figure}  
Figure \ref{fig:Ep_sym} plots $E_{p}( \rho^k_{q,\text{MSS}-1})$ as a function of the macroscopic distinctness, $k$. 
The bipartite entanglement between the target qubit and the MSS becomes more fragile to particle loss as the macroscopic distinctness in state $\ket{S_k}_{q,\text{MSS}}$ increases. At the limit of maximum macroscopic distinctness, $\ket{D_{N_h}}=\ket{\downarrow}^{\otimes N_h}$, $\ket{S_k}_{q,\text{MSS}}$ represents an overall GHZ state, and no entanglement will remain after loss of one particle from the MSS.

The entangled state of interest, $\ket{\phi^{\text{GR}}(t)}_{q,\text{MSS}}$ in Eq. (\ref{eq:micmacGR}), 
follows a similar form to
state $\ket{S_k}_{q,\text{MSS}}$ in Eq. \ref{eq:SymetricMicMac} except that $\ket{D_k}$ is replaced by $\ket{\psi_1^{\text{GR}}(t)}$. The state $\ket{\psi_1^{\text{GR}}(t)}$ is not necessarily symmetric; thus, to quantify the sensitivity of state $\ket{\phi^{\text{GR}}(t)}_{q,\text{MSS}}$ to single particle loss, we average the entanglement of projection upon losing each of the spins in the MSS, assuming that all spins have equal probabilities of being lost. 
\begin{figure}[t!h]
\centering
         \includegraphics[scale=0.39]{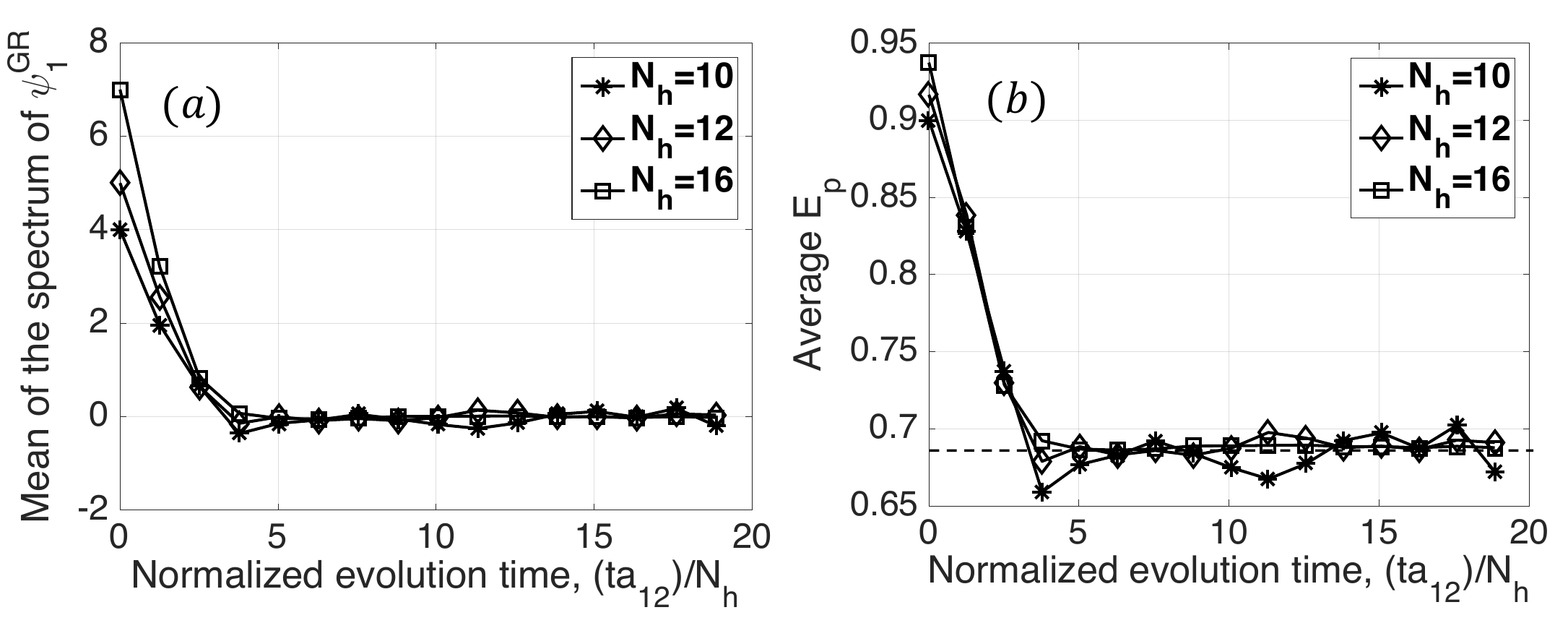}
         \caption{\label{fig:Ep_GR} (a) The mean of the spectrum of state $\ket{\psi_1^{\text{GR}}(t)}$ as a measure of macroscopic distinctness between $\ket{\psi_1^{\text{GR}}(t)}$ and $\ket{\uparrow}^{N_h}$. (b) The entanglement of projection of the state $\ket{\phi^{\text{GR}}(t)}_{q,\text{MSS}}$ upon single spin loss as a function of time. Bipartite entanglement of state $\ket{\phi^{\text{GR}}(t)}_{q,\text{MSS}}$ upon single particle loss reduces with increase in the macroscopic distinctness between $\ket{\psi_1^{\text{GR}}(t)}$ and $\ket{\uparrow}^{N_h}$ until it reaches the asymptotic value $2/3$. This asymptotic value is similar for all sizes of the MSS and corresponds to difference in the mean of the collective $J_z$ magnetization, $\left\langle J_z\right\rangle_0^{\text{GR}}- \left\langle J_z \right\rangle_1^{\text{GR}} \sim \frac{1}{2} N_h/2$, similar to symmetric bipartite entangled state $\ket{S_k}_{q,\text{MSS}}$ with $k=N_h/2$. }   
\end{figure}  

Figure \ref{fig:Ep_GR} shows simulation results for the mean of the spectrum of state $\ket{\psi_1^{\text{GR}}(t)}$ and the entanglement of projection of state $\ket{\phi^{\text{GR}}(t)}_{q,\text{MSS}}$ upon single particle loss as a function of the evolution time for different sizes of  MSS. As evolution time increases, the mean of the spectrum of state $\ket{\psi_1^{\text{GR}}(t)}$ decreases and macroscopic distinctness in the state $\ket{\phi^{\text{GR}}(t)}_{q,\text{MSS}}$ and its sensitivity to particle loss increase. 
For long evolution times, the probability of finding any of the spins in the MSS in state $\ket{\uparrow}$ is close to $1/2$. Thus, the asymptotic value of the average of entanglement of projection 
 is,
 \begin{equation}
 \sum_{j=1}^{N_h}E_{p}(\text{Tr}_j(\ketbra{\phi^{\text{GR}}(t)}{\phi^{\text{GR}}(t)}))\approx E_{p}^{r}(r_{\uparrow}=1/2)=\frac{2}{3}
 \end{equation}

Decrease in bipartite entanglement of the micro-macro entangled state $\ket{\phi^{\text{GR}}}$, upon particle loss, is reflected in the fidelity of the target qubits' entangled state.
Even with an ideal measurement on the MSS and post-selection, that perfectly selects the $\{\ket{01}_q,\ket{10}_q\}$ subspace of the qubits over the $\{\ket{00}_q,\ket{11}_q\}$ subspace, the fidelity can not be greater than 
 $F_{\max}=c_{01,01}^{\max}+c_{01,10}^{\max}=1/2+1/4=3/4$ (See Appendix \ref{ap:1}).
 This upper bound follows the reduced coherence between $\ket{01}_q$ and $\ket{10}_q$ in the state of the qubits, even when the population is preserved.

Fragility of the micro-macro entangled state and bipartite entanglement of the target qubits to particle loss illustrates 
the importance of 
shorter transient time with a MSS that has a 2D or 3D structure compared to a 1D chain, 
demonstrated in section \ref{sec:magnification}.
Since 
any loss in the MSS results in a reduction of bipartite entanglement between the target qubits,
the overall experiment time needs to be much shorter than  $T_1$ divided by the number of spins in the MSS, $t_{exp} \ll \frac{T_1}{N_h}$.

\section{Discussion and Conclusion}
\label{sec:conclusion}

We analyzed the resources required for entangling two uncoupled spin qubits through an intermediate mesoscopic spin system by indirect joint magnetization measurement. 
In contrast to direct joint measurement, that needs a high-resolution 
apparatus capable of detecting a single qubit flip to entangle two qubits, indirect joint measurement benefits from coherent magnification of the target spin qubits' state in the collective magnetization of the MSS and only requires a low-resolution collective measurement on the MSS. 
This work complements the ongoing efforts in using mesoscopic systems as coherent control elements in coupling separated qubits 
\cite{Sorensen04,Trifunovic13}
\cite{Bose3,Christandl4,Kay6,Avellino6,Burgarth7, Cappellaro07,Franco8,Ramanathan11,Cappellaro11,Ajoy13}.

A MSS consisting of two non-interacting halves, each coupled to one of the target qubits, was identified as a practically helpful geometry, that allows 
implementing the coherent magnification process with 
 experimentally available control tools; namely local interaction between each target qubit and the MSS, naturally occurring dipolar coupling among the spins in each half of the MSS and collective rotations on the MSS.
It was demonstrated that the requirements 
on the pre-measurement state of the target qubits and the MSS,  
entirely fulfill 
the 
specifications of micro-macro entanglement between each target qubit and its nearby half of the MSS. 
It has been shown that direct experimental demonstration of micro-macro entanglement is challenging \cite{Sekatski14PRL,Wang13}. 
 Verification of bipartite entanglement between the target qubits provides a means of proving micro-macro entanglement between each target qubit and half of the MSS in the pre-measurement state.

 The numerical simulations showed that available internal dipolar interaction and collective control can be used to prepare each half of the MSS in a globally correlated state, such that a one-time interaction between each target qubit and a nearby spin within the MSS suffices to magnify the qubit's state in the collective magnetization of the MSS and create a micro-macro entangled state. 
The time scale of the magnification process was discussed. 
In particular, it was demonstrated that, with long-range dipolar coupling in the MSS, the magnification time scales sub-linear with the size of the MSS regardless of the dimension of its structure. Moreover, it was shown that the magnification time is much shorter with a MSS in 2D and 3D lattices compared to a 1D spin chain.

It was shown that a low-resolution collective magnetization measurement on the MSS capable of detecting only $(1-\epsilon)N/4$ spin flips, where $N$ is the number of spins in the MSS and $1-\epsilon$ is the polarization of each spin, suffices to distinguish between the states of the MSS correlated with different magnetizations of the target qubits. 
The measurement must also probabilistically project the state of the MSS into the subspace associated with zero magnetization of the target qubits, with minimum disturbance.

\begin{figure}[t!h]
\centering
         \includegraphics[scale=0.65]{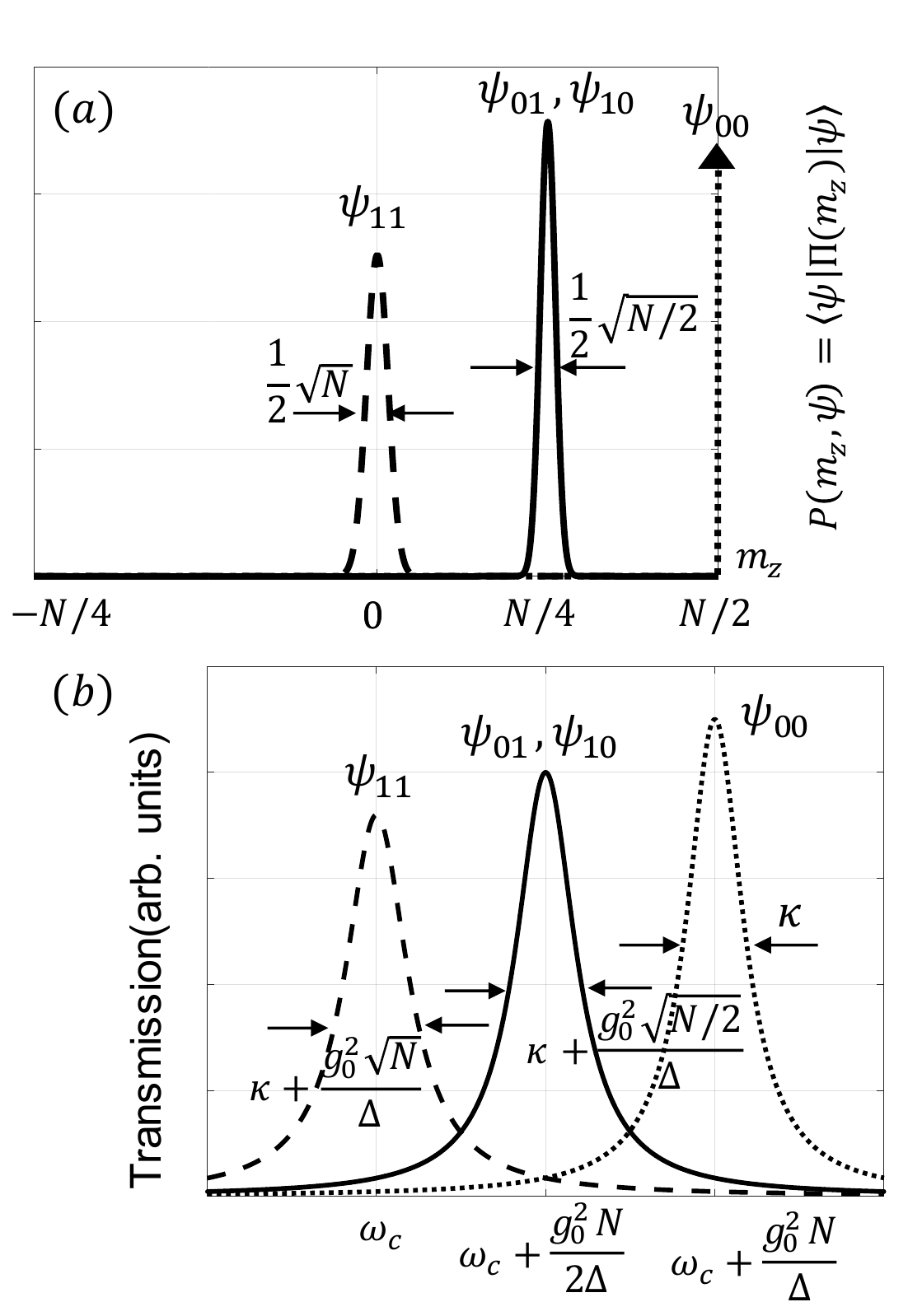}
         \caption{\label{fig:cavity_m} (a) The spectra of the MSS's states correlated to different states of the target qubits. (b) The expected transmission probability of a photon through a cavity coupled to the MSS in its dispersive regime. The unloaded resonance frequency of the cavity is $\omega_c$, $g_o$ is the coupling strength between a single spin in the MSS and the cavity, $\kappa$ is the cavity loss and $\Delta$ is the difference between the resonance frequency of the cavity and the spin system. The resolution of the measurement is high enough if the three peaks corresponding to different states of the MSS can be resolved, $\frac{g_0^2 N}{2\Delta}>\kappa+\frac{g_0^2 \sqrt{N}}{\Delta}$.}
         \end{figure}

Different scenarios can be considered for implementing such a measurement.
When a linearly-polarized photon passes through a magnetic material, its polarization   rotates
 depending on the magnetic moment of the medium, according to the Faraday rotation effect.
The Faraday effect  
follows the required collective dynamics and has been proposed as a means for implementing a quantum non-demolition  measurement on an ensemble of spins \cite{Takahashi99}. 
Strong coupling to a superconducting cavity may provide another means for implementing a 
collective measurement on the MSS that follows the required state-update-rule.
 Measurement through a cavity in the dispersive regime has been used to entangle two superconducting qubits \cite{Riste13,Roch14}
 where an incoming photon is transmitted through or reflected from the cavity depending on the joint state of the qubits.
 In these experiments, couplings between the superconducting qubits and the cavities are so strong that a single qubit flip results in a detectable shift in the resonance frequency of the cavity. 
 Couplings between spin qubits and superconducting cavities are too weak to enable direct joint measurement of two spin qubits.
Correlating different states of the spin qubits with macroscopically distinct states of the MSS can in principle compensate for this weak coupling
since the shift in the resonance frequency of the cavity corresponding to different states of the qubits scales proportionally to the size of the MSS. See FIG. \ref{fig:cavity_m}. 
Faraday rotation and state-dependent shift of a cavity's resonance frequency are examples of two phenomena that 
potentially enable measurements that are collective and update the MSS's state according to the measurement outcome. 
Evaluating the details of the measurements based on these phenomena \footnote{As an example, coupling of a spin ensemble with a cavity in the dispersive regime introduces indirect interaction between the spins mediated through the cavity. Effect of such  interactions on the measurement needs to be explored.} and their resolution 
need to be further explored.

 Indirect joint measurement through a MSS was shown to be robust to limited initial polarization of the MSS as long as $N(1-\epsilon)^2 \gg 16$ and the measurement resolution is high enough to detect $(1-\epsilon)N/4$ spin flips.
Thermal polarization of an ensemble of electron spins is close to one at low temperatures and high magnetic fields (e.g. $T\approx 1$K and $B\approx 7 $T). 
Hyperpolarization of nuclear spins may be achieved through dynamic nuclear magnetization processes that transfer polarization from electron spins to nuclear spins \cite{McCamey09, Gumann14}.

The process of entangling non-interacting qubits by indirect joint measurement 
is inevitably sensitive to  noise in the MSS. It was shown that single particle loss in the MSS reduces the upper bound on the fidelity of the target qubits' state with the intended maximally entangled state from $1$ to $3/4$.
Thus, creating highly entangled target qubits requires the relaxation time of the MSS to be long compared to the number of spins in the MSS times experiment time, $T_1\gg N t_{exp}$.

Different factors compete in determining the practical size of the MSS. 
The number of spins in the MSS needs to be large enough to satisfy the macroscopic distinctness condition, $N_1\gg 16/(1-\epsilon)^2$, and the distinguishability criteria according to the resolution of the measurement, $N_2 > 4\Delta m/ (1-\epsilon)$, where $\Delta m$ is the minimum number of spin flips the measurement apparatus can detect. 
The overall lower bound on the size of the  MSS is $N_{min}=\max(\min(N_1),\min(N_2))$.
 Upper bound on the size of the MSS is imposed by the fragility of the micro-macro entangled state between the target qubits and the MSS and as a result the fragility of the bipartite entanglement of the target qubits to noise, $N_{max}\ll T_1/t_{exp}$.

To summarize, among the required resources, 
the control tools are available, 
 highly polarized initial states of the MSS and long $T_1$ relaxation times are feasible at low temperatures. 
 The bottleneck 
 is implementing a collective  measurement on the many-body state of the MSS that follows the required state-update-rule; namely probabilistic selection of the states of the MSS correlated to zero magnetization of the qubits over the states of the MSS correlated to $\pm 1$ magnetizations of the qubits with minimum disturbance to the selected states. 

\section{Acknowledgements}
This work was supported by Industry Canada, the Canada First Research Excellence Fund (CFREF), the Canadian Excellence Research Chairs (CERC 215284) program, the Natural Sciences and Engineering Research Council of Canada (NSERC RGPIN-418579) Discovery program, the Canadian Institute for Advanced Research (CIFAR), and Province of Ontario.

\appendix

\section{Upper bound on Fidelity upon single particle loss}
\label{ap:1}

The sensitivity of the micro-macro entangled states and bipartite entanglement between the target qubits to spin loss in the MSS was discussed in section \ref{sec:noise}. In particular, it was specified that the upper bound on the fidelity of the target qubits' state with the maximally entangled state $\ket{m_0}=\frac{1}{\sqrt{2}}(\ket{01}_q+\ket{10}_q)$, upon single particle loss, reduces from one to $3/4$ and this decrease solely originates from reduction of the coherence between $\ket{01}_q$ and $\ket{10}_q$ states.
In this appendix, we prove this upper bound on the fidelity.

The state of the target qubits and the MSS after the magnification process is: $\ket{\phi^{\text{GR}}}_L\otimes \ket{\phi^{\text{GR}}}_R$ where $\ket{\phi^{\text{GR}}}_i=\dfrac{1}{\sqrt{2}}(\ket{0}_{q_i}\ket{\uparrow}^{\otimes N_h}+\ket{1}_{q_i}\ket{\psi_1^{\text{GR}}}_i$ with $i=L,R$ is the state of each qubit and its nearby half of the MSS. Let's consider that the state of one spin from the MSS with index $\mathfrak{a}$ is lost and is replaced by a state $\rho_{\mathfrak{a}}$. 
This particle loss process corresponds to a generalized amplitude damping map with the fixed point $\rho_{\mathfrak{a}}$ and the damping probability of one on spin $\mathfrak{a}$.
Without loss of generality, we assume the lost spin is in the left half of the MSS, $\mathfrak{a}\leq N/2$. The state $\ket{\psi_1^{\text{GR}}}_L$ 
can be expanded in 
the basis $\{\ket{\uparrow}_{\mathfrak{a}},\ket{\downarrow}_{\mathfrak{a}}\}$  of spin $\mathfrak{a}$ as, 
\begin{equation}
\ket{\psi_1^{\text{GR}}}_L=\alpha_{\mathfrak{a}} \ket{\psi_{\alpha}^{\mathfrak{a}}}\ket{\uparrow}_{\mathfrak{a}}+\beta_{\mathfrak{a}} \ket{\psi_{\beta}^{\mathfrak{a}}}\ket{\downarrow}_{\mathfrak{a}}
\end{equation}
 where $\alpha_{\mathfrak{a}} \ket{\psi_{\alpha}^{\mathfrak{a}}}=(\bra{\uparrow}_{\mathfrak{a}}\otimes \id_{N_h-1})\ket{\psi_1^{\text{GR}}}_L$, $\beta_{\mathfrak{a}} \ket{\psi_{\beta}^{\mathfrak{a}}}=(\bra{\downarrow}_{\mathfrak{a}}\otimes \id_{N_h-1})\ket{\psi_1^{\text{GR}}}_L$, and $|\alpha_{\mathfrak{a}}|^2+|\beta_{\mathfrak{a}}|^2=1$ following the normalization of state $\ket{\psi_1^{\text{GR}}}_L$.
 After particle loss the state of the target qubit $q_L$ and its nearby half of the MSS will be, 
\begin{equation}
\rho_{q_L,\text{MSS}_L}^{\text{GR}}=\left(p_{\uparrow}\ketbra{\psi_{\uparrow}^{\mathfrak{a}}}{\psi_{\uparrow}^{\mathfrak{a}}}+p_{\downarrow}\ketbra{\psi_{\downarrow}^{\mathfrak{a}}}{\psi_{\downarrow}^{\mathfrak{a}}}\right)\otimes \rho_{\mathfrak{a}}
\end{equation}
where the states  $\ket{\psi_{\uparrow}^{\mathfrak{a}}}$ and $\ket{\psi_{\downarrow}^{\mathfrak{a}}}$ are,
\begin{eqnarray}
\sqrt{p_{\uparrow}}\ket{\psi_{\uparrow}^{\mathfrak{a}}}&=&\frac{1}{\sqrt{2}}\left(\ket{0}_{q_L}\ket{\uparrow}^{\otimes N_h-1}+\ket{1}_{q_L}(\alpha_{\mathfrak{a}} \ket{\psi_{\alpha}^{\mathfrak{a}}})\right) \nonumber \\
\sqrt{p_{\downarrow}}\ket{\psi_{\downarrow}^{\mathfrak{a}}}&=&\frac{\ket{1}_{q_L}}{\sqrt{2}}\beta_{\mathfrak{a}} \ket{\psi_{\beta}^{\mathfrak{a}}}
\end{eqnarray}
An ideal measurement that perfectly selects the states of the MSS correlated to zero magnetization of the qubits over the states of the MSS correlated to $\pm 1$ magnetizations of the qubits, updates the state $\rho_{q_L,\text{MSS}_L}^{\text{GR}}\otimes \ketbra{\phi^{\text{GR}}}{\phi^{\text{GR}}}_R$ of the target qubits and the MSS to the state,
\begin{widetext}
\begin{eqnarray}
\rho_{q_L q_R,\text{MSS}} &=&\frac{1}{2}\left(\ketbra{01}{01}\otimes (\ketbra{\uparrow}{\uparrow}^{\otimes N_h-1}\otimes \rho_{\mathfrak{a}})\otimes \ketbra{\psi_1^{\text{GR}}}{\psi_1^{\text{GR}}}_R \right. \\
&+&  (\ketbra{10}{10}\otimes ((|\alpha_{\mathfrak{a}} |^2\ketbra{\psi_{\alpha}^{\mathfrak{a}}}{\psi_{\alpha}^{\mathfrak{a}}}+ |\beta_{\mathfrak{a}} |^2\ketbra{\psi_{\beta}^{\mathfrak{a}}}{\psi_{\beta}^{\mathfrak{a}}})\otimes \rho_{\mathfrak{a}})\otimes \ketbra{\uparrow}{\uparrow}^{\otimes N_h} \nonumber \\
&+& \ketbra{01}{10} \otimes (\alpha_{\mathfrak{a}}^{\star}\ket{\uparrow}^{N_h-1}\bra{\psi_{\alpha}^{\mathfrak{a}}}\otimes \rho_{\mathfrak{a}})\otimes \ket{\psi_1^{\text{GR}}}_R\bra{\uparrow}^{\otimes N_h} \nonumber \\
&+& \left. \ketbra{10}{01} \otimes (\alpha_{\mathfrak{a}} \ket{\psi_{\alpha}^{\mathfrak{a}}}\bra{\uparrow}^{N_h-1}\otimes \rho_{\mathfrak{a}})\otimes \ket{\uparrow}^{\otimes N_h}\bra{\psi_1^{\text{GR}}}_R \right) \nonumber 
\end{eqnarray}
\end{widetext}

Note that the off-diagonal terms of the qubits are scaled with $\alpha_{\mathfrak{a}}$ and $\alpha_{\mathfrak{a}}^{\star}$. 
This state is a correlated state between the target qubits and the MSS. The following disentangling gate needs to restore the coherence between the target qubits.
The maximum retrievable coherence between the target qubits is $|\alpha_{\mathfrak{a}}|^2$  that corresponds to the state $\rho_{\mathfrak{a}}=\ketbra{\uparrow}{\uparrow}$. With this choice of the $\rho_{\mathfrak{a}}$, after applying the disentangling gate and tracing over the MSS, the target qubits' state will be,
\begin{eqnarray}
\rho_{q_Lq_R}&=&\frac{1}{2}\left(\ketbra{01}{01}+\ketbra{01}{01} \right. \nonumber \\
&+& \left. |\alpha_{\mathfrak{a}}|^2 \ketbra{01}{10}+|\alpha_{\mathfrak{a}}|^2 \ketbra{10}{01} \right)
\end{eqnarray}
The fidelity of this state with the maximally entangled triplet zero state is $F_{\text{max}}^{\mathfrak{a}}=(1+|\alpha_{\mathfrak{a}}|^2)/2$ where $|\alpha_{\mathfrak{a}}|^2$ can be interpreted as the probability of finding spin $\mathfrak{a}$ in state $\ket{\uparrow}_{\mathfrak{a}}$ when the MSS is in state $\ket{\psi_1^{\text{GR}}}_L$. 
We know that the mean of the collective magnetization spectrum of state $\ket{\psi_1^{\text{GR}}}_L$ is zero, which is mathematically equivalent to $\sum_{\mathfrak{a}} |\alpha_{\mathfrak{a}}|^2 = \sum_{\mathfrak{a}} |\beta_{\mathfrak{a}}|^2$. Combining with the normalization condition $|\alpha_{\mathfrak{a}}|+|\beta_{\mathfrak{a}}|=1$, the average of maximum fidelity upon loss of each particle in the MSS is,
\begin{equation}
F_{\text{max}}:=\frac{1}{N}\sum_{\mathfrak{a}}F_{\text{max}}^{\mathfrak{a}}=\frac{1}{2}+\frac{\sum_{\mathfrak{a}} |\alpha_{\mathfrak{a}}|}{2N}=\frac{1}{2}+\frac{1}{4}=\frac{3}{4}
\end{equation}
This relation completes the proof for the upper bound on the fidelity of the target qubits' state.

Based on the dynamics that create state $\ket{\psi_1^{\text{GR}}}_L$, the magnetization is expected to be distributed uniformly among the spins in the MSS, thus $|\alpha_{\mathfrak{a}}|^2$ is anticipated to be close to $1/2$ for all spins and the maximum fidelity upon loss of any spin is expected to be $F_{\text{max}}^{\mathfrak{a}}\approx 3/4$.


\normalsize

\bibliographystyle{apsrev4-1} 
\bibliography{references}

\begin{thebibliography}{66}%
\makeatletter
\providecommand \@ifxundefined [1]{%
 \@ifx{#1\undefined}
}%
\providecommand \@ifnum [1]{%
 \ifnum #1\expandafter \@firstoftwo
 \else \expandafter \@secondoftwo
 \fi
}%
\providecommand \@ifx [1]{%
 \ifx #1\expandafter \@firstoftwo
 \else \expandafter \@secondoftwo
 \fi
}%
\providecommand \natexlab [1]{#1}%
\providecommand \enquote  [1]{``#1''}%
\providecommand \bibnamefont  [1]{#1}%
\providecommand \bibfnamefont [1]{#1}%
\providecommand \citenamefont [1]{#1}%
\providecommand \href@noop [0]{\@secondoftwo}%
\providecommand \href [0]{\begingroup \@sanitize@url \@href}%
\providecommand \@href[1]{\@@startlink{#1}\@@href}%
\providecommand \@@href[1]{\endgroup#1\@@endlink}%
\providecommand \@sanitize@url [0]{\catcode `\\12\catcode `\$12\catcode
  `\&12\catcode `\#12\catcode `\^12\catcode `\_12\catcode `\%12\relax}%
\providecommand \@@startlink[1]{}%
\providecommand \@@endlink[0]{}%
\providecommand \url  [0]{\begingroup\@sanitize@url \@url }%
\providecommand \@url [1]{\endgroup\@href {#1}{\urlprefix }}%
\providecommand \urlprefix  [0]{URL }%
\providecommand \Eprint [0]{\href }%
\providecommand \doibase [0]{http://dx.doi.org/}%
\providecommand \selectlanguage [0]{\@gobble}%
\providecommand \bibinfo  [0]{\@secondoftwo}%
\providecommand \bibfield  [0]{\@secondoftwo}%
\providecommand \translation [1]{[#1]}%
\providecommand \BibitemOpen [0]{}%
\providecommand \bibitemStop [0]{}%
\providecommand \bibitemNoStop [0]{.\EOS\space}%
\providecommand \EOS [0]{\spacefactor3000\relax}%
\providecommand \BibitemShut  [1]{\csname bibitem#1\endcsname}%
\let\auto@bib@innerbib\@empty
\bibitem [{\citenamefont {{S\o{}rensen, Anders S. and van der Wal, Caspar H.
  and Childress, Lilian I. and Lukin, Mikhail D.}}(2004)}]{Sorensen04}%
  \BibitemOpen
  \bibfield  {author} {\bibinfo {author} {\bibnamefont {{S\o{}rensen, Anders S.
  and van der Wal, Caspar H. and Childress, Lilian I. and Lukin, Mikhail
  D.}}},\ }\href {\doibase 10.1103/PhysRevLett.92.063601} {\bibfield  {journal}
  {\bibinfo  {journal} {Phys. Rev. Lett.}\ }\textbf {\bibinfo {volume} {92}},\
  \bibinfo {pages} {063601} (\bibinfo {year} {2004})}\BibitemShut {NoStop}%
\bibitem [{\citenamefont {Trifunovic}\ \emph {et~al.}(2013)\citenamefont
  {Trifunovic}, \citenamefont {Pedrocchi},\ and\ \citenamefont
  {Loss}}]{Trifunovic13}%
  \BibitemOpen
  \bibfield  {author} {\bibinfo {author} {\bibfnamefont {L.}~\bibnamefont
  {Trifunovic}}, \bibinfo {author} {\bibfnamefont {F.~L.}\ \bibnamefont
  {Pedrocchi}}, \ and\ \bibinfo {author} {\bibfnamefont {D.}~\bibnamefont
  {Loss}},\ }\href {\doibase 10.1103/PhysRevX.3.041023} {\bibfield  {journal}
  {\bibinfo  {journal} {Phys. Rev. X}\ }\textbf {\bibinfo {volume} {3}},\
  \bibinfo {pages} {041023} (\bibinfo {year} {2013})}\BibitemShut {NoStop}%
\bibitem [{\citenamefont {Elman}\ \emph {et~al.}(2017)\citenamefont {Elman},
  \citenamefont {Bartlett},\ and\ \citenamefont {Doherty}}]{Elman17}%
  \BibitemOpen
  \bibfield  {author} {\bibinfo {author} {\bibfnamefont {S.~J.}\ \bibnamefont
  {Elman}}, \bibinfo {author} {\bibfnamefont {S.~D.}\ \bibnamefont {Bartlett}},
  \ and\ \bibinfo {author} {\bibfnamefont {A.~C.}\ \bibnamefont {Doherty}},\
  }\href {\doibase 10.1103/PhysRevB.96.115407} {\bibfield  {journal} {\bibinfo
  {journal} {Phys. Rev. B}\ }\textbf {\bibinfo {volume} {96}},\ \bibinfo
  {pages} {115407} (\bibinfo {year} {2017})}\BibitemShut {NoStop}%
\bibitem [{\citenamefont {Benito}\ \emph {et~al.}(2016)\citenamefont {Benito},
  \citenamefont {Schuetz}, \citenamefont {Cirac}, \citenamefont {Platero},\
  and\ \citenamefont {Giedke}}]{Benito16}%
  \BibitemOpen
  \bibfield  {author} {\bibinfo {author} {\bibfnamefont {M.}~\bibnamefont
  {Benito}}, \bibinfo {author} {\bibfnamefont {M.~J.~A.}\ \bibnamefont
  {Schuetz}}, \bibinfo {author} {\bibfnamefont {J.~I.}\ \bibnamefont {Cirac}},
  \bibinfo {author} {\bibfnamefont {G.}~\bibnamefont {Platero}}, \ and\
  \bibinfo {author} {\bibfnamefont {G.}~\bibnamefont {Giedke}},\ }\href
  {\doibase 10.1103/PhysRevB.94.115404} {\bibfield  {journal} {\bibinfo
  {journal} {Phys. Rev. B}\ }\textbf {\bibinfo {volume} {94}},\ \bibinfo
  {pages} {115404} (\bibinfo {year} {2016})}\BibitemShut {NoStop}%
\bibitem [{\citenamefont {Yang}\ \emph {et~al.}(2016)\citenamefont {Yang},
  \citenamefont {Hsu}, \citenamefont {Stano}, \citenamefont {Klinovaja},\ and\
  \citenamefont {Loss}}]{Yang16}%
  \BibitemOpen
  \bibfield  {author} {\bibinfo {author} {\bibfnamefont {G.}~\bibnamefont
  {Yang}}, \bibinfo {author} {\bibfnamefont {C.-H.}\ \bibnamefont {Hsu}},
  \bibinfo {author} {\bibfnamefont {P.}~\bibnamefont {Stano}}, \bibinfo
  {author} {\bibfnamefont {J.}~\bibnamefont {Klinovaja}}, \ and\ \bibinfo
  {author} {\bibfnamefont {D.}~\bibnamefont {Loss}},\ }\href {\doibase
  10.1103/PhysRevB.93.075301} {\bibfield  {journal} {\bibinfo  {journal} {Phys.
  Rev. B}\ }\textbf {\bibinfo {volume} {93}},\ \bibinfo {pages} {075301}
  (\bibinfo {year} {2016})}\BibitemShut {NoStop}%
\bibitem [{\citenamefont {Szumniak}\ \emph {et~al.}(2015)\citenamefont
  {Szumniak}, \citenamefont {Paw\l{}owski}, \citenamefont {Bednarek},\ and\
  \citenamefont {Loss}}]{Szumniak15}%
  \BibitemOpen
  \bibfield  {author} {\bibinfo {author} {\bibfnamefont {P.}~\bibnamefont
  {Szumniak}}, \bibinfo {author} {\bibfnamefont {J.}~\bibnamefont
  {Paw\l{}owski}}, \bibinfo {author} {\bibfnamefont {S.}~\bibnamefont
  {Bednarek}}, \ and\ \bibinfo {author} {\bibfnamefont {D.}~\bibnamefont
  {Loss}},\ }\href {\doibase 10.1103/PhysRevB.92.035403} {\bibfield  {journal}
  {\bibinfo  {journal} {Phys. Rev. B}\ }\textbf {\bibinfo {volume} {92}},\
  \bibinfo {pages} {035403} (\bibinfo {year} {2015})}\BibitemShut {NoStop}%
\bibitem [{\citenamefont {Trifunovic}\ \emph {et~al.}(2012)\citenamefont
  {Trifunovic}, \citenamefont {Dial}, \citenamefont {Trif}, \citenamefont
  {Wootton}, \citenamefont {Abebe}, \citenamefont {Yacoby},\ and\ \citenamefont
  {Loss}}]{Trifunovic12}%
  \BibitemOpen
  \bibfield  {author} {\bibinfo {author} {\bibfnamefont {L.}~\bibnamefont
  {Trifunovic}}, \bibinfo {author} {\bibfnamefont {O.}~\bibnamefont {Dial}},
  \bibinfo {author} {\bibfnamefont {M.}~\bibnamefont {Trif}}, \bibinfo {author}
  {\bibfnamefont {J.~R.}\ \bibnamefont {Wootton}}, \bibinfo {author}
  {\bibfnamefont {R.}~\bibnamefont {Abebe}}, \bibinfo {author} {\bibfnamefont
  {A.}~\bibnamefont {Yacoby}}, \ and\ \bibinfo {author} {\bibfnamefont
  {D.}~\bibnamefont {Loss}},\ }\href {\doibase 10.1103/PhysRevX.2.011006}
  {\bibfield  {journal} {\bibinfo  {journal} {Phys. Rev. X}\ }\textbf {\bibinfo
  {volume} {2}},\ \bibinfo {pages} {011006} (\bibinfo {year}
  {2012})}\BibitemShut {NoStop}%
\bibitem [{\citenamefont {Friesen}\ \emph {et~al.}(2007)\citenamefont
  {Friesen}, \citenamefont {Biswas}, \citenamefont {Hu},\ and\ \citenamefont
  {Lidar}}]{Friesen07}%
  \BibitemOpen
  \bibfield  {author} {\bibinfo {author} {\bibfnamefont {M.}~\bibnamefont
  {Friesen}}, \bibinfo {author} {\bibfnamefont {A.}~\bibnamefont {Biswas}},
  \bibinfo {author} {\bibfnamefont {X.}~\bibnamefont {Hu}}, \ and\ \bibinfo
  {author} {\bibfnamefont {D.}~\bibnamefont {Lidar}},\ }\href {\doibase
  10.1103/PhysRevLett.98.230503} {\bibfield  {journal} {\bibinfo  {journal}
  {Phys. Rev. Lett.}\ }\textbf {\bibinfo {volume} {98}},\ \bibinfo {pages}
  {230503} (\bibinfo {year} {2007})}\BibitemShut {NoStop}%
\bibitem [{\citenamefont {Mirkamali}\ \emph {et~al.}(2018)\citenamefont
  {Mirkamali}, \citenamefont {Cory},\ and\ \citenamefont
  {Emerson}}]{Mirkamali18}%
  \BibitemOpen
  \bibfield  {author} {\bibinfo {author} {\bibfnamefont {M.~S.}\ \bibnamefont
  {Mirkamali}}, \bibinfo {author} {\bibfnamefont {D.~G.}\ \bibnamefont {Cory}},
  \ and\ \bibinfo {author} {\bibfnamefont {J.}~\bibnamefont {Emerson}},\ }\href
  {\doibase 10.1103/PhysRevA.98.042327} {\bibfield  {journal} {\bibinfo
  {journal} {Phys. Rev. A}\ }\textbf {\bibinfo {volume} {98}},\ \bibinfo
  {pages} {042327} (\bibinfo {year} {2018})}\BibitemShut {NoStop}%
\bibitem [{\citenamefont {Andersen}\ and\ \citenamefont
  {Neergaard-Nielsen}(2013)}]{Andersen13}%
  \BibitemOpen
  \bibfield  {author} {\bibinfo {author} {\bibfnamefont {U.~L.}\ \bibnamefont
  {Andersen}}\ and\ \bibinfo {author} {\bibfnamefont {J.~S.}\ \bibnamefont
  {Neergaard-Nielsen}},\ }\href {\doibase 10.1103/PhysRevA.88.022337}
  {\bibfield  {journal} {\bibinfo  {journal} {Phys. Rev. A}\ }\textbf {\bibinfo
  {volume} {88}},\ \bibinfo {pages} {022337} (\bibinfo {year}
  {2013})}\BibitemShut {NoStop}%
\bibitem [{\citenamefont {Lvovsky}\ \emph {et~al.}(2013)\citenamefont
  {Lvovsky}, \citenamefont {Ghobadi}, \citenamefont {Chandra}, \citenamefont
  {Prasad},\ and\ \citenamefont {Simon}}]{Lvovsky13}%
  \BibitemOpen
  \bibfield  {author} {\bibinfo {author} {\bibfnamefont {A.~I.}\ \bibnamefont
  {Lvovsky}}, \bibinfo {author} {\bibfnamefont {R.}~\bibnamefont {Ghobadi}},
  \bibinfo {author} {\bibfnamefont {A.}~\bibnamefont {Chandra}}, \bibinfo
  {author} {\bibfnamefont {A.~S.}\ \bibnamefont {Prasad}}, \ and\ \bibinfo
  {author} {\bibfnamefont {C.}~\bibnamefont {Simon}},\ }\href {\doibase
  10.1038/nphys2682} {\bibfield  {journal} {\bibinfo  {journal} {Nature
  Physics}\ }\textbf {\bibinfo {volume} {9}},\ \bibinfo {pages} {541} (\bibinfo
  {year} {2013})}\BibitemShut {NoStop}%
\bibitem [{\citenamefont {Leggett}(2002)}]{Leggett02}%
  \BibitemOpen
  \bibfield  {author} {\bibinfo {author} {\bibfnamefont {A.~J.}\ \bibnamefont
  {Leggett}},\ }\href@noop {} {\bibfield  {journal} {\bibinfo  {journal}
  {Journal of Physics: Condensed Matter}\ }\textbf {\bibinfo {volume} {14}},\
  \bibinfo {pages} {R415} (\bibinfo {year} {2002})}\BibitemShut {NoStop}%
\bibitem [{\citenamefont {Schr{\"o}dinger}(1935)}]{Schrodinger35}%
  \BibitemOpen
  \bibfield  {author} {\bibinfo {author} {\bibfnamefont {E.}~\bibnamefont
  {Schr{\"o}dinger}},\ }\href {\doibase 10.1007/BF01491891} {\bibfield
  {journal} {\bibinfo  {journal} {Naturwissenschaften}\ }\textbf {\bibinfo
  {volume} {23}},\ \bibinfo {pages} {807} (\bibinfo {year} {1935})}\BibitemShut
  {NoStop}%
\bibitem [{\citenamefont {Zurek}()}]{Zurek03}%
  \BibitemOpen
  \bibfield  {author} {\bibinfo {author} {\bibfnamefont {W.~H.}\ \bibnamefont
  {Zurek}},\ }\href@noop {} {\ }\Eprint {http://arxiv.org/abs/quant-ph/0306072}
  {arXiv:quant-ph/0306072} \BibitemShut {NoStop}%
\bibitem [{\citenamefont {Brune}\ \emph {et~al.}(1996)\citenamefont {Brune},
  \citenamefont {Hagley}, \citenamefont {Dreyer}, \citenamefont {Ma\^{\i}tre},
  \citenamefont {Maali}, \citenamefont {Wunderlich}, \citenamefont {Raimond},\
  and\ \citenamefont {Haroche}}]{Brune96}%
  \BibitemOpen
  \bibfield  {author} {\bibinfo {author} {\bibfnamefont {M.}~\bibnamefont
  {Brune}}, \bibinfo {author} {\bibfnamefont {E.}~\bibnamefont {Hagley}},
  \bibinfo {author} {\bibfnamefont {J.}~\bibnamefont {Dreyer}}, \bibinfo
  {author} {\bibfnamefont {X.}~\bibnamefont {Ma\^{\i}tre}}, \bibinfo {author}
  {\bibfnamefont {A.}~\bibnamefont {Maali}}, \bibinfo {author} {\bibfnamefont
  {C.}~\bibnamefont {Wunderlich}}, \bibinfo {author} {\bibfnamefont {J.~M.}\
  \bibnamefont {Raimond}}, \ and\ \bibinfo {author} {\bibfnamefont
  {S.}~\bibnamefont {Haroche}},\ }\href {\doibase 10.1103/PhysRevLett.77.4887}
  {\bibfield  {journal} {\bibinfo  {journal} {Phys. Rev. Lett.}\ }\textbf
  {\bibinfo {volume} {77}},\ \bibinfo {pages} {4887} (\bibinfo {year}
  {1996})}\BibitemShut {NoStop}%
\bibitem [{\citenamefont {Haroche}(1998)}]{Haroche98}%
  \BibitemOpen
  \bibfield  {author} {\bibinfo {author} {\bibfnamefont {S.}~\bibnamefont
  {Haroche}},\ }\href@noop {} {\bibfield  {journal} {\bibinfo  {journal}
  {Physica Scripta}\ }\textbf {\bibinfo {volume} {1998}},\ \bibinfo {pages}
  {159} (\bibinfo {year} {1998})}\BibitemShut {NoStop}%
\bibitem [{\citenamefont {Vlastakis}\ \emph {et~al.}(2013)\citenamefont
  {Vlastakis}, \citenamefont {Kirchmair}, \citenamefont {Leghtas},
  \citenamefont {Nigg}, \citenamefont {Frunzio}, \citenamefont {Girvin},
  \citenamefont {Mirrahimi}, \citenamefont {Devoret},\ and\ \citenamefont
  {Schoelkopf}}]{Vlastakis13}%
  \BibitemOpen
  \bibfield  {author} {\bibinfo {author} {\bibfnamefont {B.}~\bibnamefont
  {Vlastakis}}, \bibinfo {author} {\bibfnamefont {G.}~\bibnamefont
  {Kirchmair}}, \bibinfo {author} {\bibfnamefont {Z.}~\bibnamefont {Leghtas}},
  \bibinfo {author} {\bibfnamefont {S.~E.}\ \bibnamefont {Nigg}}, \bibinfo
  {author} {\bibfnamefont {L.}~\bibnamefont {Frunzio}}, \bibinfo {author}
  {\bibfnamefont {S.~M.}\ \bibnamefont {Girvin}}, \bibinfo {author}
  {\bibfnamefont {M.}~\bibnamefont {Mirrahimi}}, \bibinfo {author}
  {\bibfnamefont {M.~H.}\ \bibnamefont {Devoret}}, \ and\ \bibinfo {author}
  {\bibfnamefont {R.~J.}\ \bibnamefont {Schoelkopf}},\ }\href {\doibase
  10.1126/science.1243289} {\bibfield  {journal} {\bibinfo  {journal}
  {Science}\ }\textbf {\bibinfo {volume} {342}},\ \bibinfo {pages} {607}
  (\bibinfo {year} {2013})}\BibitemShut {NoStop}%
\bibitem [{\citenamefont {Bruno}\ \emph {et~al.}(2013)\citenamefont {Bruno},
  \citenamefont {Martin}, \citenamefont {Sekatski}, \citenamefont {Sangouard},
  \citenamefont {Thew},\ and\ \citenamefont {Gisin}}]{Bruno13}%
  \BibitemOpen
  \bibfield  {author} {\bibinfo {author} {\bibfnamefont {N.}~\bibnamefont
  {Bruno}}, \bibinfo {author} {\bibfnamefont {A.}~\bibnamefont {Martin}},
  \bibinfo {author} {\bibfnamefont {P.}~\bibnamefont {Sekatski}}, \bibinfo
  {author} {\bibfnamefont {N.}~\bibnamefont {Sangouard}}, \bibinfo {author}
  {\bibfnamefont {R.~T.}\ \bibnamefont {Thew}}, \ and\ \bibinfo {author}
  {\bibfnamefont {N.}~\bibnamefont {Gisin}},\ }\href {\doibase
  10.1038/nphys2681} {\bibfield  {journal} {\bibinfo  {journal} {Nature
  Physics}\ }\textbf {\bibinfo {volume} {9}},\ \bibinfo {pages} {545} (\bibinfo
  {year} {2013})}\BibitemShut {NoStop}%
\bibitem [{\citenamefont {Monroe}\ \emph {et~al.}(1996)\citenamefont {Monroe},
  \citenamefont {Meekhof}, \citenamefont {King},\ and\ \citenamefont
  {Wineland}}]{Monroe96}%
  \BibitemOpen
  \bibfield  {author} {\bibinfo {author} {\bibfnamefont {C.}~\bibnamefont
  {Monroe}}, \bibinfo {author} {\bibfnamefont {D.~M.}\ \bibnamefont {Meekhof}},
  \bibinfo {author} {\bibfnamefont {B.~E.}\ \bibnamefont {King}}, \ and\
  \bibinfo {author} {\bibfnamefont {D.~J.}\ \bibnamefont {Wineland}},\ }\href
  {\doibase 10.1126/science.272.5265.1131} {\bibfield  {journal} {\bibinfo
  {journal} {Science}\ }\textbf {\bibinfo {volume} {272}},\ \bibinfo {pages}
  {1131} (\bibinfo {year} {1996})},\ \Eprint
  {http://arxiv.org/abs/http://science.sciencemag.org/content/272/5265/1131.full.pdf}
  {http://science.sciencemag.org/content/272/5265/1131.full.pdf} \BibitemShut
  {NoStop}%
\bibitem [{\citenamefont {Johnson}\ \emph {et~al.}(2017)\citenamefont
  {Johnson}, \citenamefont {Wong-Campos}, \citenamefont {Neyenhuis},
  \citenamefont {Mizrahi},\ and\ \citenamefont {Monroe}}]{Johnson17}%
  \BibitemOpen
  \bibfield  {author} {\bibinfo {author} {\bibfnamefont {K.~G.}\ \bibnamefont
  {Johnson}}, \bibinfo {author} {\bibfnamefont {J.~D.}\ \bibnamefont
  {Wong-Campos}}, \bibinfo {author} {\bibfnamefont {B.}~\bibnamefont
  {Neyenhuis}}, \bibinfo {author} {\bibfnamefont {J.}~\bibnamefont {Mizrahi}},
  \ and\ \bibinfo {author} {\bibfnamefont {C.}~\bibnamefont {Monroe}},\ }\href
  {\doibase 10.1038/s41467-017-00682-6x} {\bibfield  {journal} {\bibinfo
  {journal} {Nature Communications}\ }\textbf {\bibinfo {volume} {8}},\
  \bibinfo {pages} {697} (\bibinfo {year} {2017})}\BibitemShut {NoStop}%
\bibitem [{\citenamefont {Bennett}\ and\ \citenamefont
  {DiVincenzo}(2000)}]{Bennett00}%
  \BibitemOpen
  \bibfield  {author} {\bibinfo {author} {\bibfnamefont {C.~H.}\ \bibnamefont
  {Bennett}}\ and\ \bibinfo {author} {\bibfnamefont {D.~P.}\ \bibnamefont
  {DiVincenzo}},\ }\href {\doibase 10.1038/35005001} {\bibfield  {journal}
  {\bibinfo  {journal} {Nature}\ }\textbf {\bibinfo {volume} {404}},\ \bibinfo
  {pages} {247} (\bibinfo {year} {2000})}\BibitemShut {NoStop}%
\bibitem [{\citenamefont {Bose}(2003)}]{Bose3}%
  \BibitemOpen
  \bibfield  {author} {\bibinfo {author} {\bibfnamefont {S.}~\bibnamefont
  {Bose}},\ }\href {\doibase 10.1103/PhysRevLett.91.207901} {\bibfield
  {journal} {\bibinfo  {journal} {Phys. Rev. Lett.}\ }\textbf {\bibinfo
  {volume} {91}},\ \bibinfo {pages} {207901} (\bibinfo {year}
  {2003})}\BibitemShut {NoStop}%
\bibitem [{\citenamefont {Bennett}\ \emph {et~al.}(1993)\citenamefont
  {Bennett}, \citenamefont {Brassard}, \citenamefont {Cr\'epeau}, \citenamefont
  {Jozsa}, \citenamefont {Peres},\ and\ \citenamefont {Wootters}}]{Bennett93}%
  \BibitemOpen
  \bibfield  {author} {\bibinfo {author} {\bibfnamefont {C.~H.}\ \bibnamefont
  {Bennett}}, \bibinfo {author} {\bibfnamefont {G.}~\bibnamefont {Brassard}},
  \bibinfo {author} {\bibfnamefont {C.}~\bibnamefont {Cr\'epeau}}, \bibinfo
  {author} {\bibfnamefont {R.}~\bibnamefont {Jozsa}}, \bibinfo {author}
  {\bibfnamefont {A.}~\bibnamefont {Peres}}, \ and\ \bibinfo {author}
  {\bibfnamefont {W.~K.}\ \bibnamefont {Wootters}},\ }\href {\doibase
  10.1103/PhysRevLett.70.1895} {\bibfield  {journal} {\bibinfo  {journal}
  {Phys. Rev. Lett.}\ }\textbf {\bibinfo {volume} {70}},\ \bibinfo {pages}
  {1895} (\bibinfo {year} {1993})}\BibitemShut {NoStop}%
\bibitem [{\citenamefont {Christandl}\ \emph {et~al.}(2004)\citenamefont
  {Christandl}, \citenamefont {Datta}, \citenamefont {Ekert},\ and\
  \citenamefont {Landahl}}]{Christandl4}%
  \BibitemOpen
  \bibfield  {author} {\bibinfo {author} {\bibfnamefont {M.}~\bibnamefont
  {Christandl}}, \bibinfo {author} {\bibfnamefont {N.}~\bibnamefont {Datta}},
  \bibinfo {author} {\bibfnamefont {A.}~\bibnamefont {Ekert}}, \ and\ \bibinfo
  {author} {\bibfnamefont {A.~J.}\ \bibnamefont {Landahl}},\ }\href {\doibase
  10.1103/PhysRevLett.92.187902} {\bibfield  {journal} {\bibinfo  {journal}
  {Phys. Rev. Lett.}\ }\textbf {\bibinfo {volume} {92}},\ \bibinfo {pages}
  {187902} (\bibinfo {year} {2004})}\BibitemShut {NoStop}%
\bibitem [{\citenamefont {Kay}(2006)}]{Kay6}%
  \BibitemOpen
  \bibfield  {author} {\bibinfo {author} {\bibfnamefont {A.}~\bibnamefont
  {Kay}},\ }\href {\doibase 10.1103/PhysRevA.73.032306} {\bibfield  {journal}
  {\bibinfo  {journal} {Phys. Rev. A}\ }\textbf {\bibinfo {volume} {73}},\
  \bibinfo {pages} {032306} (\bibinfo {year} {2006})}\BibitemShut {NoStop}%
\bibitem [{\citenamefont {Avellino}\ \emph {et~al.}(2006)\citenamefont
  {Avellino}, \citenamefont {Fisher},\ and\ \citenamefont {Bose}}]{Avellino6}%
  \BibitemOpen
  \bibfield  {author} {\bibinfo {author} {\bibfnamefont {M.}~\bibnamefont
  {Avellino}}, \bibinfo {author} {\bibfnamefont {A.~J.}\ \bibnamefont
  {Fisher}}, \ and\ \bibinfo {author} {\bibfnamefont {S.}~\bibnamefont
  {Bose}},\ }\href {\doibase 10.1103/PhysRevA.74.012321} {\bibfield  {journal}
  {\bibinfo  {journal} {Phys. Rev. A}\ }\textbf {\bibinfo {volume} {74}},\
  \bibinfo {pages} {012321} (\bibinfo {year} {2006})}\BibitemShut {NoStop}%
\bibitem [{\citenamefont {Burgarth}\ \emph {et~al.}(2007)\citenamefont
  {Burgarth}, \citenamefont {Giovannetti},\ and\ \citenamefont
  {Bose}}]{Burgarth7}%
  \BibitemOpen
  \bibfield  {author} {\bibinfo {author} {\bibfnamefont {D.}~\bibnamefont
  {Burgarth}}, \bibinfo {author} {\bibfnamefont {V.}~\bibnamefont
  {Giovannetti}}, \ and\ \bibinfo {author} {\bibfnamefont {S.}~\bibnamefont
  {Bose}},\ }\href {\doibase 10.1103/PhysRevA.75.062327} {\bibfield  {journal}
  {\bibinfo  {journal} {Phys. Rev. A}\ }\textbf {\bibinfo {volume} {75}},\
  \bibinfo {pages} {062327} (\bibinfo {year} {2007})}\BibitemShut {NoStop}%
\bibitem [{\citenamefont {Cappellaro}\ \emph {et~al.}(2007)\citenamefont
  {Cappellaro}, \citenamefont {Ramanathan},\ and\ \citenamefont
  {Cory}}]{Cappellaro07}%
  \BibitemOpen
  \bibfield  {author} {\bibinfo {author} {\bibfnamefont {P.}~\bibnamefont
  {Cappellaro}}, \bibinfo {author} {\bibfnamefont {C.}~\bibnamefont
  {Ramanathan}}, \ and\ \bibinfo {author} {\bibfnamefont {D.~G.}\ \bibnamefont
  {Cory}},\ }\href {\doibase 10.1103/PhysRevLett.99.250506} {\bibfield
  {journal} {\bibinfo  {journal} {Phys. Rev. Lett.}\ }\textbf {\bibinfo
  {volume} {99}},\ \bibinfo {pages} {250506} (\bibinfo {year}
  {2007})}\BibitemShut {NoStop}%
\bibitem [{\citenamefont {Di~Franco}\ \emph {et~al.}(2008)\citenamefont
  {Di~Franco}, \citenamefont {Paternostro},\ and\ \citenamefont
  {Kim}}]{Franco8}%
  \BibitemOpen
  \bibfield  {author} {\bibinfo {author} {\bibfnamefont {C.}~\bibnamefont
  {Di~Franco}}, \bibinfo {author} {\bibfnamefont {M.}~\bibnamefont
  {Paternostro}}, \ and\ \bibinfo {author} {\bibfnamefont {M.~S.}\ \bibnamefont
  {Kim}},\ }\href {\doibase 10.1103/PhysRevLett.101.230502} {\bibfield
  {journal} {\bibinfo  {journal} {Phys. Rev. Lett.}\ }\textbf {\bibinfo
  {volume} {101}},\ \bibinfo {pages} {230502} (\bibinfo {year}
  {2008})}\BibitemShut {NoStop}%
\bibitem [{\citenamefont {Ramanathan}\ \emph {et~al.}(2011)\citenamefont
  {Ramanathan}, \citenamefont {Cappellaro}, \citenamefont {Viola},\ and\
  \citenamefont {Cory}}]{Ramanathan11}%
  \BibitemOpen
  \bibfield  {author} {\bibinfo {author} {\bibfnamefont {C.}~\bibnamefont
  {Ramanathan}}, \bibinfo {author} {\bibfnamefont {P.}~\bibnamefont
  {Cappellaro}}, \bibinfo {author} {\bibfnamefont {L.}~\bibnamefont {Viola}}, \
  and\ \bibinfo {author} {\bibfnamefont {D.~G.}\ \bibnamefont {Cory}},\
  }\href@noop {} {\bibfield  {journal} {\bibinfo  {journal} {New Journal of
  Physics}\ }\textbf {\bibinfo {volume} {13}},\ \bibinfo {pages} {103015}
  (\bibinfo {year} {2011})}\BibitemShut {NoStop}%
\bibitem [{\citenamefont {Cappellaro}\ \emph {et~al.}(2011)\citenamefont
  {Cappellaro}, \citenamefont {Viola},\ and\ \citenamefont
  {Ramanathan}}]{Cappellaro11}%
  \BibitemOpen
  \bibfield  {author} {\bibinfo {author} {\bibfnamefont {P.}~\bibnamefont
  {Cappellaro}}, \bibinfo {author} {\bibfnamefont {L.}~\bibnamefont {Viola}}, \
  and\ \bibinfo {author} {\bibfnamefont {C.}~\bibnamefont {Ramanathan}},\
  }\href {\doibase 10.1103/PhysRevA.83.032304} {\bibfield  {journal} {\bibinfo
  {journal} {Phys. Rev. A}\ }\textbf {\bibinfo {volume} {83}},\ \bibinfo
  {pages} {032304} (\bibinfo {year} {2011})}\BibitemShut {NoStop}%
\bibitem [{\citenamefont {Yao}\ \emph {et~al.}(2011)\citenamefont {Yao},
  \citenamefont {Jiang}, \citenamefont {Gorshkov}, \citenamefont {Gong},
  \citenamefont {Zhai}, \citenamefont {Duan},\ and\ \citenamefont
  {Lukin}}]{Yao11}%
  \BibitemOpen
  \bibfield  {author} {\bibinfo {author} {\bibfnamefont {N.~Y.}\ \bibnamefont
  {Yao}}, \bibinfo {author} {\bibfnamefont {L.}~\bibnamefont {Jiang}}, \bibinfo
  {author} {\bibfnamefont {A.~V.}\ \bibnamefont {Gorshkov}}, \bibinfo {author}
  {\bibfnamefont {Z.-X.}\ \bibnamefont {Gong}}, \bibinfo {author}
  {\bibfnamefont {A.}~\bibnamefont {Zhai}}, \bibinfo {author} {\bibfnamefont
  {L.-M.}\ \bibnamefont {Duan}}, \ and\ \bibinfo {author} {\bibfnamefont
  {M.~D.}\ \bibnamefont {Lukin}},\ }\href {\doibase
  10.1103/PhysRevLett.106.040505} {\bibfield  {journal} {\bibinfo  {journal}
  {Phys. Rev. Lett.}\ }\textbf {\bibinfo {volume} {106}},\ \bibinfo {pages}
  {040505} (\bibinfo {year} {2011})}\BibitemShut {NoStop}%
\bibitem [{\citenamefont {Ajoy}\ and\ \citenamefont
  {Cappellaro}(2013)}]{Ajoy13}%
  \BibitemOpen
  \bibfield  {author} {\bibinfo {author} {\bibfnamefont {A.}~\bibnamefont
  {Ajoy}}\ and\ \bibinfo {author} {\bibfnamefont {P.}~\bibnamefont
  {Cappellaro}},\ }\href {\doibase 10.1103/PhysRevB.87.064303} {\bibfield
  {journal} {\bibinfo  {journal} {Phys. Rev. B}\ }\textbf {\bibinfo {volume}
  {87}},\ \bibinfo {pages} {064303} (\bibinfo {year} {2013})}\BibitemShut
  {NoStop}%
\bibitem [{\citenamefont {Jordan}\ and\ \citenamefont
  {Wigner}(1928)}]{Jordan28}%
  \BibitemOpen
  \bibfield  {author} {\bibinfo {author} {\bibfnamefont {P.}~\bibnamefont
  {Jordan}}\ and\ \bibinfo {author} {\bibfnamefont {E.}~\bibnamefont
  {Wigner}},\ }\href {\doibase 10.1007/BF01331938} {\bibfield  {journal}
  {\bibinfo  {journal} {Zeitschrift f{\"u}r Physik}\ }\textbf {\bibinfo
  {volume} {47}},\ \bibinfo {pages} {631} (\bibinfo {year} {1928})}\BibitemShut
  {NoStop}%
\bibitem [{\citenamefont {Garisto}\ and\ \citenamefont
  {Hardy}(1999)}]{Garisto99}%
  \BibitemOpen
  \bibfield  {author} {\bibinfo {author} {\bibfnamefont {R.}~\bibnamefont
  {Garisto}}\ and\ \bibinfo {author} {\bibfnamefont {L.}~\bibnamefont
  {Hardy}},\ }\href {\doibase 10.1103/PhysRevA.60.827} {\bibfield  {journal}
  {\bibinfo  {journal} {Phys. Rev. A}\ }\textbf {\bibinfo {volume} {60}},\
  \bibinfo {pages} {827} (\bibinfo {year} {1999})}\BibitemShut {NoStop}%
\bibitem [{\citenamefont {Bennett}\ \emph
  {et~al.}(1996{\natexlab{a}})\citenamefont {Bennett}, \citenamefont
  {Brassard}, \citenamefont {Popescu}, \citenamefont {Schumacher},
  \citenamefont {Smolin},\ and\ \citenamefont {Wootters}}]{Bennett96PRL}%
  \BibitemOpen
  \bibfield  {author} {\bibinfo {author} {\bibfnamefont {C.~H.}\ \bibnamefont
  {Bennett}}, \bibinfo {author} {\bibfnamefont {G.}~\bibnamefont {Brassard}},
  \bibinfo {author} {\bibfnamefont {S.}~\bibnamefont {Popescu}}, \bibinfo
  {author} {\bibfnamefont {B.}~\bibnamefont {Schumacher}}, \bibinfo {author}
  {\bibfnamefont {J.~A.}\ \bibnamefont {Smolin}}, \ and\ \bibinfo {author}
  {\bibfnamefont {W.~K.}\ \bibnamefont {Wootters}},\ }\href {\doibase
  10.1103/PhysRevLett.76.722} {\bibfield  {journal} {\bibinfo  {journal} {Phys.
  Rev. Lett.}\ }\textbf {\bibinfo {volume} {76}},\ \bibinfo {pages} {722}
  (\bibinfo {year} {1996}{\natexlab{a}})}\BibitemShut {NoStop}%
\bibitem [{\citenamefont {Bennett}\ \emph
  {et~al.}(1996{\natexlab{b}})\citenamefont {Bennett}, \citenamefont
  {Bernstein}, \citenamefont {Popescu},\ and\ \citenamefont
  {Schumacher}}]{Bennett96PRA}%
  \BibitemOpen
  \bibfield  {author} {\bibinfo {author} {\bibfnamefont {C.~H.}\ \bibnamefont
  {Bennett}}, \bibinfo {author} {\bibfnamefont {H.~J.}\ \bibnamefont
  {Bernstein}}, \bibinfo {author} {\bibfnamefont {S.}~\bibnamefont {Popescu}},
  \ and\ \bibinfo {author} {\bibfnamefont {B.}~\bibnamefont {Schumacher}},\
  }\href {\doibase 10.1103/PhysRevA.53.2046} {\bibfield  {journal} {\bibinfo
  {journal} {Phys. Rev. A}\ }\textbf {\bibinfo {volume} {53}},\ \bibinfo
  {pages} {2046} (\bibinfo {year} {1996}{\natexlab{b}})}\BibitemShut {NoStop}%
\bibitem [{\citenamefont {Clauser}\ \emph {et~al.}(1969)\citenamefont
  {Clauser}, \citenamefont {Horne}, \citenamefont {Shimony},\ and\
  \citenamefont {Holt}}]{CHSH69}%
  \BibitemOpen
  \bibfield  {author} {\bibinfo {author} {\bibfnamefont {J.~F.}\ \bibnamefont
  {Clauser}}, \bibinfo {author} {\bibfnamefont {M.~A.}\ \bibnamefont {Horne}},
  \bibinfo {author} {\bibfnamefont {A.}~\bibnamefont {Shimony}}, \ and\
  \bibinfo {author} {\bibfnamefont {R.~A.}\ \bibnamefont {Holt}},\ }\href
  {\doibase 10.1103/PhysRevLett.23.880} {\bibfield  {journal} {\bibinfo
  {journal} {Phys. Rev. Lett.}\ }\textbf {\bibinfo {volume} {23}},\ \bibinfo
  {pages} {880} (\bibinfo {year} {1969})}\BibitemShut {NoStop}%
\bibitem [{\citenamefont {\ifmmode~\dot{Z}\else \.{Z}\fi{}ukowski}\ \emph
  {et~al.}(1993)\citenamefont {\ifmmode~\dot{Z}\else \.{Z}\fi{}ukowski},
  \citenamefont {Zeilinger}, \citenamefont {Horne},\ and\ \citenamefont
  {Ekert}}]{swapping93}%
  \BibitemOpen
  \bibfield  {author} {\bibinfo {author} {\bibfnamefont {M.}~\bibnamefont
  {\ifmmode~\dot{Z}\else \.{Z}\fi{}ukowski}}, \bibinfo {author} {\bibfnamefont
  {A.}~\bibnamefont {Zeilinger}}, \bibinfo {author} {\bibfnamefont {M.~A.}\
  \bibnamefont {Horne}}, \ and\ \bibinfo {author} {\bibfnamefont {A.~K.}\
  \bibnamefont {Ekert}},\ }\href {\doibase 10.1103/PhysRevLett.71.4287}
  {\bibfield  {journal} {\bibinfo  {journal} {Phys. Rev. Lett.}\ }\textbf
  {\bibinfo {volume} {71}},\ \bibinfo {pages} {4287} (\bibinfo {year}
  {1993})}\BibitemShut {NoStop}%
\bibitem [{\citenamefont {Warren}\ \emph {et~al.}(1979)\citenamefont {Warren},
  \citenamefont {Sinton}, \citenamefont {Weitekamp},\ and\ \citenamefont
  {Pines}}]{Warren79}%
  \BibitemOpen
  \bibfield  {author} {\bibinfo {author} {\bibfnamefont {W.~S.}\ \bibnamefont
  {Warren}}, \bibinfo {author} {\bibfnamefont {S.}~\bibnamefont {Sinton}},
  \bibinfo {author} {\bibfnamefont {D.~P.}\ \bibnamefont {Weitekamp}}, \ and\
  \bibinfo {author} {\bibfnamefont {A.}~\bibnamefont {Pines}},\ }\href
  {\doibase 10.1103/PhysRevLett.43.1791} {\bibfield  {journal} {\bibinfo
  {journal} {Phys. Rev. Lett.}\ }\textbf {\bibinfo {volume} {43}},\ \bibinfo
  {pages} {1791} (\bibinfo {year} {1979})}\BibitemShut {NoStop}%
\bibitem [{\citenamefont {Lieb}\ and\ \citenamefont {Robinson}(1972)}]{Lieb72}%
  \BibitemOpen
  \bibfield  {author} {\bibinfo {author} {\bibfnamefont {E.~H.}\ \bibnamefont
  {Lieb}}\ and\ \bibinfo {author} {\bibfnamefont {D.~W.}\ \bibnamefont
  {Robinson}},\ }\href {\doibase 10.1007/BF01645779} {\bibfield  {journal}
  {\bibinfo  {journal} {Communications in Mathematical Physics}\ }\textbf
  {\bibinfo {volume} {28}},\ \bibinfo {pages} {251} (\bibinfo {year}
  {1972})}\BibitemShut {NoStop}%
\bibitem [{\citenamefont {Hastings}\ and\ \citenamefont
  {Koma}(2006)}]{Hastings06}%
  \BibitemOpen
  \bibfield  {author} {\bibinfo {author} {\bibfnamefont {M.~B.}\ \bibnamefont
  {Hastings}}\ and\ \bibinfo {author} {\bibfnamefont {T.}~\bibnamefont
  {Koma}},\ }\href {\doibase 10.1007/s00220-006-0030-4} {\bibfield  {journal}
  {\bibinfo  {journal} {Communications in Mathematical Physics}\ }\textbf
  {\bibinfo {volume} {265}},\ \bibinfo {pages} {781} (\bibinfo {year}
  {2006})}\BibitemShut {NoStop}%
\bibitem [{\citenamefont {Hauke}\ and\ \citenamefont
  {Tagliacozzo}(2013)}]{Hauke13}%
  \BibitemOpen
  \bibfield  {author} {\bibinfo {author} {\bibfnamefont {P.}~\bibnamefont
  {Hauke}}\ and\ \bibinfo {author} {\bibfnamefont {L.}~\bibnamefont
  {Tagliacozzo}},\ }\href {\doibase 10.1103/PhysRevLett.111.207202} {\bibfield
  {journal} {\bibinfo  {journal} {Phys. Rev. Lett.}\ }\textbf {\bibinfo
  {volume} {111}},\ \bibinfo {pages} {207202} (\bibinfo {year}
  {2013})}\BibitemShut {NoStop}%
\bibitem [{\citenamefont {Gong}\ \emph {et~al.}(2014)\citenamefont {Gong},
  \citenamefont {Foss-Feig}, \citenamefont {Michalakis},\ and\ \citenamefont
  {Gorshkov}}]{Gong14}%
  \BibitemOpen
  \bibfield  {author} {\bibinfo {author} {\bibfnamefont {Z.-X.}\ \bibnamefont
  {Gong}}, \bibinfo {author} {\bibfnamefont {M.}~\bibnamefont {Foss-Feig}},
  \bibinfo {author} {\bibfnamefont {S.}~\bibnamefont {Michalakis}}, \ and\
  \bibinfo {author} {\bibfnamefont {A.~V.}\ \bibnamefont {Gorshkov}},\ }\href
  {\doibase 10.1103/PhysRevLett.113.030602} {\bibfield  {journal} {\bibinfo
  {journal} {Phys. Rev. Lett.}\ }\textbf {\bibinfo {volume} {113}},\ \bibinfo
  {pages} {030602} (\bibinfo {year} {2014})}\BibitemShut {NoStop}%
\bibitem [{\citenamefont {Hazzard}\ \emph {et~al.}(2014)\citenamefont
  {Hazzard}, \citenamefont {van~den Worm}, \citenamefont {Foss-Feig},
  \citenamefont {Manmana}, \citenamefont {Dalla~Torre}, \citenamefont {Pfau},
  \citenamefont {Kastner},\ and\ \citenamefont {Rey}}]{Hazzard14}%
  \BibitemOpen
  \bibfield  {author} {\bibinfo {author} {\bibfnamefont {K.~R.~A.}\
  \bibnamefont {Hazzard}}, \bibinfo {author} {\bibfnamefont {M.}~\bibnamefont
  {van~den Worm}}, \bibinfo {author} {\bibfnamefont {M.}~\bibnamefont
  {Foss-Feig}}, \bibinfo {author} {\bibfnamefont {S.~R.}\ \bibnamefont
  {Manmana}}, \bibinfo {author} {\bibfnamefont {E.~G.}\ \bibnamefont
  {Dalla~Torre}}, \bibinfo {author} {\bibfnamefont {T.}~\bibnamefont {Pfau}},
  \bibinfo {author} {\bibfnamefont {M.}~\bibnamefont {Kastner}}, \ and\
  \bibinfo {author} {\bibfnamefont {A.~M.}\ \bibnamefont {Rey}},\ }\href
  {\doibase 10.1103/PhysRevA.90.063622} {\bibfield  {journal} {\bibinfo
  {journal} {Phys. Rev. A}\ }\textbf {\bibinfo {volume} {90}},\ \bibinfo
  {pages} {063622} (\bibinfo {year} {2014})}\BibitemShut {NoStop}%
\bibitem [{\citenamefont {Jurcevic}\ \emph {et~al.}(2014)\citenamefont
  {Jurcevic}, \citenamefont {Lanyon}, \citenamefont {Hauke}, \citenamefont
  {Hempel}, \citenamefont {Zoller}, \citenamefont {Blatt},\ and\ \citenamefont
  {Roos}}]{Jurcevic14}%
  \BibitemOpen
  \bibfield  {author} {\bibinfo {author} {\bibfnamefont {P.}~\bibnamefont
  {Jurcevic}}, \bibinfo {author} {\bibfnamefont {B.~P.}\ \bibnamefont
  {Lanyon}}, \bibinfo {author} {\bibfnamefont {P.}~\bibnamefont {Hauke}},
  \bibinfo {author} {\bibfnamefont {C.}~\bibnamefont {Hempel}}, \bibinfo
  {author} {\bibfnamefont {P.}~\bibnamefont {Zoller}}, \bibinfo {author}
  {\bibfnamefont {R.}~\bibnamefont {Blatt}}, \ and\ \bibinfo {author}
  {\bibfnamefont {C.~F.}\ \bibnamefont {Roos}},\ }\href {\doibase
  10.1038/nature13461} {\bibfield  {journal} {\bibinfo  {journal} {Nature}\
  }\textbf {\bibinfo {volume} {511}},\ \bibinfo {pages} {202} (\bibinfo {year}
  {2014})}\BibitemShut {NoStop}%
\bibitem [{\citenamefont {Richerme}\ \emph {et~al.}(2014)\citenamefont
  {Richerme}, \citenamefont {Gong}, \citenamefont {Lee}, \citenamefont {Senko},
  \citenamefont {Smith}, \citenamefont {Foss-Feig}, \citenamefont {Michalakis},
  \citenamefont {Gorshkov},\ and\ \citenamefont {Monroe}}]{Richerme14}%
  \BibitemOpen
  \bibfield  {author} {\bibinfo {author} {\bibfnamefont {P.}~\bibnamefont
  {Richerme}}, \bibinfo {author} {\bibfnamefont {Z.-X.}\ \bibnamefont {Gong}},
  \bibinfo {author} {\bibfnamefont {A.}~\bibnamefont {Lee}}, \bibinfo {author}
  {\bibfnamefont {C.}~\bibnamefont {Senko}}, \bibinfo {author} {\bibfnamefont
  {J.}~\bibnamefont {Smith}}, \bibinfo {author} {\bibfnamefont
  {M.}~\bibnamefont {Foss-Feig}}, \bibinfo {author} {\bibfnamefont
  {S.}~\bibnamefont {Michalakis}}, \bibinfo {author} {\bibfnamefont {A.~V.}\
  \bibnamefont {Gorshkov}}, \ and\ \bibinfo {author} {\bibfnamefont
  {C.}~\bibnamefont {Monroe}},\ }\href {\doibase 10.1038/nature13450}
  {\bibfield  {journal} {\bibinfo  {journal} {Nature}\ }\textbf {\bibinfo
  {volume} {511}},\ \bibinfo {pages} {198} (\bibinfo {year}
  {2014})}\BibitemShut {NoStop}%
\bibitem [{\citenamefont {Foss-Feig}\ \emph {et~al.}(2015)\citenamefont
  {Foss-Feig}, \citenamefont {Gong}, \citenamefont {Clark},\ and\ \citenamefont
  {Gorshkov}}]{Foss-Feig15}%
  \BibitemOpen
  \bibfield  {author} {\bibinfo {author} {\bibfnamefont {M.}~\bibnamefont
  {Foss-Feig}}, \bibinfo {author} {\bibfnamefont {Z.-X.}\ \bibnamefont {Gong}},
  \bibinfo {author} {\bibfnamefont {C.~W.}\ \bibnamefont {Clark}}, \ and\
  \bibinfo {author} {\bibfnamefont {A.~V.}\ \bibnamefont {Gorshkov}},\ }\href
  {\doibase 10.1103/PhysRevLett.114.157201} {\bibfield  {journal} {\bibinfo
  {journal} {Phys. Rev. Lett.}\ }\textbf {\bibinfo {volume} {114}},\ \bibinfo
  {pages} {157201} (\bibinfo {year} {2015})}\BibitemShut {NoStop}%
\bibitem [{\citenamefont {Eldredge}\ \emph {et~al.}(2017)\citenamefont
  {Eldredge}, \citenamefont {Gong}, \citenamefont {Young}, \citenamefont
  {Moosavian}, \citenamefont {Foss-Feig},\ and\ \citenamefont
  {Gorshkov}}]{Eldredge17}%
  \BibitemOpen
  \bibfield  {author} {\bibinfo {author} {\bibfnamefont {Z.}~\bibnamefont
  {Eldredge}}, \bibinfo {author} {\bibfnamefont {Z.-X.}\ \bibnamefont {Gong}},
  \bibinfo {author} {\bibfnamefont {J.~T.}\ \bibnamefont {Young}}, \bibinfo
  {author} {\bibfnamefont {A.~H.}\ \bibnamefont {Moosavian}}, \bibinfo {author}
  {\bibfnamefont {M.}~\bibnamefont {Foss-Feig}}, \ and\ \bibinfo {author}
  {\bibfnamefont {A.~V.}\ \bibnamefont {Gorshkov}},\ }\href {\doibase
  10.1103/PhysRevLett.119.170503} {\bibfield  {journal} {\bibinfo  {journal}
  {Phys. Rev. Lett.}\ }\textbf {\bibinfo {volume} {119}},\ \bibinfo {pages}
  {170503} (\bibinfo {year} {2017})}\BibitemShut {NoStop}%
\bibitem [{\citenamefont {von Neumann}(1932)}]{Neumann32}%
  \BibitemOpen
  \bibfield  {author} {\bibinfo {author} {\bibfnamefont {J.}~\bibnamefont {von
  Neumann}},\ }\href@noop {} {\emph {\bibinfo {title} {Mathematische Grundlagen
  der Quantenmechanik}}}\ (\bibinfo  {publisher} {Springer Verlag, Berlin},\
  \bibinfo {year} {1932})\ \bibinfo {note} {[English translation (by R. Beyer):
  Mathematical Foundations of Quantum Mechanics (Princeton University Press,
  Princeton 1955)]}\BibitemShut {NoStop}%
\bibitem [{\citenamefont {Neumark}(1943)}]{Neumark43}%
  \BibitemOpen
  \bibfield  {author} {\bibinfo {author} {\bibfnamefont {M.~A.}\ \bibnamefont
  {Neumark}},\ }\href@noop {} {\bibfield  {journal} {\bibinfo  {journal} {C. R.
  (Dokl.) Acad. Sci. URSS}\ }\textbf {\bibinfo {volume} {41}},\ \bibinfo
  {pages} {359} (\bibinfo {year} {1943})}\BibitemShut {NoStop}%
\bibitem [{\citenamefont {Giedke}\ \emph {et~al.}(2006)\citenamefont {Giedke},
  \citenamefont {Taylor}, \citenamefont {D'Alessandro}, \citenamefont {Lukin},\
  and\ \citenamefont {Imamo\ifmmode~\breve{g}\else \u{g}\fi{}lu}}]{Giedke06}%
  \BibitemOpen
  \bibfield  {author} {\bibinfo {author} {\bibfnamefont {G.}~\bibnamefont
  {Giedke}}, \bibinfo {author} {\bibfnamefont {J.~M.}\ \bibnamefont {Taylor}},
  \bibinfo {author} {\bibfnamefont {D.}~\bibnamefont {D'Alessandro}}, \bibinfo
  {author} {\bibfnamefont {M.~D.}\ \bibnamefont {Lukin}}, \ and\ \bibinfo
  {author} {\bibfnamefont {A.}~\bibnamefont {Imamo\ifmmode~\breve{g}\else
  \u{g}\fi{}lu}},\ }\href {\doibase 10.1103/PhysRevA.74.032316} {\bibfield
  {journal} {\bibinfo  {journal} {Phys. Rev. A}\ }\textbf {\bibinfo {volume}
  {74}},\ \bibinfo {pages} {032316} (\bibinfo {year} {2006})}\BibitemShut
  {NoStop}%
\bibitem [{\citenamefont {Sekatski}\ \emph {et~al.}(2014)\citenamefont
  {Sekatski}, \citenamefont {Gisin},\ and\ \citenamefont
  {Sangouard}}]{Sekatski14PRL}%
  \BibitemOpen
  \bibfield  {author} {\bibinfo {author} {\bibfnamefont {P.}~\bibnamefont
  {Sekatski}}, \bibinfo {author} {\bibfnamefont {N.}~\bibnamefont {Gisin}}, \
  and\ \bibinfo {author} {\bibfnamefont {N.}~\bibnamefont {Sangouard}},\ }\href
  {\doibase 10.1103/PhysRevLett.113.090403} {\bibfield  {journal} {\bibinfo
  {journal} {Phys. Rev. Lett.}\ }\textbf {\bibinfo {volume} {113}},\ \bibinfo
  {pages} {090403} (\bibinfo {year} {2014})}\BibitemShut {NoStop}%
\bibitem [{\citenamefont {Wang}\ \emph {et~al.}(2013)\citenamefont {Wang},
  \citenamefont {Ghobadi}, \citenamefont {Raeisi},\ and\ \citenamefont
  {Simon}}]{Wang13}%
  \BibitemOpen
  \bibfield  {author} {\bibinfo {author} {\bibfnamefont {T.}~\bibnamefont
  {Wang}}, \bibinfo {author} {\bibfnamefont {R.}~\bibnamefont {Ghobadi}},
  \bibinfo {author} {\bibfnamefont {S.}~\bibnamefont {Raeisi}}, \ and\ \bibinfo
  {author} {\bibfnamefont {C.}~\bibnamefont {Simon}},\ }\href {\doibase
  10.1103/PhysRevA.88.062114} {\bibfield  {journal} {\bibinfo  {journal} {Phys.
  Rev. A}\ }\textbf {\bibinfo {volume} {88}},\ \bibinfo {pages} {062114}
  (\bibinfo {year} {2013})}\BibitemShut {NoStop}%
\bibitem [{\citenamefont {Vidal}\ and\ \citenamefont {Werner}(2002)}]{Vidal02}%
  \BibitemOpen
  \bibfield  {author} {\bibinfo {author} {\bibfnamefont {G.}~\bibnamefont
  {Vidal}}\ and\ \bibinfo {author} {\bibfnamefont {R.~F.}\ \bibnamefont
  {Werner}},\ }\href {\doibase 10.1103/PhysRevA.65.032314} {\bibfield
  {journal} {\bibinfo  {journal} {Phys. Rev. A}\ }\textbf {\bibinfo {volume}
  {65}},\ \bibinfo {pages} {032314} (\bibinfo {year} {2002})}\BibitemShut
  {NoStop}%
\bibitem [{\citenamefont {Fr\"owis}\ \emph {et~al.}(2018)\citenamefont
  {Fr\"owis}, \citenamefont {Sekatski}, \citenamefont {D\"ur}, \citenamefont
  {Gisin},\ and\ \citenamefont {Sangouard}}]{Frowis18}%
  \BibitemOpen
  \bibfield  {author} {\bibinfo {author} {\bibfnamefont {F.}~\bibnamefont
  {Fr\"owis}}, \bibinfo {author} {\bibfnamefont {P.}~\bibnamefont {Sekatski}},
  \bibinfo {author} {\bibfnamefont {W.}~\bibnamefont {D\"ur}}, \bibinfo
  {author} {\bibfnamefont {N.}~\bibnamefont {Gisin}}, \ and\ \bibinfo {author}
  {\bibfnamefont {N.}~\bibnamefont {Sangouard}},\ }\href {\doibase
  10.1103/RevModPhys.90.025004} {\bibfield  {journal} {\bibinfo  {journal}
  {Rev. Mod. Phys.}\ }\textbf {\bibinfo {volume} {90}},\ \bibinfo {pages}
  {025004} (\bibinfo {year} {2018})}\BibitemShut {NoStop}%
\bibitem [{\citenamefont {D\"ur}\ \emph {et~al.}(2002)\citenamefont {D\"ur},
  \citenamefont {Simon},\ and\ \citenamefont {Cirac}}]{Dur02}%
  \BibitemOpen
  \bibfield  {author} {\bibinfo {author} {\bibfnamefont {W.}~\bibnamefont
  {D\"ur}}, \bibinfo {author} {\bibfnamefont {C.}~\bibnamefont {Simon}}, \ and\
  \bibinfo {author} {\bibfnamefont {J.~I.}\ \bibnamefont {Cirac}},\ }\href
  {\doibase 10.1103/PhysRevLett.89.210402} {\bibfield  {journal} {\bibinfo
  {journal} {Phys. Rev. Lett.}\ }\textbf {\bibinfo {volume} {89}},\ \bibinfo
  {pages} {210402} (\bibinfo {year} {2002})}\BibitemShut {NoStop}%
\bibitem [{\citenamefont {Kwon}\ \emph {et~al.}(2017)\citenamefont {Kwon},
  \citenamefont {Park}, \citenamefont {Tan},\ and\ \citenamefont
  {Jeong}}]{Kwon17}%
  \BibitemOpen
  \bibfield  {author} {\bibinfo {author} {\bibfnamefont {H.}~\bibnamefont
  {Kwon}}, \bibinfo {author} {\bibfnamefont {C.-Y.}\ \bibnamefont {Park}},
  \bibinfo {author} {\bibfnamefont {K.~C.}\ \bibnamefont {Tan}}, \ and\
  \bibinfo {author} {\bibfnamefont {H.}~\bibnamefont {Jeong}},\ }\href@noop {}
  {\bibfield  {journal} {\bibinfo  {journal} {New Journal of Physics}\ }\textbf
  {\bibinfo {volume} {19}},\ \bibinfo {pages} {043024} (\bibinfo {year}
  {2017})}\BibitemShut {NoStop}%
\bibitem [{\citenamefont {Hill}\ and\ \citenamefont {Wootters}(1997)}]{Hill97}%
  \BibitemOpen
  \bibfield  {author} {\bibinfo {author} {\bibfnamefont {S.}~\bibnamefont
  {Hill}}\ and\ \bibinfo {author} {\bibfnamefont {W.~K.}\ \bibnamefont
  {Wootters}},\ }\href {\doibase 10.1103/PhysRevLett.78.5022} {\bibfield
  {journal} {\bibinfo  {journal} {Phys. Rev. Lett.}\ }\textbf {\bibinfo
  {volume} {78}},\ \bibinfo {pages} {5022} (\bibinfo {year}
  {1997})}\BibitemShut {NoStop}%
\bibitem [{\citenamefont {Wootters}(1998)}]{Wootters98}%
  \BibitemOpen
  \bibfield  {author} {\bibinfo {author} {\bibfnamefont {W.~K.}\ \bibnamefont
  {Wootters}},\ }\href {\doibase 10.1103/PhysRevLett.80.2245} {\bibfield
  {journal} {\bibinfo  {journal} {Phys. Rev. Lett.}\ }\textbf {\bibinfo
  {volume} {80}},\ \bibinfo {pages} {2245} (\bibinfo {year}
  {1998})}\BibitemShut {NoStop}%
\bibitem [{\citenamefont {Takahashi}\ \emph {et~al.}(1999)\citenamefont
  {Takahashi}, \citenamefont {Honda}, \citenamefont {Tanaka}, \citenamefont
  {Toyoda}, \citenamefont {Ishikawa},\ and\ \citenamefont
  {Yabuzaki}}]{Takahashi99}%
  \BibitemOpen
  \bibfield  {author} {\bibinfo {author} {\bibfnamefont {Y.}~\bibnamefont
  {Takahashi}}, \bibinfo {author} {\bibfnamefont {K.}~\bibnamefont {Honda}},
  \bibinfo {author} {\bibfnamefont {N.}~\bibnamefont {Tanaka}}, \bibinfo
  {author} {\bibfnamefont {K.}~\bibnamefont {Toyoda}}, \bibinfo {author}
  {\bibfnamefont {K.}~\bibnamefont {Ishikawa}}, \ and\ \bibinfo {author}
  {\bibfnamefont {T.}~\bibnamefont {Yabuzaki}},\ }\href {\doibase
  10.1103/PhysRevA.60.4974} {\bibfield  {journal} {\bibinfo  {journal} {Phys.
  Rev. A}\ }\textbf {\bibinfo {volume} {60}},\ \bibinfo {pages} {4974}
  (\bibinfo {year} {1999})}\BibitemShut {NoStop}%
\bibitem [{\citenamefont {Riste}\ \emph {et~al.}(2013)\citenamefont {Riste},
  \citenamefont {Dukalski}, \citenamefont {Watson}, \citenamefont {de~Lange},
  \citenamefont {Tiggelman}, \citenamefont {Blanter}, \citenamefont {Lehnert},
  \citenamefont {Schouten},\ and\ \citenamefont {DiCarlo}}]{Riste13}%
  \BibitemOpen
  \bibfield  {author} {\bibinfo {author} {\bibfnamefont {D.}~\bibnamefont
  {Riste}}, \bibinfo {author} {\bibfnamefont {M.}~\bibnamefont {Dukalski}},
  \bibinfo {author} {\bibfnamefont {C.~A.}\ \bibnamefont {Watson}}, \bibinfo
  {author} {\bibfnamefont {G.}~\bibnamefont {de~Lange}}, \bibinfo {author}
  {\bibfnamefont {M.~J.}\ \bibnamefont {Tiggelman}}, \bibinfo {author}
  {\bibfnamefont {Y.~M.}\ \bibnamefont {Blanter}}, \bibinfo {author}
  {\bibfnamefont {K.~W.}\ \bibnamefont {Lehnert}}, \bibinfo {author}
  {\bibfnamefont {R.~N.}\ \bibnamefont {Schouten}}, \ and\ \bibinfo {author}
  {\bibfnamefont {L.}~\bibnamefont {DiCarlo}},\ }\href {\doibase
  10.1038/nature12513} {\bibfield  {journal} {\bibinfo  {journal} {Nature}\
  }\textbf {\bibinfo {volume} {502}},\ \bibinfo {pages} {350} (\bibinfo {year}
  {2013})}\BibitemShut {NoStop}%
\bibitem [{\citenamefont {Roch}\ \emph {et~al.}(2014)\citenamefont {Roch},
  \citenamefont {Schwartz}, \citenamefont {Motzoi}, \citenamefont {Macklin},
  \citenamefont {Vijay}, \citenamefont {Eddins}, \citenamefont {Korotkov},
  \citenamefont {Whaley}, \citenamefont {Sarovar},\ and\ \citenamefont
  {Siddiqi}}]{Roch14}%
  \BibitemOpen
  \bibfield  {author} {\bibinfo {author} {\bibfnamefont {N.}~\bibnamefont
  {Roch}}, \bibinfo {author} {\bibfnamefont {M.~E.}\ \bibnamefont {Schwartz}},
  \bibinfo {author} {\bibfnamefont {F.}~\bibnamefont {Motzoi}}, \bibinfo
  {author} {\bibfnamefont {C.}~\bibnamefont {Macklin}}, \bibinfo {author}
  {\bibfnamefont {R.}~\bibnamefont {Vijay}}, \bibinfo {author} {\bibfnamefont
  {A.~W.}\ \bibnamefont {Eddins}}, \bibinfo {author} {\bibfnamefont {A.~N.}\
  \bibnamefont {Korotkov}}, \bibinfo {author} {\bibfnamefont {K.~B.}\
  \bibnamefont {Whaley}}, \bibinfo {author} {\bibfnamefont {M.}~\bibnamefont
  {Sarovar}}, \ and\ \bibinfo {author} {\bibfnamefont {I.}~\bibnamefont
  {Siddiqi}},\ }\href {\doibase 10.1103/PhysRevLett.112.170501} {\bibfield
  {journal} {\bibinfo  {journal} {Phys. Rev. Lett.}\ }\textbf {\bibinfo
  {volume} {112}},\ \bibinfo {pages} {170501} (\bibinfo {year}
  {2014})}\BibitemShut {NoStop}%
\bibitem [{\citenamefont {McCamey}\ \emph {et~al.}(2009)\citenamefont
  {McCamey}, \citenamefont {van Tol}, \citenamefont {Morley},\ and\
  \citenamefont {Boehme}}]{McCamey09}%
  \BibitemOpen
  \bibfield  {author} {\bibinfo {author} {\bibfnamefont {D.~R.}\ \bibnamefont
  {McCamey}}, \bibinfo {author} {\bibfnamefont {J.}~\bibnamefont {van Tol}},
  \bibinfo {author} {\bibfnamefont {G.~W.}\ \bibnamefont {Morley}}, \ and\
  \bibinfo {author} {\bibfnamefont {C.}~\bibnamefont {Boehme}},\ }\href
  {\doibase 10.1103/PhysRevLett.102.027601} {\bibfield  {journal} {\bibinfo
  {journal} {Phys. Rev. Lett.}\ }\textbf {\bibinfo {volume} {102}},\ \bibinfo
  {pages} {027601} (\bibinfo {year} {2009})}\BibitemShut {NoStop}%
\bibitem [{\citenamefont {Gumann}\ \emph {et~al.}(2014)\citenamefont {Gumann},
  \citenamefont {Patange}, \citenamefont {Ramanathan}, \citenamefont {Haas},
  \citenamefont {Moussa}, \citenamefont {Thewalt}, \citenamefont {Riemann},
  \citenamefont {Abrosimov}, \citenamefont {Becker}, \citenamefont {Pohl},
  \citenamefont {Itoh},\ and\ \citenamefont {Cory}}]{Gumann14}%
  \BibitemOpen
  \bibfield  {author} {\bibinfo {author} {\bibfnamefont {P.}~\bibnamefont
  {Gumann}}, \bibinfo {author} {\bibfnamefont {O.}~\bibnamefont {Patange}},
  \bibinfo {author} {\bibfnamefont {C.}~\bibnamefont {Ramanathan}}, \bibinfo
  {author} {\bibfnamefont {H.}~\bibnamefont {Haas}}, \bibinfo {author}
  {\bibfnamefont {O.}~\bibnamefont {Moussa}}, \bibinfo {author} {\bibfnamefont
  {M.~L.~W.}\ \bibnamefont {Thewalt}}, \bibinfo {author} {\bibfnamefont
  {H.}~\bibnamefont {Riemann}}, \bibinfo {author} {\bibfnamefont {N.~V.}\
  \bibnamefont {Abrosimov}}, \bibinfo {author} {\bibfnamefont {P.}~\bibnamefont
  {Becker}}, \bibinfo {author} {\bibfnamefont {H.-J.}\ \bibnamefont {Pohl}},
  \bibinfo {author} {\bibfnamefont {K.~M.}\ \bibnamefont {Itoh}}, \ and\
  \bibinfo {author} {\bibfnamefont {D.~G.}\ \bibnamefont {Cory}},\ }\href
  {\doibase 10.1103/PhysRevLett.113.267604} {\bibfield  {journal} {\bibinfo
  {journal} {Phys. Rev. Lett.}\ }\textbf {\bibinfo {volume} {113}},\ \bibinfo
  {pages} {267604} (\bibinfo {year} {2014})}\BibitemShut {NoStop}%
\bibitem [{\citenamefont {Sidje}(1998)}]{Sidje98}%
  \BibitemOpen
  \bibfield  {author} {\bibinfo {author} {\bibfnamefont {R.~B.}\ \bibnamefont
  {Sidje}},\ }\href {\doibase 10.1145/285861.285868} {\bibfield  {journal}
  {\bibinfo  {journal} {ACM Trans. Math. Softw.}\ }\textbf {\bibinfo {volume}
  {24}},\ \bibinfo {pages} {130} (\bibinfo {year} {1998})}\BibitemShut
  {NoStop}%
\end{thebibliography}%




\end{document}